\documentclass[aps,prd,twocolumn,showpacs,groupedaddress]{revtex4}

\usepackage{epsfig}
\usepackage{graphicx}% Include figure files

\newcommand{\pt} {\ensuremath{p_{_T}}}
\newcommand{\et} {\ensuremath{E_{T}}}
\newcommand{\WW} {\mbox{$W^{+}W^{-}$}}
\newcommand{\pbp}    {\mbox{$\overline{p}p$}}
\newcommand{\ttb}    {\mbox{$t\overline{t}$}}
\newcommand{\bbb}    {\mbox{$b\overline{b}$}}
\newcommand{\qqb}    {\mbox{$q\overline{q}$}}

\newcommand{\elel}   {\mbox{$e^{+}e^{-}$}}

\newcommand{\afb}    {\mbox{$A_{FB}$}}
\newcommand{\afbr}    {\mbox{$A_{FB}^{raw}$}}
\newcommand{\afbp}    {\mbox{$A_{FB}^{phys}$}}
\newcommand{\cost}    {\mbox{$\cos\theta^{*}$}}

\newcommand{\mee}    {\mbox{$M_{ee}$}}
\newcommand{\met}    {\mbox{$\not\!\!E_{t}$}}

\newcommand{\wgenu}   {\mbox{$W\gamma \rightarrow\ e \nu\gamma$}}

\newcommand{\wenu}   {\mbox{$W \rightarrow\ e \nu$}}

\newcommand{\zee}    {\mbox{$Z \rightarrow\ e^+ e^-$}}
\newcommand{\zgee}   {\mbox{$Z/\gamma^* \rightarrow\ e^+ e^-$}}
\newcommand{\zg}   {\mbox{$Z/\gamma^*$}}
\newcommand{\zgtautau}    {\mbox{$Z/\gamma^* \rightarrow\ \tau^+ \tau^-$}}

\begin{document}

% Use the 'preprintnumbers' class option to override journal defaults
% to display numbers if necessary
%\preprint{FERMILAB-PUB-04-275-E}
%\preprint{CDF-6815}
\clearpage
%\flushright{FERMILAB-PUB-XX-XXX-X}
%\vspace{-0.3cm}
%\flushright{CDF-6815} \\
%\flushleft
\onecolumngrid  
\font\eightit=cmti8
\def\r#1{\ignorespaces $^{#1}$}
\hfilneg
\begin{sloppypar}
\noindent
\noindent
\center{\large \bf{Measurement of the Forward-Backward Charge Asymmetry of Electron-Positron
    Pairs in \pbp~Collisions at $\sqrt{s}=1.96$~TeV \\ }}
\vspace{0.4cm}

D.~Acosta,\r {16} T.~Affolder,\r 9 T.~Akimoto,\r {54}
M.G.~Albrow,\r {15} D.~Ambrose,\r {43} S.~Amerio,\r {42}  
D.~Amidei,\r {33} A.~Anastassov,\r {50} K.~Anikeev,\r {31} A.~Annovi,\r {44}
% ...
\end{sloppypar}
\vskip .026in
\begin{center}
(CDF Collaboration)
\end{center}
\vskip .026in
\begin{center}
\r 1  {\eightit Institute of Physics, Academia Sinica, Taipei, Taiwan 11529, 
Republic of China} \\
\r 2  {\eightit Argonne National Laboratory, Argonne, Illinois 60439} \\
\r 3  {\eightit Institut de Fisica d'Altes Energies, Universitat Autonoma
de Barcelona, E-08193, Bellaterra (Barcelona), Spain} \\
\r 4  {\eightit Istituto Nazionale di Fisica Nucleare, University of Bologna,
I-40127 Bologna, Italy} \\
\r 5  {\eightit Brandeis University, Waltham, Massachusetts 02254} \\
\r 6  {\eightit University of California at Davis, Davis, California  95616} \\
\r 7  {\eightit University of California at Los Angeles, Los 
Angeles, California  90024} \\
\r 8  {\eightit University of California at San Diego, La Jolla, California  92093} \\ 
\r 9  {\eightit University of California at Santa Barbara, Santa Barbara, California 
93106} \\ 
\r {10} {\eightit Instituto de Fisica de Cantabria, CSIC-University of Cantabria, 
39005 Santander, Spain} \\
\r {11} {\eightit Carnegie Mellon University, Pittsburgh, PA  15213} \\
\r {12} {\eightit Enrico Fermi Institute, University of Chicago, Chicago, 
Illinois 60637} \\
\r {13}  {\eightit Joint Institute for Nuclear Research, RU-141980 Dubna, Russia}
\\
\r {14} {\eightit Duke University, Durham, North Carolina  27708} \\
\r {15} {\eightit Fermi National Accelerator Laboratory, Batavia, Illinois 
60510} \\
\r {16} {\eightit University of Florida, Gainesville, Florida  32611} \\
\r {17} {\eightit Laboratori Nazionali di Frascati, Istituto Nazionale di Fisica
               Nucleare, I-00044 Frascati, Italy} \\
\r {18} {\eightit University of Geneva, CH-1211 Geneva 4, Switzerland} \\
\r {19} {\eightit Glasgow University, Glasgow G12 8QQ, United Kingdom}\\
\r {20} {\eightit Harvard University, Cambridge, Massachusetts 02138} \\
\r {21} {\eightit The Helsinki Group: Helsinki Institute of Physics; and Division of
High Energy Physics, Department of Physical Sciences, University of Helsinki, FIN-00044, Helsinki, Finland}\\
\r {22} {\eightit Hiroshima University, Higashi-Hiroshima 724, Japan} \\
\r {23} {\eightit University of Illinois, Urbana, Illinois 61801} \\
\r {24} {\eightit The Johns Hopkins University, Baltimore, Maryland 21218} \\
\r {25} {\eightit Institut f\"{u}r Experimentelle Kernphysik, 
Universit\"{a}t Karlsruhe, 76128 Karlsruhe, Germany} \\
\r {26} {\eightit High Energy Accelerator Research Organization (KEK), Tsukuba, 
Ibaraki 305, Japan} \\
\r {27} {\eightit Center for High Energy Physics: Kyungpook National
University, Taegu 702-701; Seoul National University, Seoul 151-742; and
SungKyunKwan University, Suwon 440-746; Korea} \\
\r {28} {\eightit Ernest Orlando Lawrence Berkeley National Laboratory, 
Berkeley, California 94720} \\
\r {29} {\eightit University of Liverpool, Liverpool L69 7ZE, United Kingdom} \\
\r {30} {\eightit University College London, London WC1E 6BT, United Kingdom} \\
\r {31} {\eightit Massachusetts Institute of Technology, Cambridge,
Massachusetts  02139} \\   
\r {32} {\eightit Institute of Particle Physics: McGill University,
Montr\'{e}al, Canada H3A~2T8; and University of Toronto, Toronto, Canada
M5S~1A7} \\
\r {33} {\eightit University of Michigan, Ann Arbor, Michigan 48109} \\
\r {34} {\eightit Michigan State University, East Lansing, Michigan  48824} \\
\r {35} {\eightit Institution for Theoretical and Experimental Physics, ITEP,
Moscow 117259, Russia} \\
\r {36} {\eightit University of New Mexico, Albuquerque, New Mexico 87131} \\
\r {37} {\eightit Northwestern University, Evanston, Illinois  60208} \\
\r {38} {\eightit The Ohio State University, Columbus, Ohio  43210} \\  
\r {39} {\eightit Okayama University, Okayama 700-8530, Japan}\\  
\r {40} {\eightit Osaka City University, Osaka 588, Japan} \\
\r {41} {\eightit University of Oxford, Oxford OX1 3RH, United Kingdom} \\
\r {42} {\eightit University of Padova, Istituto Nazionale di Fisica 
          Nucleare, Sezione di Padova-Trento, I-35131 Padova, Italy} \\
\r {43} {\eightit University of Pennsylvania, Philadelphia, 
        Pennsylvania 19104} \\   
\r {44} {\eightit Istituto Nazionale di Fisica Nucleare, University and Scuola
               Normale Superiore of Pisa, I-56100 Pisa, Italy} \\
\r {45} {\eightit University of Pittsburgh, Pittsburgh, Pennsylvania 15260} \\
\r {46} {\eightit Purdue University, West Lafayette, Indiana 47907} \\
\r {47} {\eightit University of Rochester, Rochester, New York 14627} \\
\r {48} {\eightit The Rockefeller University, New York, New York 10021} \\
\r {49} {\eightit Istituto Nazionale di Fisica Nucleare, Sezione di Roma 1,
University di Roma ``La Sapienza," I-00185 Roma, Italy}\\
\r {50} {\eightit Rutgers University, Piscataway, New Jersey 08855} \\
\r {51} {\eightit Texas A\&M University, College Station, Texas 77843} \\
\r {52} {\eightit Texas Tech University, Lubbock, Texas 79409} \\
\r {53} {\eightit Istituto Nazionale di Fisica Nucleare, University of Trieste/\
Udine, Italy} \\
\r {54} {\eightit University of Tsukuba, Tsukuba, Ibaraki 305, Japan} \\
\r {55} {\eightit Tufts University, Medford, Massachusetts 02155} \\
\r {56} {\eightit Waseda University, Tokyo 169, Japan} \\
\r {57} {\eightit Wayne State University, Detroit, Michigan  48201} \\
\r {58} {\eightit University of Wisconsin, Madison, Wisconsin 53706} \\
\r {59} {\eightit Yale University, New Haven, Connecticut 06520} \\
\end{center}

\begin{abstract}
We present a measurement of the mass dependence of the forward-backward 
charge asymmetry (\afb) for $e^+e^-$ pairs produced via an intermediate $\zg$ 
with mass $M_{ee} > 40$~GeV/$c^2$.
%The \afb~is measured with and without Standard Model assumptions about the \afb~distribution.  
We study the constraints on the $Z$-quark couplings imposed
by our measurement. We analyze
an integrated luminosity of 72~pb$^{-1}$ collected by the CDF II 
detector in $\pbp$ collisions at $\sqrt s$ = 1.96~TeV at the Fermilab 
Tevatron. A comparison of the uncorrected $\afb$ between data
and Standard Model Monte Carlo gives good agreement with a $\chi^2$/DOF of 15.7/15.  
The couplings measurements are also consistent with Standard Model predictions.
\end{abstract}

% insert suggested PACS numbers in braces on next line
% PACS, the Physics and Astronomy
                             % Classification Scheme.

%\maketitle must follow title, authors, abstract, \pacs, and \keywords
\maketitle

\section{Introduction}\label{s_Intro}

The reaction $\pbp \rightarrow \ell^+\ell^-$, where 
$\ell$ is an isolated high-\pt~lepton, is mediated
primarily by virtual photons at low values of dilepton invariant mass 
($M_{\ell^+\ell^-}$)~\cite{Drell:1970wh,Drell:1970yt},
primarily by the $Z$ at $M_{\ell^+\ell^-}
\sim M_Z$, and by a combination of photons and $Z$ bosons outside these regions.
The presence of 
both vector and axial-vector couplings of electroweak 
bosons to fermions in the process $\qqb \rightarrow Z/\gamma^*
\rightarrow \ell^+\ell^-$
gives rise to an asymmetry in the polar angle of the 
lepton momentum relative to the incoming quark momentum
in the rest frame of the lepton pair.
\par 
At tree level, the process $\qqb \rightarrow \ell^{\rm +}\ell^{\rm -}$ proceeds 
via an $s$-channel exchange of either a virtual photon or a $Z$ boson. The 
neutral current coupling of a fermion $f$ to the $Z$ boson has vector and 
axial-vector components: $J^{Zf} \sim \bar{f}(g_V^f+g_A^f \gamma_5)f$, where 
$g_V^f$ and $g_A^f$ are the vector and axial-vector couplings of the fermion 
to the $Z$ respectively. The coupling of the same fermion to the photon is 
purely a vector coupling and its strength is proportional to the charge of the
fermion $Q_f$. The differential cross section for $\qqb \rightarrow \ell^{\rm +}\ell^{\rm -}$  
is obtained by squaring the matrix element, integrating over the azimuthal 
angle, averaging over the polarization of the incoming particles, and
summing over the spin and polarization of the final state particles:
\begin{widetext}
\begin{eqnarray}
\frac{d\sigma(q\bar{q} \rightarrow \ell^{\rm +}\ell^{\rm -})}{d \cos\theta} 
& = & C\frac{\pi\alpha^2}{2s} [ Q_\ell^2 Q_q^2 (1+\cos^2\theta) \nonumber 
 +   Q_{\ell} Q_q Re(\chi(s)) ( 2 g_V^q g_V^{\ell} (1+\cos^2\theta) +  4 g_A^q g_A^{\ell} \cos\theta) \nonumber \\
& + & |\chi(s)|^2 \left( ({g_V^q}^2+{g_A^q}^2) ({g_V^{\ell}}^2+{g_A^{\ell}}^2) (1+\cos^2\theta)
+ 8 g_V^q g_A^q g_V^{\ell} g_A^{\ell} \cos\theta \right) ]
\label{e_diff_x-section}
\end{eqnarray}
\end{widetext}
where $C$ is the color factor, $\theta$ is the emission angle of the lepton (anti-lepton) relative to 
the quark (anti-quark) in the rest frame of the lepton pair, and 
\begin{eqnarray}
\chi(s) = \frac{1}{\cos^2\theta_W \sin^2\theta_W} \frac{s}{s - M_Z^2 + i \Gamma_Z M_Z}. 
\end{eqnarray}
The first and the third terms in Eq.~(\ref{e_diff_x-section}) correspond 
to the pure $\gamma^*$ and $Z$ exchange respectively while the second term 
corresponds to the $Z/\gamma^* $ interference. The angular dependence of the 
various terms is either $\cos\theta$ or $(1+\cos^2\theta)$.
The $\cos\theta$ terms integrate to zero in the total cross section but induce the 
forward-backward asymmetry.
\par
Let
\begin{eqnarray}
R^{VV} & \equiv & Q_\ell^2 Q_q^2 + 2 Q_{\ell} Q_q  g_V^q g_V^{\ell} Re(\chi(s)) \\
& + & {g_V^{\ell}}^2 ({g_V^q}^2+{g_A^q}^2) |\chi(s)|^2 \nonumber \\
R^{AA} & \equiv & {g_A^{\ell}}^2 ({g_V^q}^2+{g_A^q}^2) |\chi(s)|^2 \\
R^{VA} & \equiv & \frac{3}{2} g_A^q g_A^{\ell} ( Q_{\ell} Q_q Re(\chi(s)) + 2 g_V^q g_V^{\ell} |\chi(s)|^2).
\end{eqnarray}
Then the differential cross section is reduced to the following simple expression:
\begin{eqnarray}
\frac{d\sigma}{d \cos\theta} & = & 
C \frac{4}{3} \frac{\pi \alpha^2}{s}R_f[\frac{3}{8}(1+\cos^2\theta)+A_{FB} \cos\theta] 
\label{e_x-section}
\end{eqnarray}
where $R_f= R^{VV}+R^{AA}$ and $A_{FB} = \frac{R^{VA}}{R_f}$. The meaning of the 
quantities $R_f$ and $A_{FB}$ can be clearly seen. Integrating Eq.~(\ref{e_x-section})
 over 
$\cos\theta$, the first term in the square brackets integrates to unity, the 
second integrates to 0. Therefore $\sigma_{total} = C R_f \sigma_0^{QED}$, 
where $\sigma_0^{QED}$ is the total QED cross section (the cross section if the $Z^0$ 
exchange amplitude were absent). 
The quantity $A_{FB}$ can be written as
\begin{eqnarray}
A_{FB} & = & \frac{\int_0^{+1} \frac{d\sigma}{d \cos\theta}{d \cos\theta} 
+ \int_0^{-1} \frac{d\sigma}{d \cos\theta}{d \cos\theta}}
{\int_{-1}^{+1}\frac{d\sigma}{d \cos\theta}{d \cos\theta}} \nonumber\\
&=& \frac{\sigma_F - \sigma_B}{\sigma_F + \sigma_B} \nonumber \\
&=& \frac{N^F - N^B}{N^F + N^B} 
\label{e_afbdef}
\end{eqnarray}
and is identified as the forward-backward asymmetry, where
$N^F$ is the number of forward ($\cos\theta>0$) events and 
$N^B$ is the number of backward ($\cos\theta<0$) events.

\par

A measurement of $A_{FB}$ can constrain the properties 
of any additional non--Standard Model amplitudes contributing 
to $q\bar q \rightarrow \ell^+\ell^-$~\cite{Rosner:1995ft}, and is complementary 
to direct searches for non--Standard Model amplitudes that look for an excess
in the total cross section. 
This is particularly interesting for $M_{ee}$ above LEP II energies,
where the 
measurement is unique to the Tevatron. 

In addition, Eq.~(\ref{e_diff_x-section}) shows that depending on 
the invariant mass, a different combination of vector and axial-vector 
couplings contribute to the
differential cross section. Consequently,
\afb~is a direct probe of the
relative strengths of the coupling constants between the $Z$ boson and the 
quarks. The invariant-mass dependence of $\afb$ is also sensitive to $u$ and $d$ quarks
separately, unlike other precise measurements of light quark $Z$ couplings
in $\nu N$ scattering~\cite{Zeller:2001hh} and atomic parity violation~\cite{Bennett:1999pd}
on heavy nuclei.

	In this paper, a number of different comparisons between the data and
Standard Model expectations are presented. The uncorrected
$\mee$, $\cost$, and $\afb$ data distributions are compared with the output of our Monte
Carlo and detector simulation. The first principal result is a measurement 
of $\afb$ in 15 $\mee$ bins using an unfolding analysis that doesn't 
assume a prior Standard Model $\afb$ distribution. The second principal 
result is a measurement of three sets of parameters:  the $Z$-quark couplings, 
the $Z$-electron couplings, and $\sin^2\theta_W$. In making each of 
these three measurements, the other parameters are held fixed with 
the values given by the Standard Model.  
The measured $Z$-quark couplings are then used to determine
experimental correction factors for acceptance and efficiency of 
dielectron events. The correction factors are used for the 
measurement of $\afb$ with Standard Model assumptions.

	The previous measurement from the Collider Detector at Fermilab 
(CDF)~\cite{Affolder:2001ha} 
was made with the data taken between 1992 and 1995 using the Run I detector. The
Run I measurement assumed a Standard Model \afb~distribution for calculating 
efficiency and acceptance experimental correction factors.
The present measurement uses the dielectron data taken between March 2002 and January 
2003 with the CDF II detector, corresponding to an integrated luminosity 
of 72 pb$^{-1}$.

The paper is structured as follows. 
A description of the detector and an overview of the 
analysis are given in Sec.~\ref{s_overview}.  Event selection and candidate events are discussed in 
Sec.~\ref{s_ElectronMeasurement}.  The estimation and characteristics of the backgrounds are described
in Sec.~\ref{s_backgrounds}. The acceptance and corrections for detector effects 
are described in Sec.~\ref{s_acceptance}. The systematic uncertainties are summarized
in Sec.~\ref{s_sys}.  Finally, the results of the forward-backward 
asymmetry and coupling measurements are presented in Sec.~\ref{s_results}.

%
%***********************************************************************************
\section{Overview}
\label{s_overview}
	This section begins with a discussion of aspects of the detector, 
triggers and data samples that are relevant to this measurement. The nature of 
$\zgee$ events in a hadron collider and the overall strategy of the analysis are presented. 
This paper uses a cylindrical coordinate system, with the positive $z$ axis oriented along 
the beamline in the direction of the proton's momentum.

%***********************************************************************************
\subsection{Detector and triggers}
\begin{figure}
\begin{center}
\includegraphics[width=8cm]{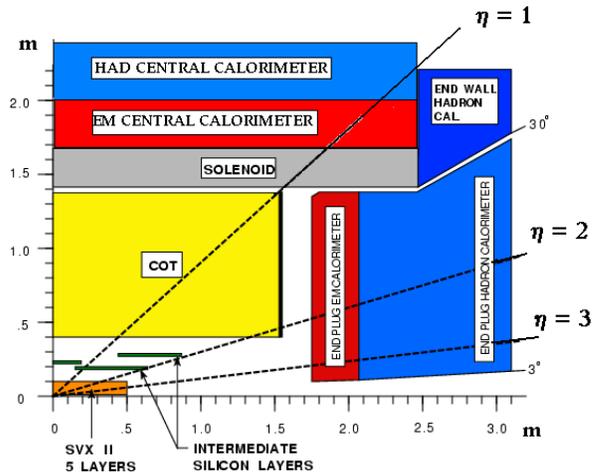}
\caption{\emph{One quadrant of the CDF II tracking and calorimetric detectors.  The detectors 
have axial and reflective symmetry about z=0.
CDF uses a cylindrical coordinate system with the $z$ (longitudinal) axis
along the proton-beam direction; $r$ is the transverse coordinate, and $\phi$ is the azimuthal
angle.  
The detector pseudorapidity is defined as 
$\eta_{det} \equiv {\rm -ln(tan}{\theta_{det} \over 2})$, where
$\theta_{det}$ is the polar angle relative to the proton-beam direction
measured from $z = 0$. 
The event pseudorapidity is defined as 
$\eta_{evt} \equiv {\rm -ln(tan}{\theta_{evt} \over 2})$, where
$\theta_{evt}$ is the polar angle measured from the nominal \pbp~collision point in $z$.
The transverse momentum (\pt) 
and energy (\et) are the components projected onto the plane perpendicular to the beam axis 
($\pt \equiv p\cdot\sin\theta$ ; $\et \equiv E\cdot\sin\theta$). 
The missing transverse energy, $\met$,
is defined as the magnitude of $-\Sigma_i E^i_T \hat{n}_i$, where $\hat{n}_i$ is a 
unit vector in the perpendicular plane that points from the beamline to the $i$th 
calorimeter tower.}}
\label{f_detector}
\end{center}
\end{figure}

	The Collider Detector at Fermilab (CDF II) is a general-purpose detector designed 
to study the physics of~$\pbp$ collisions at $\sqrt s=1.96$~TeV at the Fermilab Tevatron 
collider. Like most detectors used at high-energy colliders, it has a cylindrical geometry 
with axial and forward-backward symmetry. A diagram of the inner part of the CDF~II 
detector is shown in Fig.~\ref{f_detector}. The innermost part of the detector contains 
an integrated tracking system with a silicon detector and an open-cell drift chamber.
A solenoidal magnet surrounding the tracking chambers provides a 1.4~T field aligned
with the proton beam axis. The integrated tracking system is 
surrounded by calorimeters which cover $2\pi$ in azimuth and from $-$3.6 to 3.6 in 
pseudorapidity, $\eta_{det}$ (see Fig.~\ref{f_detector}). Outside of the calorimeters is a muon 
system with coverage from $-1.5$ to 1.5 in $\eta_{det}$. The CDF II detector is a major 
upgrade to the detector that took data until 1996. The entire tracking system subtending 
$|\eta_{det}| < 2$ and the plug calorimeter subtending $1.1 < |\eta_{det}| < 3.6$ have been 
replaced to handle the higher rate of collisions and increase the capabilities for physics 
analyses in Run~II. This analysis uses the open-cell drift chamber called the central outer tracker 
(COT) and the calorimeters.  A more detailed 
detector description can be found in Refs. \cite{Abe:1988me,Abe:1994st},
and a description of the upgraded detector can be found in Ref. \cite{Blair:1996kx}.

\begin{figure}
\begin{center}
\includegraphics[width=6.0cm]{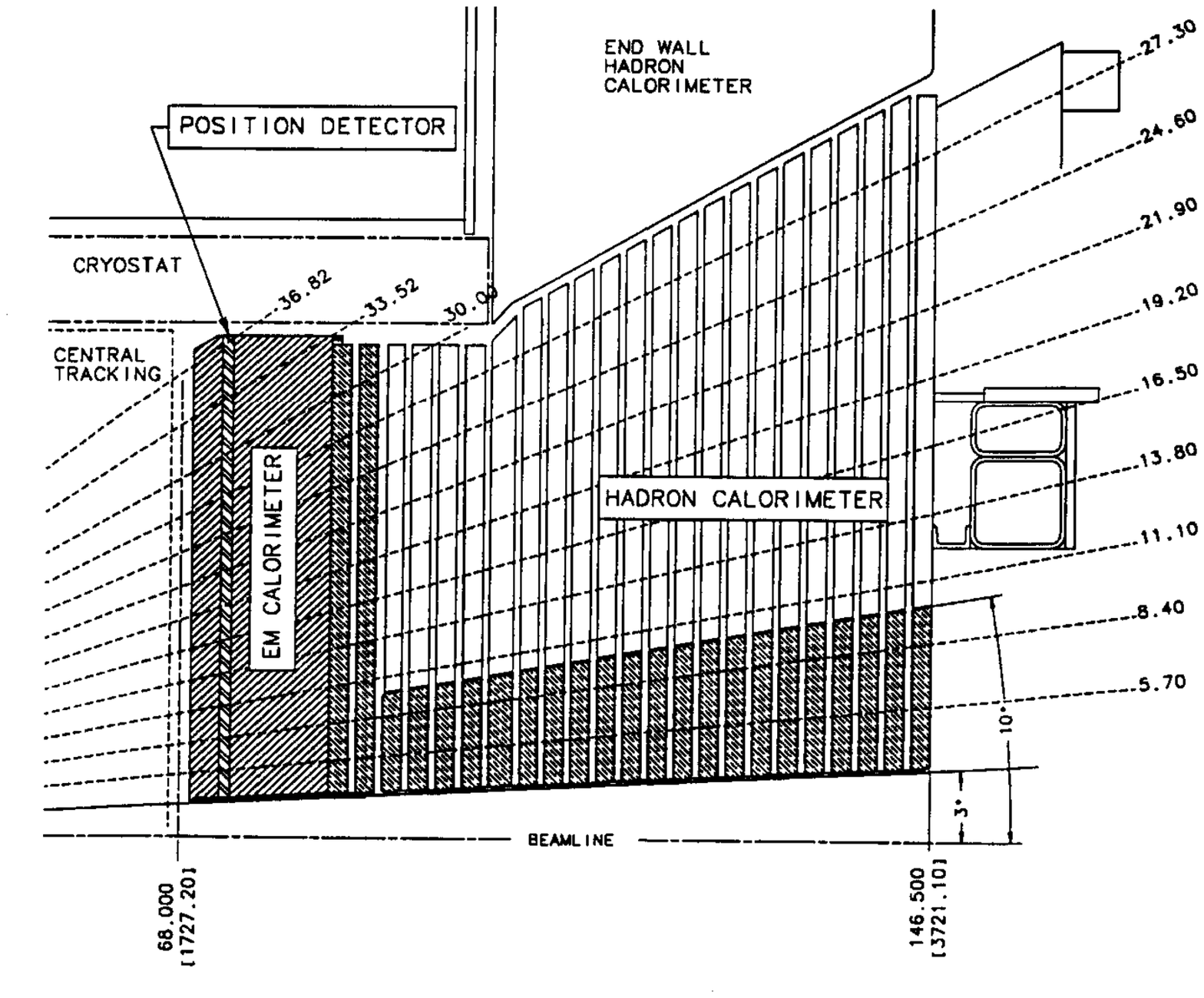}
\includegraphics[width=6.0cm]{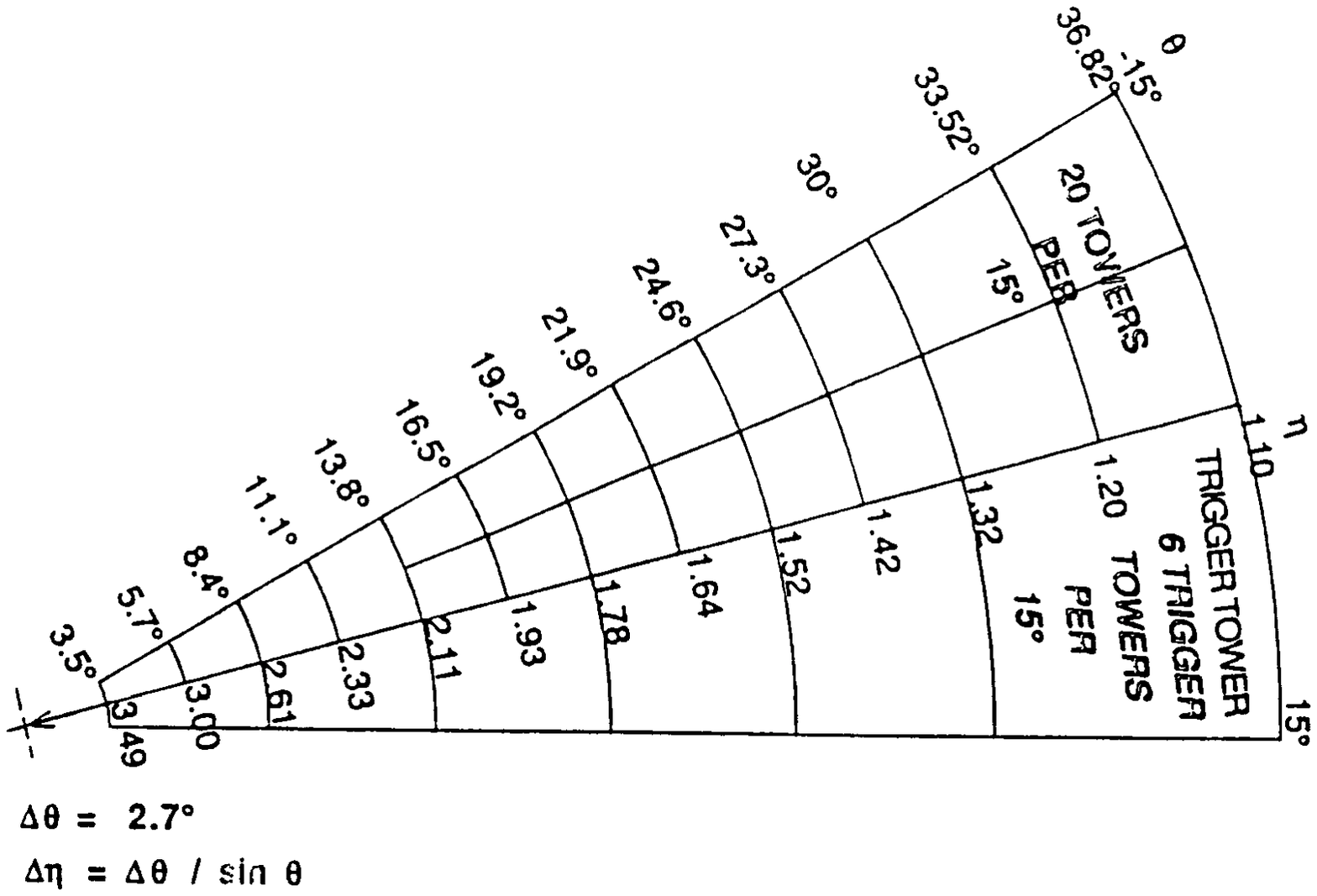}
\caption{\emph{Cross section of upper part of plug calorimeter (top),
	  and transverse segmentation, showing physical and trigger towers in 
	  a $30^{\circ}~\phi$ section (bottom).
	  The logical segmentation for
	  clustering purposes is the same except in the outer two rings
	  ($\theta>30^{\circ}$), where two neighboring (in azimuth) 
	  $7.5^{\circ}$ towers are
	  merged to match the $15^{\circ}$ segmentation of the central and wall
	  calorimeters behind them.}}
\label{f_pluggeom}
\end{center}
\end{figure}

The COT detector~\cite{Affolder:2003ep} is a 96 layer, 3.2~m long open-cell drift 
chamber which, combined with the solenoid, is used to measure the momenta of 
charged particles with $|\eta_{det}| < 1$. The detector extends from a radius of 
40~cm to a radius of 137~cm. The 96 layers are divided into 8 ``super-layers'', 
which alternate between super-layers where the wires are axial (i.e., parallel to the $z$
axis) and super-layers where the wires have a $\pm2^{\circ}$ stereo angle, 
providing three-dimensional tracking. 
The calorimeter consists of a lead-scintillator electromagnetic (EM) compartment 
with shower position detection backed by an iron-scintillator hadronic compartment. The 
calorimeters are segmented in projective $\eta_{det}$--$\phi$ towers pointing to the nominal 
interaction point, at $z=0$. While the central calorimeter ($|\eta_{det}|  < 1.1$) 
is retained mostly unchanged from 
Run I \cite{Balka:1987ty,Hahn:1987tx,Yasuoka:1987ar,Wagner:1987mf,Devlin:1987mf,Bertolucci:1987zn}, the plug calorimeter~\cite{deBarbaro:1995ci} with $1.1 < |\eta_{det}| < 3.6$
is a major component of the Run~II upgrade (Fig.~\ref{f_pluggeom}), and largely 
follows the design of the central detector.
Since the calorimeter segmentation is coarse compared to the dimensions of an 
electron shower, position detectors (shower maximum detectors, CES in the central
region and PES in the plug region) are placed at a depth of approximately 
six radiation lengths, roughly the position of the shower maximum, 
inside the EM calorimeters.  These detectors measure the position and profile of the 
showers and help differentiate electrons from hadrons.

The trigger system has undergone a complete redesign as a result 
of the accelerator and detector upgrades.  The CDF trigger is a three-level 
system that selects events out of a 2.5~MHz crossing rate to be written 
to magnetic tape at a rate of $\sim$75~Hz.  
The first two levels~\cite{Winer:2001gj} are composed of 
custom electronics with data paths separate from the data acquisition 
system.  The third level~\cite{Anikeev:2001pc} receives the complete detector information from the 
data acquisition system and runs a version of the reconstruction software optimized 
for speed on a farm of commercial computers.  The $Z/\gamma^*$ events used 
in this measurement are selected by high-$E_T$ central electron triggers.  
At Level~1, electrons are selected by the presence of an energy deposition
of $E_T > 8$~GeV in an EM calorimeter tower and a matching two-dimensional 
($r$--$\phi$ plane) track with $p_T > 8$~GeV/$c$ reconstructed in the COT by the 
eXtremely Fast Tracker (XFT)~\cite{Thomson:2002xp}.
At Level~2, the electron energy is reconstructed in a cluster of EM towers 
and is required to have $E_T > 16$~GeV.  
At Level~3, a reconstructed EM cluster with $E_T > 18$~GeV 
and a matching three-dimensional track with $p_T > 9$~GeV/$c$ are required.  
At each level, the energy in the hadronic towers just behind the EM tower or cluster 
is required to be less than 12.5\% of the EM energy.  Another trigger path that 
does not require a hadronic energy fraction is added to this measurement in 
order to improve the trigger efficiency for central electrons with very high $E_T$.  
The integrated luminosity of the data sample for this analysis is 
($72\pm 4$)~pb$^{-1}$.

%***********************************************************************************
\subsection{Data samples}
\label{s_DataSamples}
Four data samples are employed in this analysis.  These are described briefly below 
and in more detail in subsequent sections.

\begin{itemize}
\item {\it $Z/\gamma^* \rightarrow e^+e^-$ sample:} A sample of 5,200 dielectron 
candidates is used to measure $A_{FB}$, calibrate the energy scale 
and resolution of the EM calorimeter, and study the material in the tracking volume.

\item {\it $W\rightarrow e\nu$ sample:} A sample of 38,000 $W\rightarrow e\nu$ 
candidates, where the electron is reconstructed in the central calorimeter, 
is used to study the material in the tracking volume, to calibrate the 
relative calorimeter response within a central tower, and to check charge biases in 
measuring electrons.

\item	{\it Inclusive electron sample:} A sample of 3 million central electron 
candidates with $E_T>8$ GeV is used to calibrate the relative response of the 
central EM calorimeter towers. 

\item	{\it Dijet samples:} A sample of 1 million dijet events (events with at 
least two jets, each with $E_T> 20$ GeV) is used to measure the rate at which 
jets fake an electron signature and to estimate the dijet background. 
A jet is defined as a cluster of energy reconstructed in the calorimeter.
Triggers for the sample require a calorimeter tower with $E_T > 5$~GeV at 
Level~1, a calorimeter cluster with $E_T > 15$~GeV at Level~2, and a reconstructed jet  
with $E_T > 20$~GeV at Level~3.  Due to the high cross section, only 1 in roughly 400 events 
on average were randomly selected to be recorded for this trigger.  Jet samples with 
higher trigger thresholds (50 GeV, 70 GeV, and 100 GeV at Level~3) are also used to 
cross-check the fake rate for a trigger bias.
\end{itemize}

%***********************************************************************************
\subsection{Monte Carlo samples}
\label{s_mc}

	Monte Carlo generation and detector simulation are used to measure the 
acceptance for the Drell-Yan process, model the effect of radiation
and detector resolution, determine the characteristics 
and amount of background in the data sample, and understand 
systematic uncertainties on the $\afb$ measurement.  
PYTHIA~\cite{Sjostrand:2000wi} and HERWIG~\cite{Corcella:2000bw} generators with
CTEQ5L PDFs~\cite{Lai:1999wy}  are used for most of the samples. These generate 
processes at leading order and incorporate initial- and final-state QCD and 
QED radiation via their parton shower algorithms (HERWIG does not include final-state QED radiation).
PYTHIA is tuned so that the underlying event (the remaining \pbp~fragments
from the collision) and the $\pt$ spectrum of $Z$ bosons 
agree with the CDF data~\cite{Affolder:1999jh}.
Two matrix element generators, WGAMMA~\cite{Baur:1989gk} and
ALPGEN~\cite{Mangano:2002ea}, are used to check the $W + X \rightarrow e\nu + X$ background 
estimate from PYTHIA, where $X$ is a photon or hadronic jet. 
The generator WGAMMA calculates the cross section of the $\pbp \rightarrow W\gamma$ 
process.  It uses electroweak helicity amplitudes for $W\gamma$ production and 
radiative $W$ boson decays, including all interference terms. ALPGEN
performs the calculation of the matrix element for the
production of $W$+quark and $W$+gluon final states.
The detector simulation models the decay of generated particles and 
their interactions with the various elements of the CDF detector. A
full GEANT3~\cite{Agostinelli:2002hh} simulation is used to simulate the tracking
volume. A parameterized simulation is used for the calorimeters and
muon chambers.
Comparisons between the data and Monte 
Carlo simulation are discussed in Sec.~\ref{s_simulation}.

There are nine Monte Carlo samples used in this analysis, which are briefly described below.

\begin{itemize}
\item	{\it $\zgee$ sample:} A sample generated with PYTHIA is used 
to calculate corrections due to acceptance, bremsstrahlung, and energy resolution, and 
to estimate the systematic uncertainties due to the energy scale and resolution. 
A quarter of the sample was generated with 
$M_{Z/\gamma*} > 105$~GeV/$c^2$ to reduce the statistical uncertainties associated with 
the Monte Carlo sample in the high-mass region.

\item	{\it $\zgee$ sample for material systematics:} Three PYTHIA samples 
are used to estimate the change in the measured $\afb$ between the default simulation and 
a modified simulation which adds or subtracts $1.5\%$ 
radiation length ($X_0$) of copper in a cylinder in the central region and ${1 \over 6} X_0$ of 
iron on the face of the plug calorimeter. QCD fragmentation is turned off for these samples in
order to save CPU time.
	
\item	{\it $\zgtautau$ sample:} A PYTHIA sample is 
used to estimate the background due to $\zgtautau$. TAUOLA~\cite{Was:2000st} is used
to decay $\tau$'s.

\item	{\it Dijet sample:} A PYTHIA sample with all $2 \rightarrow 2$ processes
is used to understand the characteristics of the dijet background.  
A lower limit of $\pt > 18$~GeV on the transverse momentum 
in the rest frame of the hard interaction is applied.

\item	{\it $\ttb$ sample:} A HERWIG sample is used 
to estimate the background due to $\ttb$ production.

\item	{\it Diboson samples:} A sample with $WW$ production 
and a sample with $WZ$ production are generated using PYTHIA 
and used to estimate the diboson backgrounds.

\item	{\it $\wenu$ sample:} A PYTHIA sample is 
used to estimate the background due to the inclusive $W$ production.

\item	{\it $\wgenu$ sample:} A WGAMMA sample is 
used to cross-check the background due to $W+\gamma$ production.

\item	{\it $W + q/g \rightarrow e\nu + q/g$ sample:} An ALPGEN sample is 
used to estimate the background due to $W$ + quark or gluon production.
\end{itemize}

%***********************************************************************************
\subsection{Strategy of analysis}
\label{s_strategy}
This analysis focuses on $e^+ e^-$ pair production via an intermediate \zg.
The goal of this 
analysis is to measure $A_{FB}$ as a function of the invariant mass of the 
dielectron pair. The dielectron sample is chosen because of the low backgrounds and 
the good polar angle coverage of electrons in the CDF II detector.

\begin{figure}
\begin{center}
\includegraphics[width=5cm]{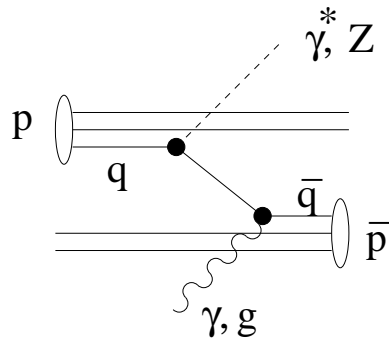}
\caption{\emph{Diagram of $\pbp \rightarrow \zg$, where one of the
quarks radiates a gluon or photon imparting transverse momentum to the quark.
Once the quarks annihilate to $\zg$, the source of the transverse momentum is 
ambiguous.}}
\label{f_Zpt_diagram}
\end{center}
\end{figure}

	When the incoming quarks participating 
in the Drell-Yan process have no transverse momentum relative to their parent baryons, 
$\theta$ in Eq.~(\ref{e_x-section}) is determined unambiguously from 
the four-momenta of the 
electrons by calculating the angle that the electron makes with the proton 
beam in the center-of-mass frame of the electron-positron pair.  
When either of the incoming quarks has significant transverse momentum, however, 
there exists an ambiguity in the four-momenta of the incoming quarks in the frame 
of the electron-positron pair, since one cannot determine the four-momenta
of the quark and anti-quark individually (see Fig.~\ref{f_Zpt_diagram}).
The Collins-Soper formalism~\cite{Collins:1977iv} is adopted to minimize 
the effects of the transverse momentum of the incoming quarks in 
Eq.~(\ref{e_x-section}). (The magnitude of the effect in the Collins-Soper
frame is discussed in Sec.~\ref{s_theory_band}.)
In this formalism, the polar axis is defined as the bisector of the proton
beam momentum and the negative of the anti-proton beam momentum when
they are boosted into the center-of-mass frame of the electron-positron pair.
The variable $\theta^*$ is defined as the angle between the electron and
the polar axis.
Let $Q$ ($Q_T$) be the four-momentum (transverse momentum) 
of the electron-positron pair,
$P_1$ be the four-momentum of the electron, and $P_2$ be the 
four-momentum of the positron, all measured in the lab frame.
Then $\cost$ is given by 
\begin{eqnarray}
\cost = {2 \over \sqrt{Q^2(Q^2 + Q_T^2)}}
        ~(P_1^+ P_2^- - P_1^- P_2^+),
\label{e_collinSoper}
\end{eqnarray}
where $P_i^\pm = {1\over \sqrt 2} (P_i^0 \pm P_i^3)$, and $P^0$ and $P^3$ 
represent the energy and the longitudinal component of the momentum, respectively.
Forward events are defined as having $\cost>0$ and backward events are defined
as having $\cost<0$.

	The $\afb$ measurement is made in 15 bins of $\mee$ between 40 GeV/$c^2$ 
and 600 GeV/$c^2$. The bin sizes
are chosen based on the detector $\mee$ resolution and the relative Drell-Yan 
cross sections in each bin. Once the event selection has been made, 
the estimated number of forward and backward background 
events are subtracted from the candidate sample in each $\mee$ bin. The raw 
forward-backward asymmetry (\afbr) is calculated by applying 
Eq.~(\ref{e_afbdef}) to the background subtracted sample. 

	The goal is to connect what we measure in the detector after 
background subtraction ($\afbr$) to theoretical calculations ($\afbp$). If
the detector simulation is available, a direct comparison can be made
between the uncorrected data and a model prediction. These comparisons
are made in Sec.~\ref{s_rawAfb}. If the simulation is not available,
the true $\afbp$ must be unfolded from $\afbr$. $\afbr$ must
be corrected for detector acceptances and efficiencies (which sculpt the 
$\cost$ distribution) and for smearing effects (which move events 
between $M_{ee}$ bins) to obtain \afbp, which can be compared directly to 
theoretical calculations.
The $\mee$ bin migration near the $Z$ pole is not negligible, which
makes it difficult to unfold \afbp~in those bins. Performing an
unconstrained unfolding analysis in those bins results in large uncertainties,
while applying Standard Model or other constraints can bias the result. 
Two different unfolding analyses 
are performed to reconstruct \afbp~with different constraints. Both unfolding
analyses assume the Standard Model $\frac{d\sigma}{dM_{ee}}$ to be able to calculate 
the event migration effects. The uncertainties associated with systematic effects are 
estimated by varying the magnitude of these effects in the Monte Carlo simulation 
and recalculating the results for each analysis.

	The first and least constrained analysis is performed by fitting for $\afbp$
with a smoothing constraint. The acceptance and event migration are
parameterized (Sec.~\ref{s_smear}) to transform $\afbp$ into $\afbr$, 
\begin{eqnarray}
(\afbr)_i &=& g({\textbf A},i), \\
\mathrm{where}~{\textbf A}&=&\{(\afbp)_1,(\afbp)_2,\ldots,(\afbp)_{15}\}.
\end{eqnarray}
The probability, for a given $\afb$, to find a number of forward and 
backward events is the binomial probability ($P[\afb,N^F,N^B]$). The maximum likelihood
method is used to obtain the 15 values of $\afbp$ or $\textbf{A}$ (Sec.~\ref{s_fitAfb}), where the
negative log likelihood with a smoothing constraint for bins near the $Z$ pole 
is defined as
\begin{eqnarray}
\alpha  = {\sum_{i=1}^{15}-log(P[g({\textbf A},i),N^F_i,N^B_i])} 
        + {\lambda\cdot\sum_{i=2}^{9}S[{\textbf A}_i]}, \\
\mathrm{where}~{\textbf A}_i = \{(\afbp)_{i-1},(\afbp)_i,(\afbp)_{i+1}\},
\label{e_loglikelihood}
\end{eqnarray}
$S[{\textbf A}_i]$ is a regularization function, and $\lambda$ is
the regularization parameter. This analysis makes no 
assumptions about the shape of $\afb$ (aside from it being smooth near the $Z$ pole), 
but it has the largest uncertainties.

	In a second analysis, the parameterized acceptance and 
event migration are also used for the
measurement of the $Z$-quark and $Z$-electron coupling
constants, and $\sin^2\theta_W^{eff}$. We vary the couplings at generator 
level and perform a $\chi^2$ fit between the smeared theoretical 
calculations and $\afbr$ to extract the couplings. This analysis is 
described in detail in Sec.~\ref{s_afb_couplings}. 

	Continuing with the second analysis, $\afb$ can be measured 
by assuming the Standard Model $\afb$ shape with
the measured $Z$-quark couplings. In this method, acceptance correction factors 
($a_{cor}^\pm$) are used to 
correct the number of forward and backward events in each $\mee$
bin. The correction factor accounts for the effects of detector acceptance
and efficiency, and bin migration (Sec.~\ref{s_corrfac}). These correction 
factors depend on the input 
$\afbp$ that is used to calculate them. To constrain the Standard Model bias, 
the input $\afbp$ is constrained to the prediction with uncertainties 
from the $Z$-quark coupling analysis. Then $\afbp$ can be 
calculated directly using the correction factors 
(Sec.~\ref{s_afb_measurement}). Equation~(\ref{e_afbdef}) can be rewritten:
\begin{eqnarray}
A_{FB}^i  & = & {\left(\frac{d\sigma}{dM}\right)^{+}_{i} - 
		 \left(\frac{d\sigma}{dM}\right)^{-}_{i} \over
	         \left(\frac{d\sigma}{dM}\right)^{+}_{i} + 	
		 \left(\frac{d\sigma}{dM}\right)^{-}_{i} } ,
\end{eqnarray}
where
\begin{eqnarray}
\left(\frac{d\sigma}{dM}\right)^{\pm}_{i} & = & 
	{\frac{(N^{\pm}_{i} - B^{\pm}_{i})}
		{\Delta M_{i}L(a_{cor}^{\pm})_{i}} }.
\label{e_dsigma_dm}
\end{eqnarray}
In Eq.~(\ref{e_dsigma_dm}), $\Delta M_i$ is the size of the {\it i}-th mass bin, 
and $L$ is the integrated luminosity. 
$N^+ (N^-)$ and $B^+(B^-)$ are the number of candidate and 
estimated background events in the forward (backward) region, respectively. 
Canceling common factors, the forward-backward asymmetry can be written as
 \begin{eqnarray}
A_{FB}^i  & = & 
	{\frac{\frac{N^{+}_i - B^{+}_i}{(a_{cor}^{+})_{i}} 
	         - \frac{N^{-}_i - B^{-}_i}{(a_{cor}^{-})_{i}}}
              {\frac{N^{+}_i - B^{+}_i}{(a_{cor}^{+})_{i}} 
	         + \frac{N^{-}_i - B^{-}_i}{(a_{cor}^{-})_{i}}}}.
\label{e_afbcalc}		
\end{eqnarray}
The analysis using correction factors has smaller uncertainties than
the unconstrained method, but it assumes no non-standard physics 
outside of deviations in $Z$-quark couplings. This analysis technique
is similar to the Run I measurement.

	While these analyses make different constraints on the form 
of $\afb$, it is important to note that they should give similar results
in bins where there is a negligible amount of event migration between $\mee$
bins. At high mass, where the bin sizes are very large compared to the
detector resolution, the results are independent of the unfolding method.

%***********************************************************************************
%***********************************************************************************
\section{Electron measurement}
\label{s_ElectronMeasurement}
%***********************************************************************************
This analysis requires two electrons ($e^+,e^-$) in the event, one in the central 
region, and the other in the central or plug region.  This section describes the 
identification of central and plug electrons, the event selection criteria, the 
electron energy scale and resolution, and the charge identification of electrons. In most 
cases, discussion of electrons refer to both electrons and positrons.

\subsection{Central electron identification}
\label{s_CentEleID}
	Electron identification in the central calorimeter is almost identical 
to the algorithm used in Run I, since the calorimeter is unchanged and the
new drift chamber has a geometry very similar to the previous one. For a more
detailed description of the central electron reconstruction variables see 
Ref.~\cite{Abe:1994st}. An electron candidate is reconstructed if there is a 
central tower with $E_T > 2$~GeV and a charged drift chamber track that 
extrapolates to the tower. 
The adjacent towers on either side in $\eta_{det}$ are added to the cluster, 
and the cluster is not accepted if the energy in the 
hadronic part is more than 12.5\% of the energy in the EM part.
An electron is considered within the fiducial region of the detector if 
its track points within 60~cm in $z$ of the center of the detector and 
extrapolates to the calorimeter away from any wedge boundaries. (A
wedge consists of those towers with the same value of $\phi$.) 
The polar range of electrons in the central region is $|\eta_{det}| < 1.0$. 
The energy of the electron is determined by the total energy 
it deposits in the EM calorimeter.  The momentum (\pt) of the electron is determined by
the highest-$\pt$ COT track associated with the EM cluster.  The track is constrained 
to the position of the beamline in $r$--$\phi$. The direction of the electron's momentum is taken
from the direction of the track and is used in the calculations of the 
transverse component of the energy ($E_T$) and the invariant mass of the electron pairs.
The charge of the electron ($q$) is determined from the curvature of the track. 
The variables that are used to discriminate electrons from hadrons are:
(1) the ratio of the hadronic energy to the electromagnetic energy ($E_{had}/E_{em}$); 
(2) the total transverse energy within a radius of 0.4 in 
$\Delta R=\sqrt{\Delta\eta_{evt}^2+\Delta\phi^2}$ of the cluster centroid, 
excluding the cluster energy itself ($E_T^{iso}$);(3) the ratio of the calorimeter 
energy to the momentum of the track ($E/p$); (4)  the comparison of the 
lateral sharing of energy among the calorimeter towers with that of
test beam electron data ($L_{shr}$); (6) a $\chi^2$ comparison of the shower 
profile measured by the
shower maximum detector with the shower profile measured from test beam
electrons ($\chi^2_{strip}$); (5)  the distance in $r$--$\phi$ and $z$ 
between the electron shower position measured by the shower maximum 
detector and the extrapolated track position ($q \cdot \Delta x$ and $\Delta z$). 
An asymmetric cut is made on $q \cdot \Delta x$ 
because bremsstrahlung distorts the shower shape in the $r$--$\phi$ direction. Since the
magnetic field bends an electron's trajectory, but not a photon's, the 
bremsstrahlung photons tend to enter the calorimeter 
to the side of the primary electron opposite the bending, which is determined by
the electron's charge.  By multiplying the charge by $\Delta x$, most of
the distortion from bremsstrahlung photons is isolated to 
$q \cdot \Delta x < 0$.

%
%***********************************************************************************
\subsection{Plug electron identification}
	The electron clusters in the plug region, subtending $ 1.2 < |\eta_{det}| < 3.0$, 
are limited to $2 \times 2$ towers (two towers in pseudorapidity by two towers in azimuth).
Since the Moliere radius of a typical electron shower is significantly smaller than the size of 
the plug EM towers, the clusters fully contain electron energies.  As with the central
clusters, plug electron clusters are accepted if $E_{had}/E_{em} < 12.5\%$. The major 
difference between central and plug electrons is the tracking.  In the central region, 
the COT tracking is very efficient 
($99.6\%$), whereas in the plug region the efficiency rapidly falls off as $|\eta_{det}|$ increases 
due to the geometrical acceptance of the COT.   In this analysis no tracking is used for
plug electrons.  The $z$ position of the collision for the event ($z_{vertex}$) is provided by 
the $z$ position of the central electron's track.
The electron's shower centroid is determined 
from a fit of the energy distribution among the calorimeter towers. 
The direction of the plug electron is determined by extrapolating from $z_{vertex}$ 
to the shower centroid.
The unmeasured charge of the plug electron is assumed to be the opposite of
the central electron. The following variables are used to discriminate
electrons from hadrons in the plug region 
: (1) the ratio of the 
hadronic energy to the EM energy ($E_{had}/E_{em}$); (2) the total 
transverse energy within a radius of 0.4 in 
$\Delta R=\sqrt{\Delta\eta_{evt}^2+\Delta\phi^2}$ of the cluster centroid, 
excluding the cluster energy itself ($E_T^{iso}$); (3) a $\chi^2$ comparison
of the energy distribution in $3 \times 3$ EM towers around the seed tower
to the energy distributions from test beam electrons 
($PEM \chi_{3\times 3}^2$).  The selection criteria for plug electrons are summarized 
in Table~\ref{t_electron}.

%***********************************************************************************
\subsection{Event selection}
\label{s_event_sel}
$\zgee$ candidate events are required to have two electrons with $E_{T} > 20$ GeV. 
One of the electrons is required to be in the central region and to pass the full set 
of identification cuts (see Table~\ref{t_electron}).  This electron
is called the central-tight electron. The second loose electron is allowed
to be in either the central or plug region and has relaxed identification cuts for 
higher efficiency (see Table~\ref{t_electron}). 
Based on these selection criteria, two topologies are defined for dielectron events: 
central-central topology, where one central-tight and one central-loose electron 
are required, and central-plug topology, where one central-tight electron and one 
plug-loose electron are required. In the central-central 
topology, the two electrons are required to have opposite charge. In the central-plug 
topology, no charge requirement is made since the plug electron's charge is not measured.

\begin{table}
\begin{center}
\begin{tabular}{|l||l|l|l|}\hline
Variable 	    & Central-Tight         & Central-Loose     & Plug-Loose \\ \hline
$E_T$ 		    & $> 20$ GeV 	        	&  $> 20$ GeV 	        	& $> 20$ GeV \\ \hline
$p_T$ 		    & $>10$ GeV 		&  $>10$ GeV     	& N/A \\ \hline
%$E_{had}/E_{em}$& $< 0.055 + 0.00045*E$     	&  $< 0.055 + 0.00045*E$ & $< 0.05$\\
$E_{had}/E_{em}$& $< 0.055$     	&  $< 0.055$     & $< 0.05$\\
                & $~~+ 0.00045*E$     	&  $~~+ 0.00045*E$ & \\ \hline
$E_T^{iso}$ 	    & $< 4$ GeV 	        	&  $< 4$ GeV  	& $< 4$ GeV \\ \hline
%$E/p$ 		    & $< 2$ for $E_T < 50$~GeV  &      no cut 		& N/A \\ \hline
$E/p$ 		    & $< 2$   &      no cut 		& N/A\\
 		    & for $E_T < 50$~GeV        &      		        &  \\ \hline
$L_{shr}$	    & $< 0.2$		 	&      no cut		& N/A \\ \hline
$\chi_{strip}^2$    & $< 10$ 			&      no cut		& N/A \\ \hline
$q*\Delta x$	    & $> -3$~cm,		&      no cut		& N/A \\
		    & $<1.5$~cm			&  			& \\ \hline
$|\Delta z|$	    & $< 3$~cm			&      no cut		& N/A \\ \hline  
$PEM \chi_{3\times3}^2$  &  N/A   			&	N/A		& $<10$\\ \hline
\end{tabular}
\caption{\emph{\label{t_electron} Criteria for electron candidates. 
$\zgee$ candidates require at least one  
central-tight electron and at least one additional loose electron (central or plug)
in the event.}}
\end{center}
\end{table}

\begin{figure}
\begin{center}
\includegraphics[width=8cm]{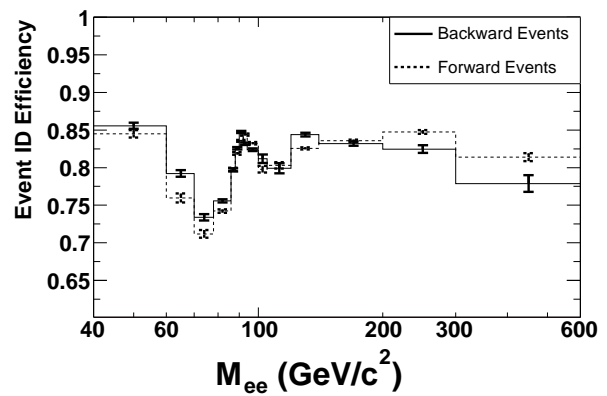}
\caption{\emph{ Event electron identification efficiency, $(\epsilon_i^{\pm})_{ID}$, 
as a function of $M_{ee}$ measured from the $\zgee$ simulation. The dashed line is for 
forward events and the solid line is for backward events. The dip in efficiency
below 90 GeV/$c^2$ and the differences between forward and backward efficiencies
are due to radiation effects (see Sec.~\ref{s_event_sel}).}}
\label{f_IDeff}
\end{center}
\end{figure}

	The absolute identification efficiencies are measured from the $\zgee$ data
where one electron is selected using the central-tight electron cuts, 
and the second electron is used as a probe to measure the identification efficiencies.
The total identification efficiencies are found to be $(83.4 \pm 0.8)\%$ for 
central-tight electrons, $(94.3 \pm 0.5)\%$ for central-loose electrons, 
and $(87 \pm 2)\%$ for plug-loose electrons. 
Since we measure the ratios of forward and backward events to the total events,
only the relative difference in efficiencies between forward
and backward events affects the $\afb$ measurement. 
The relative efficiencies and their dependence
on $\mee$ are estimated from the Monte Carlo simulation (see Fig.~\ref{f_IDeff}).
The difference between the forward and backward event efficiencies is largest 
just below the $Z$ pole. The dip in efficiency below the $Z$ pole is due 
to contamination from mis-measured events from the pole. The events are
mis-measured due to photon radiation which also lowers the electron
identification efficiency. The forward efficiency is lower because
the events are more forward at the $Z$ pole, resulting in more
contamination for forward events.

\begin{figure}
\begin{center}
\includegraphics[width=7.5cm]{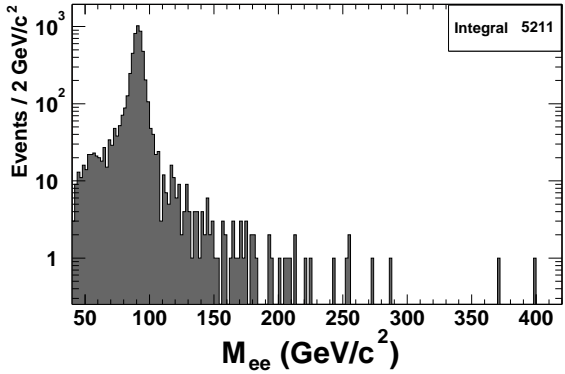}
\caption{\emph{Distribution of dielectron invariant mass from 
$\zgee$ candidates in 72 pb$^{-1}$ of Run II data.}}
\label{f_signal}
\end{center}
\end{figure}

	Based on these selection criteria, we find 1,892 central-central 
and 3,319 central-plug $\zgee$ candidates.
The invariant-mass distribution is shown in Fig.~\ref{f_signal}
and the number of events in each mass bin is given in Table~\ref{t_zgsel}.

%%%%%%%%%%% Summary of Dielectron Sample %%%%%%%%%%%%%%%%%%%%%%%%%%%%%%%%%%%
\begin{table}
\begin{center}
\begin{tabular}{|c||r|r|}\hline
Mass Region	 	& C-C & C-P  \\ \hline\hline
$ 40 \leq \mee < 60$ GeV/$c^2$ & 69  &  85      \\ \hline
$ 60 \leq \mee < 70$ GeV/$c^2$ & 42  &  72      \\ \hline
$ 70 \leq \mee < 78$ GeV/$c^2$ & 48  &  119     \\ \hline
$ 78 \leq \mee < 86$ GeV/$c^2$ & 204  &  329     \\ \hline
$ 86 \leq \mee < 88$ GeV/$c^2$ & 151  &  299     \\ \hline
$ 88 \leq \mee < 90$ GeV/$c^2$ & 301  &  512     \\ \hline
$ 90 \leq \mee < 92$ GeV/$c^2$ & 416  &  610     \\ \hline
$ 92 \leq \mee < 94$ GeV/$c^2$ & 330  &  543     \\ \hline
$ 94 \leq \mee < 100$ GeV/$c^2$ & 243  &  545    \\ \hline
$ 100 \leq \mee < 105$ GeV/$c^2$ & 30  &  68    \\ \hline
$ 105 \leq \mee < 120$ GeV/$c^2$ & 29  &  61    \\ \hline
$ 120 \leq \mee < 140$ GeV/$c^2$ & 13  &  31    \\ \hline
$ 140 \leq \mee < 200$ GeV/$c^2$ & 9  &  36     \\ \hline
$ 200 \leq \mee < 300$ GeV/$c^2$ & 6  &  8    \\ \hline
$ 300 \leq \mee < 600$ GeV/$c^2$ & 1  &  1    \\ \hline

\end{tabular}
\caption{\emph{\label{t_zgsel} The number of central-central (C-C) and central-plug (C-P)
 $\zgee$ candidates in each mass region.}}
\end{center}
\end{table}

%***********************************************************************************
\subsection{Detector simulation of \zgee}
\label{s_simulation}
The simulation is used to understand the detector's geometric 
acceptance for electrons, bremsstrahlung radiation from 
interactions with the detector material, and the energy
scale and resolution for electrons in the tracking chambers
and calorimeters. In this section, comparisons of the 
kinematic and geometric distributions between the data
and simulation are discussed, and any discrepancies are
presented as a possible systematic uncertainty.

\begin{figure}
\begin{center}
\includegraphics[width=4cm]{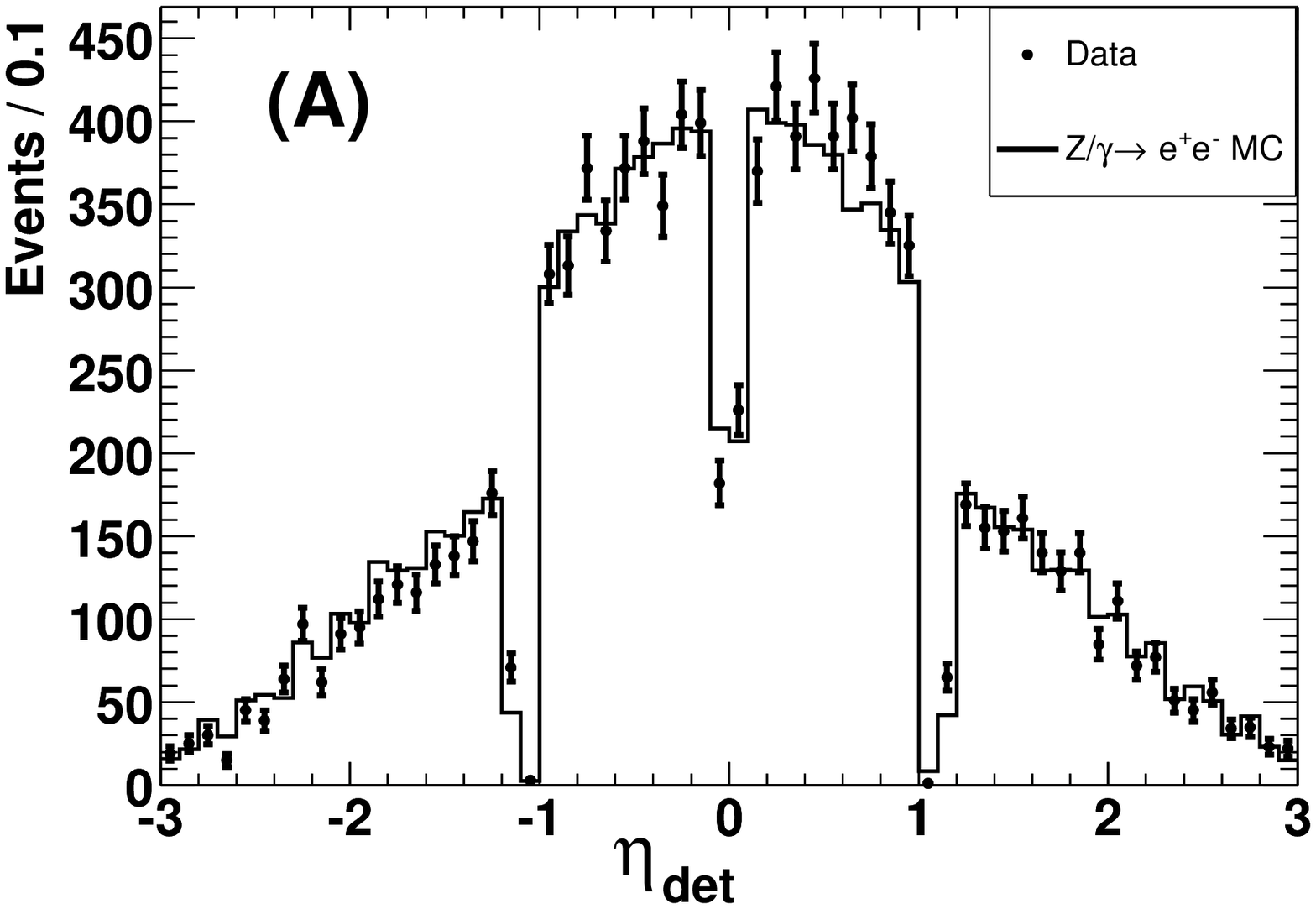}
\includegraphics[width=4cm]{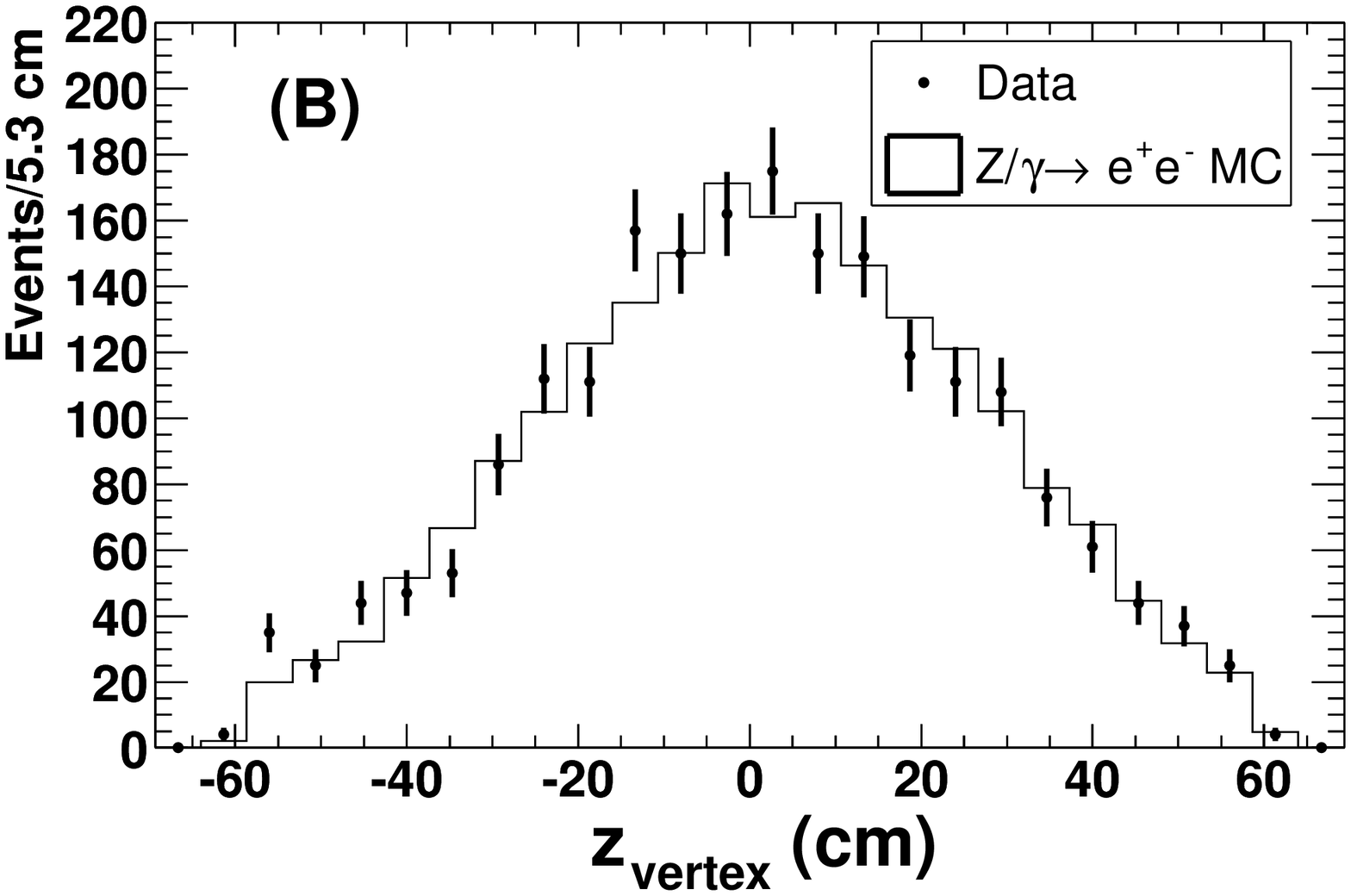}
\includegraphics[width=4cm]{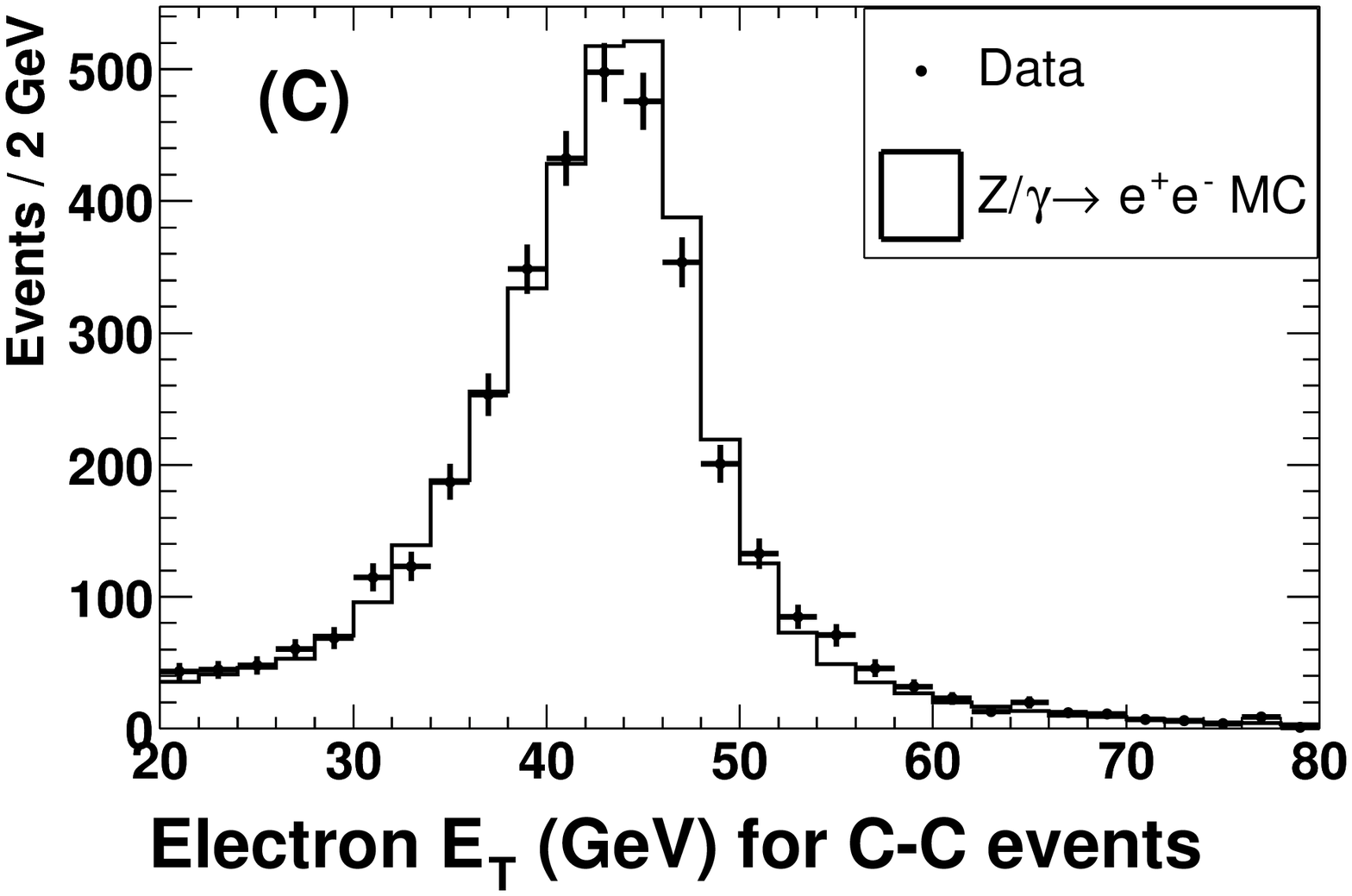}
\includegraphics[width=4cm]{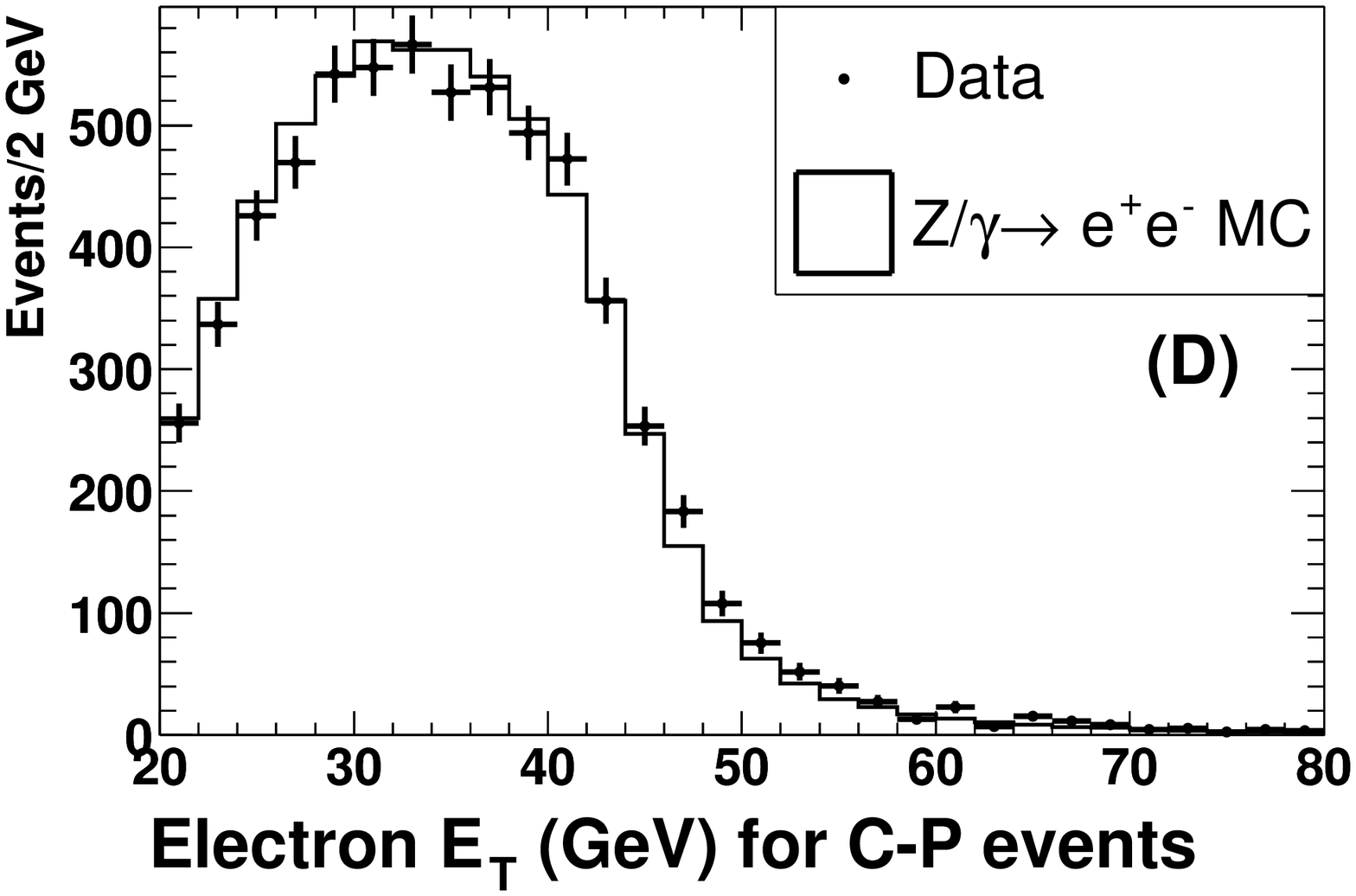}
\caption{\emph{The electron $\eta_{det}$(A) and event vertex $z$ position (B) are
important for understanding the geometric acceptance in the simulation.
The electron $E_T$ distribution for central-central(C), and 
central-plug(D) are used to the kinematic distributions. The Monte Carlo 
histograms are normalized to the number of entries in the background-subtracted data. 
The detector simulation for electrons has been tuned so that the $Z$ peak and 
width match the data (Sec.~\ref{s_ecorr}). The discrepancies
are used to estimate systematic uncertainties on the Monte Carlo simulation
(see Sec.~\ref{s_sys}).}}
\label{f_datasim}
\end{center}
\end{figure}

The geometric acceptance is largely dependent on the 
location of the $\pbp$~interaction vertex ($z_{vertex}$ distribution)
and the geometric description of the detector.
Figure~\ref{f_datasim}A shows the distribution of $\eta_{det}$ 
for electrons in the $\zgee$ data sample compared with the prediction 
from the Monte Carlo simulation. The discrepancies shown in 
Fig.~\ref{f_datasim}A are used to estimate a systematic uncertainty 
on the fiducial acceptance (see Sec.~\ref{s_sys}).
The Monte Carlo simulation 
correctly models the observed $z_{vertex}$ distribution
(Fig.~\ref{f_datasim}B). Fitting the $z_{vertex}$ distribution to 
a Gaussian function yields an average position in $z$ of $+2.5$~cm and a width of 28~cm. 

	The calorimeter energy scale and resolution 
in the simulation are tuned so that the mean and width of the $Z \rightarrow e^+e^-$ peak 
in the simulation are consistent with those from the data (see Sec.~\ref{s_ecorr}). 
The $\et$ distribution of the electrons for central-central and
central-plug events after the simulation tuning are shown 
in Figs.~\ref{f_datasim}C~and~\ref{f_datasim}D.
(The background prediction included in these plots is discussed 
in Sec.~\ref{s_backgrounds}.)

\begin{figure}
\begin{center}
\includegraphics[width=8cm]{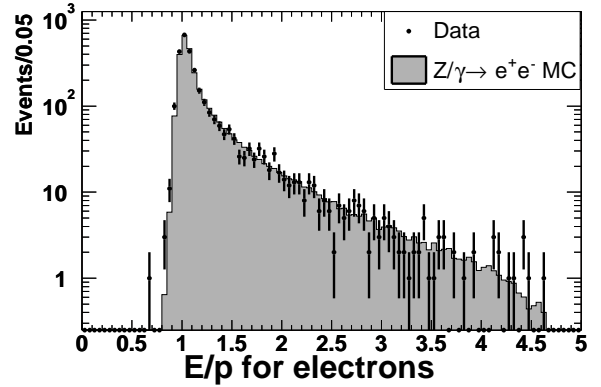}
\vspace{0in}
\caption{\emph{The $E/p$ distribution for central electrons in the $\zgee$ data sample. 
Points and histograms are data and Monte Carlo simulation, respectively. The Monte Carlo 
histogram is normalized to the number of entries in the data.
The high tail ($E/p>2$) of this distribution is used to calibrate the amount of material in the
Monte Carlo simulation between the interaction point and the tracking volume.}}
\label{f_eop}
\end{center}
\end{figure}

	The amount of material between the collision point and the outer cylinder 
of the COT is tuned so that the electron identification variables sensitive to 
external bremsstrahlung in the data match with those in the simulation 
(see Sec.~\ref{s_echarge}). 
The material in the Monte Carlo simulation between the interaction point and 
the tracking volume 
is tuned using the $E/p$ distribution as shown in Fig.~\ref{f_eop}. 
The ratio between the number of events in the high tail ($E/p>2$) and that in
the peak is used to calibrate and determine an uncertainty due to the modeling 
of the material in the detector simulation.

%***********************************************************************************
\subsection{Electron energy scale and resolution}
\label{s_ecorr}
\begin{figure}
\begin{center}
\includegraphics[width=8cm]{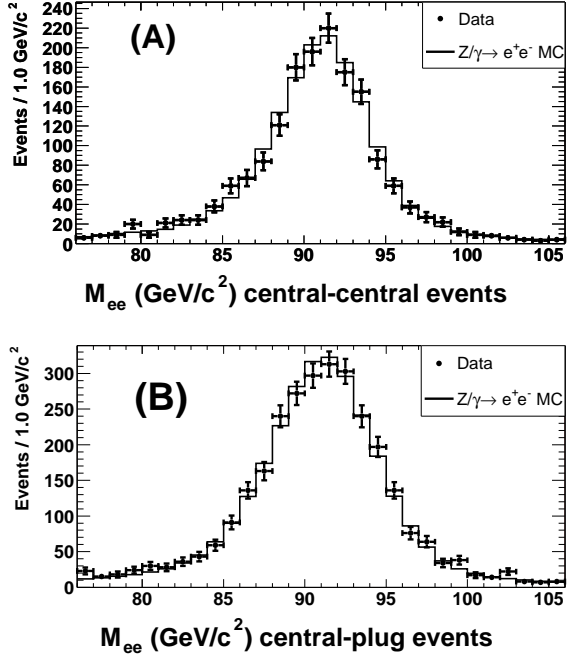}
\caption{\emph{Invariant-mass distributions of $\zgee$ candidates: two 
electrons in the central region (top), 
one electron in the central region and the other in the plug region
(bottom). Points and histograms are data and Monte Carlo simulation, respectively.
Energy scale corrections and extra energy smearing are
applied to the simulation so that the Gaussian widths and peaks match.}}
\label{f_mass}
\end{center}
\end{figure}

	Both local and global energy scale corrections are applied 
to the electron energy. Local corrections are applied to improve 
resolution by correcting for variations in the energy response of the 
calorimeter. They include corrections for time dependence, variations in the
response at different points within a calorimeter tower~\cite{Yasuoka:1987ar}, and variations in 
the gains of the different calorimeter tower channels. Electrons from the
$W$ sample and the inclusive electron sample are used to calibrate these variations.
The reference for correcting the electron energy is the track momentum as
measured by the COT. Uniformity is achieved by adjusting the tower energy
responses (gains) until the mean $E/p$ is flat as a function of time and
$\phi$, and agrees with the Monte Carlo simulation as a function of $\eta$.
Figure~\ref{f_mass} shows the invariant-mass distributions 
near the $Z$ peak for central-central and central-plug events for 
data and Monte Carlo simulation. High-energy electrons are measured
with a resolution of $\frac{\sigma(E_T)}{E_T}=\frac{13.5\%}{\sqrt{E_T}}\oplus 1.5\%$~\cite{Balka:1987ty} in
the central calorimeters and $\frac{\sigma(E)}{E}=\frac{14.4\%}{\sqrt{E}}\oplus 0.7\%$~\cite{Albrow:2001jw}
in the plug calorimeters.

%***********************************************************************************
\subsection{Electron charge identification}
\label{s_echarge}
	The charge measurement of electrons is essential for this analysis because 
the charge determines the sign of $\cost$~(see Eq.~(\ref{e_collinSoper})), and
therefore determines whether the event is forward or backward.
In the central-central topology, we measure the charge of both electrons and require
that they have opposite sign. 
The central-central opposite-sign requirement removes events where one of the electrons 
has a misidentified charge from the event sample, leaving very little ambiguity 
on the forward-backward measurement due to charge misidentification.  
However, in events with a central-plug topology, the charge of the two electrons is 
determined solely from the central electron. So a misidentification
of the central electron's charge will switch the sign on $\cost$, and result
in a misassignment of the event as forward or backward.
For this reason, the charge misidentification rate
needs to be properly understood, and well modeled by the simulation. 

\begin{figure}
\begin{center}
\includegraphics[width=8cm]{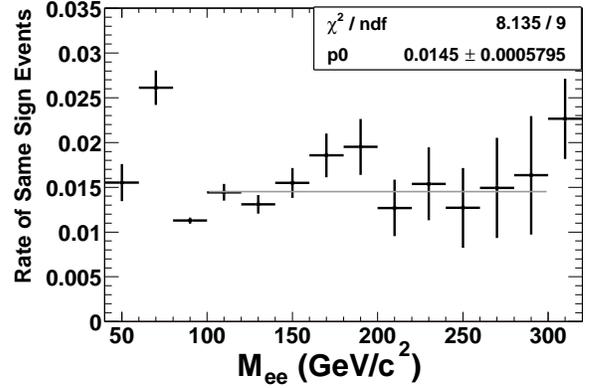}
\caption{\emph{The rate of same-sign events as a function of the dielectron invariant mass 
from the $\zgee$ Monte Carlo sample. The
rate of charge misidentification is half of the same-sign event rate. 
The last bin shows the rate for all events with $\mee>300$~GeV/$c^2$.
A fit to a constant between 100 GeV/$c^2 < \mee< 300$ GeV/$c^2$ has been made with 
the result that \zg events have an average same-sign rate (p0) of 1.45\%.}}
\label{f_chargeID}
\end{center}
\end{figure}

\begin{figure}
\begin{center}\includegraphics[width=8cm]{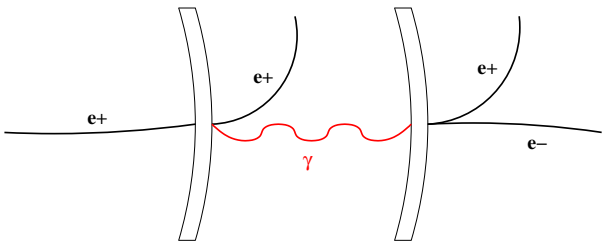}
\caption{\emph{A schematic diagram of a ``trident'' electron where a positron radiates a hard 
bremsstrahlung photon in the material and the photon converts into an electron-positron 
pair in the material.  The electron from the photon conversion carries the highest momentum, 
thus the charge of the primary electron ($e^+$) is assigned to be negative $(e^-)$.}}
\label{f_trident}
\end{center}
\end{figure}

	We use the $\zgee$ Monte Carlo sample with the central-central topology 
to study the sources of charge misidentification 
and to measure the misidentification rate. The misidentification rate is 
determined by counting the number 
of events where both electrons have the same charge. If the 
rate is small, the probability of having same-sign events is approximately 
twice that of misidentifying the charge of a single electron. 
Figure~\ref{f_chargeID} shows that above the $Z$ pole ($\mee>100$~GeV/$c^2$), the rate of 
events with the same sign is approximately flat up 
to $\mee \simeq 300$~GeV/$c^2$.  The drop in the rate at the $Z$ pole and
the increase in the rate in the next lower bin are attributable to 
hard bremsstrahlung followed by $e^+e^-$ pair production in the material.  This
process is referred to as charge misidentification due to ``trident'' electrons.  Figure~\ref{f_trident} 
shows a ``trident'' electron where a positron radiates a hard 
bremsstrahlung photon in the material, and the photon subsequently converts into an 
electron-positron pair in the material.  
The electron from the photon conversion carries the highest momentum, 
and the charge of the primary electron ($e^+$) is assigned to be negative $(e^-)$.
The Monte Carlo sample shows that the charge misidentification rate coming from ``trident'' electrons
is $(0.7 \pm 0.3)\%$.  The other source of charge misidentification is the drift chamber tracking resolution 
of $\delta({\frac{1}{\pt}}) \simeq 0.001$~GeV$^{-1}$. 
At higher energies, the tracks become almost straight ($\frac{1}{\pt} \sim 0$), and the charge 
determination has a higher probability of being wrong.  The last bin, which includes all events 
with $\mee>300$~GeV/$c^2$, has a misidentification rate measured from the
simulation of $(1.1 \pm 0.2)\%$.
Corrections for charge misidentification are included as part of the
acceptance calculation via simulation.  The dominant systematic uncertainty comes from the uncertainty 
in the amount of material between the interaction point and the tracking volume.
A comparison of the same-sign events between data and Monte Carlo simulation is used to get
an estimate on the background. The comparison shows agreement at the $Z$ pole where there is little
background (Fig.~\ref{f_InvMass_SS_MC}).

\subsection{Charge dependence of electron efficiencies}
\begin{figure}
\begin{center}
\includegraphics[width=8cm]{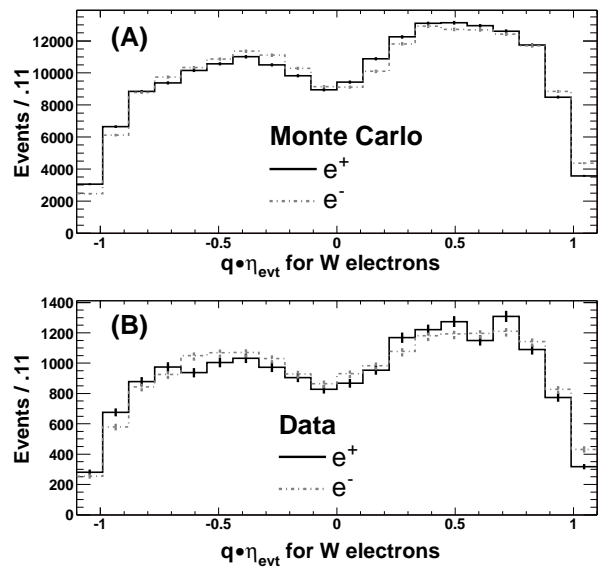}
\caption{\emph{The $q \cdot \eta_{evt}$ distribution of the electrons 
in $W$ Monte Carlo (A) and data (B) events.
The asymmetry about 0 is due to the charge asymmetry in the $W$ production.
The difference between $e^+$ and $e^-$ is largely due to the asymmetric 
distribution of the event vertex ($z_{vertex}$) distribution. 
}}
\label{f_Wcharge}
\end{center}
\end{figure}

A systematic bias in the forward-backward asymmetry may occur 
if the detector response to electrons differs from that to positrons. 
We compare acceptances and efficiencies for electrons and positrons
using the $W^\pm \rightarrow e^\pm \nu$ Monte Carlo and data samples.  
In Fig.~\ref{f_Wcharge}, the number of $e^+$ events and 
$e^-$ events are plotted as a function of $q \cdot \eta_{det}$ of the electrons.  The difference between 
the $q \cdot \eta_{det} < 0$ region and  the $q \cdot \eta_{det} > 0$ region comes 
from the intrinsic charge asymmetry in $W^\pm$ production in the
proton direction and the anti-proton direction.  Because the average momentum 
of $u$ ($\bar u$) quarks is larger than that of $d$ ($\bar d$) 
quarks in the proton (anti-proton), $W^+$ ($W^-$) event production is enhanced in the proton 
(anti-proton) direction.  

Small differences are observed in the data
between the $e^+$ and $e^-$ yields (Fig.~\ref{f_Wcharge}B). These could be caused 
by differences in the detector response between electrons and positrons, 
differences between the $|\eta_{det}| > 0$ and $|\eta_{det}| < 0$ detectors, or 
an asymmetric $z_{vertex}$ distribution about 0.  
Similar differences are seen in the Monte Carlo sample 
(Fig.~\ref{f_Wcharge}A) where the differences 
come purely from the asymmetric $z_{vertex}$ distribution (see Sec.~\ref{s_mc}).  
The asymmetry between the $e^+$ and $e^-$ yields due to the asymmetric 
$z_{vertex}$ distribution is measured to be $(0.7 \pm 0.2)$\% from the simulation.
After the effect of the $z_{vertex}$ offset is taken into account, the observed asymmetry 
in the data is reduced to $(0.1 \pm 0.7)$\%. The impact of this asymmetry on the forward-backward 
charge asymmetry measurement is therefore estimated to be small compared to other sources of 
systematic uncertainties.

%\clearpage
%***********************************************************************************
%*********************************************************************************** 
\section{Backgrounds}
\label{s_backgrounds}

The dominant sources of background to the process
$\pbp \rightarrow \zgee + X$ are:
\begin{enumerate}
\item Dijet events where the jets are misidentified as electrons, 
\item	$W + X \rightarrow e\nu + X$, where $X$ is a photon or a jet misidentified as an electron,
\item	$\zgtautau\rightarrow\elel\nu_{\tau}\nu_{e}\bar\nu_{\tau}\bar\nu_{e}$, 
\item $\WW \rightarrow \elel \nu_e\bar\nu_e$, 
\item $W^\pm Z$ where $\zee$, 
\item $\ttb \rightarrow \elel\nu_e\bar\nu_e \bbb$.
\end{enumerate}
The determination of $\afb$ requires
knowledge of the number of background events and the forward-backward charge asymmetry 
of the background events in each mass bin.

%*********************************************************************************** 
\subsection{Dijet background}\label{s_dijet}
Dijet events are the dominant source of background in our sample.  
This background category includes events with two high-$\pt$ jets and those with one jet 
and one photon.  Hadrons in jets can be misidentified as electrons, and jets can contain real electrons 
produced in semi-leptonic heavy-flavor decays (from $c$ or $b$ quarks).  Photons from initial- or 
final-state bremsstrahlung 
are identified as electrons if they are in the plug region. In the central region, photons must
convert in order to be identified as electrons because of the tracking requirement.
The physics objects producing electron candidates in dijet events are discussed in 
Sec.~\ref{s_dijet_MC}.  The forward-backward charge asymmetry from the dijet background is 
discussed in Sec.~\ref{s_qcd_mass_asymm}.  In Sec.~\ref{s_dijet_fake}, the rate of jets faking 
electrons is measured, the dijet background in the data is estimated, and the invariant-mass 
distribution of the dijet background is extracted.
In Sec.~\ref{s_dijet_check}, the amount of dijet background
in the central-central topology is cross-checked using same-sign events in the $Z/\gamma^*$ 
data and Monte Carlo samples.

%*********************************************************************************** 
\subsubsection{Sources of electron fakes from dijet Monte Carlo}
\label{s_dijet_MC}
	The Monte Carlo sample of $\sim$750,000 dijet events 
	is used to understand the sources of high-$\pt$ 
fake electrons. 
We find 47 central-tight electrons, 179 central-loose electrons, and 
1702 plug-loose electrons in this sample.  We then look for hadron jets or photons 
with $E_T > 15$~GeV that match the direction of these fake electrons.
The dominant sources for central-tight electrons are photons which have
converted 
and light quark jets (those originating from $u$, $d$, and $s$ quarks).  Light quark jets 
are the dominant source for the loose electrons, especially the plug electrons.  
The fraction of events found in each category is shown in Table~\ref{t_fakeparton}. 

%\begin{widetext}
\begin{table*}
\begin{center}
Dijet Events \\
\begin{tabular}{|l||c|c|c|c|}\hline
Electron Type & Light-Quark Jets & Heavy-Quark Jets & Gluon Jets & Photons \\ \hline \hline
Central-Tight &   0.38 $\pm$ 0.09 &   0.15 $\pm$ 0.06 &   0.13 $\pm$ 0.05 &   0.34 $\pm$ 0.09    \\ \hline
Central-Loose &   0.63 $\pm$ 0.06 &   0.08 $\pm$ 0.02 &   0.20 $\pm$ 0.03 &   0.09 $\pm$ 0.02    \\ \hline
Plug-Loose    &   0.82 $\pm$ 0.02 &   0.01 $\pm$ 0.00 &   0.07 $\pm$ 0.01 &   0.10 $\pm$ 0.01    \\ \hline \hline
\end{tabular} \\ 
\vspace{0.15cm}
\wenu~Events \\
\vspace{0.05cm}
\begin{tabular}{|l||c|c|c|c|}\hline
Electron Type & Light-Quark Jets & Heavy-Quark Jets & Gluon Jets & Photons \\ \hline \hline
Central-Tight &   0.28 $\pm$ 0.09 &   0.00 $\pm$ 0.00 &   0.00 $\pm$ 0.00 &   0.72 $\pm$ 0.15    \\ \hline
Central-Loose &   0.63 $\pm$ 0.09 &   0.01 $\pm$ 0.01 &   0.01 $\pm$ 0.01 &   0.35 $\pm$ 0.07    \\ \hline
Plug-Loose    &   0.20 $\pm$ 0.03 &   0.01 $\pm$ 0.01 &   0.01 $\pm$ 0.01 &   0.78 $\pm$ 0.07    \\ \hline \hline
\end{tabular}
\caption{\emph{\label{t_fakeparton} The physics objects producing electron candidates in Monte Carlo 
dijet and \wenu~events and their fractional contributions for central-tight electrons, central-loose 
electrons, and plug-loose electrons. The fractional contribution of the different physics objects producing
electron candidates varies depending on the type of electron and background type. Light-quark jets
are those originating from $u$, $d$, and $s$ quarks, and 
heavy-quark jets are those originating from to $b$ and $c$ quarks. 
Uncertainties include only Monte Carlo statistics.}}
\end{center}
\end{table*}
%\end{widetext}
%*********************************************************************************** 
\subsubsection{Charge correlation and $A_{FB}$ distribution of dijet events}
\label{s_qcd_mass_asymm}

We do not expect any correlation between the charges of the two fake electrons in the 
dijet events.  This expectation is  
checked using the hadron-enriched data sample, where both electrons are ``jet-like'';
events collected with the high-$\pt$ central electron triggers that 
have two electromagnetic clusters passing the kinematic cuts for electrons 
in this analysis and a significant amount of energy near 
them ($E_{T}^{iso} > 4$~GeV). An additional cut of $\met~<~10$~GeV eliminates 
a large fraction of the remaining $\wenu$ contamination in the sample.
The selection criteria are summarized in Table~\ref{t_qcdsel2}.
The sample contains 8595 (8797) events where the two electron candidates have the 
same (opposite) charge.  Although the difference is statistically significant 
(2.2$\sigma$), there are only 2\% more opposite-sign events than same-sign 
events. This demonstrates that the charges of the two electrons identified in dijet 
events are nearly uncorrelated. 

\begin{figure}
\begin{center}
\includegraphics[width=8cm]{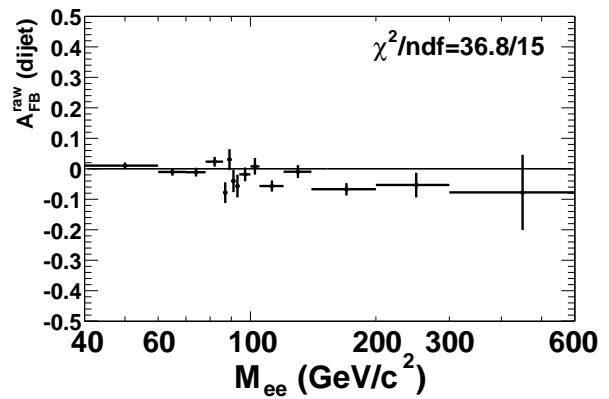}
\caption{\emph{$\afbr$ of dijet background estimated from
	 the hadron-enriched data sample.}}
\label{f_Afb_qcd}
\end{center}
\end{figure}

The raw forward-backward asymmetry of the dijet background is measured with the 
same hadron-enriched sample in the 15 mass regions.  As shown in Fig.~\ref{f_Afb_qcd},
the $\afbr$ values are close to zero, consistent with 
the symmetric angular distribution expected 
for dijet events. This analysis assumes $\afbr=0$ for the dijet background,
where the events are split evenly between forward and backward categories
when subtracted from the candidate sample. The measured asymmetries listed in 
Table~\ref{t_qcdafb} are used for estimating a systematic uncertainty on
this assumption.

%%%%%%%%%%% QCD Control Sample TABLE %%%%%%%%%%%%%%%%%%%%%%%%%%%%%%%%%%%
\begin{table}
\begin{center}
\begin{tabular}{|c|}\hline
Hadron-enriched selection  \\ \hline\hline
$\pt > 10$ GeV/$c$ (central only)\\ \hline
$ E_T^{corrected} > 20$ GeV \\ \hline
$E_T^{iso} > 4$ GeV \\ \hline
$\met < 10$ GeV\\ \hline
\end{tabular}
\caption{\emph{\label{t_qcdsel2} Selection criteria for the hadron-enriched data sample.}}
\end{center}
\end{table}

%%%%%%%%%%% QCD Control Sample TABLE %%%%%%%%%%%%%%%%%%%%%%%%%%%%%%%%%%%
\begin{table}
\begin{center}
\begin{tabular}{|c|c|c|c|}\hline
Mass Region & \multicolumn{2}{|c|}{\# Events} & $\afbr$(dijet) \\ \cline{2-3}
 GeV/$c^2$			& C-C		& C-P & 	\\ \hline\hline
$ 40  \leq M_{jj} < 60$ &  $5.4\pm 1.9$  &  $  51\pm 18 $ &  0.00 $\pm$ 0.01   \\ \hline
$ 60  \leq M_{jj} < 70$ &  $1.5\pm 0.5$  &  $25.7\pm 9.0$  &  0.00 $\pm$ 0.01   \\ \hline
$ 70  \leq M_{jj} < 78$ &  $0.7\pm 0.2$  &  $15.3\pm 5.4$  &  -0.01 $\pm$ 0.01   \\ \hline
$ 78  \leq M_{jj} < 86$ &  $0.4\pm 0.1$  &  $11.1\pm 3.9$  &  0.01 $\pm$ 0.01   \\ \hline
$ 86  \leq M_{jj} < 88$ &  $0.1\pm 0.04$ &  $ 2.2\pm 0.8$  &  -0.09 $\pm$ 0.03   \\ \hline
$ 88  \leq M_{jj} < 90$ &  $0.1\pm 0.04$ &  $ 1.9\pm 0.7$  &  0.00 $\pm$ 0.03   \\ \hline
$ 90  \leq M_{jj} < 92$ &  $0.1\pm 0.04$ &  $ 1.8\pm 0.6$  &  -0.02 $\pm$ 0.03   \\ \hline
$ 92  \leq M_{jj} < 94$ &  $0.1\pm 0.04$ &  $ 1.6\pm 0.6$  &  -0.02 $\pm$ 0.03   \\ \hline
$ 94  \leq M_{jj} < 100 $  & $0.1\pm 0.04$ &  $ 4.1\pm 1.4$  &  -0.03 $\pm$ 0.02   \\ \hline
$ 100  \leq M_{jj} < 105$ &$0.1\pm 0.04$ &  $ 2.6\pm 0.9$  &  -0.02 $\pm$ 0.02   \\ \hline
$ 105  \leq M_{jj} < 120$ &$0.2\pm 0.07$ &  $ 4.8\pm 1.7$  &  -0.04 $\pm$ 0.01   \\ \hline
$ 120  \leq M_{jj} < 140$ &$0.1\pm 0.04$ &  $ 2.9\pm 1.0$  &  -0.02 $\pm$ 0.02   \\ \hline
$ 140  \leq M_{jj} < 200$ &$0.1\pm 0.04$ &  $ 1.9\pm 0.7$  &  -0.05 $\pm$ 0.02   \\ \hline
$ 200  \leq M_{jj} < 300$ &$0.0\pm 0.00$ &  $ 0.2\pm 0.07$  &  -0.01 $\pm$ 0.03   \\ \hline
$ 300  \leq M_{jj} < 600$ &$0.0\pm 0.00$ &  $ 0.0\pm 0.00$  &  -0.07 $\pm$ 0.11   \\ \hline \hline
Total	                        &$9.0\pm 3.2$ &   $ 128\pm 45  $  &     N/A            \\ \hline
\end{tabular}
\caption{\emph{\label{t_qcdafb}The estimated number of events
and measured $\afbr$ of dijet background in each invariant-mass bin.
The number of events is estimated using the rate of jets faking electrons. 
The $\afbr$(dijet) values 
are measured from the hadron-enriched data sample.  When calculating 
$\afb(\zgee)$, $\afbr$(dijet) = 0 is assumed. The measured $\afbr$(dijet) is
used to estimate a systematic uncertainty on the measurement.}}
\end{center}
\end{table}
%*********************************************************************************** 
\subsubsection{Estimation and $\mee$ distribution of the dijet background}
\label{s_dijet_fake}

\begin{figure}
\begin{center}
\includegraphics[width=8cm]{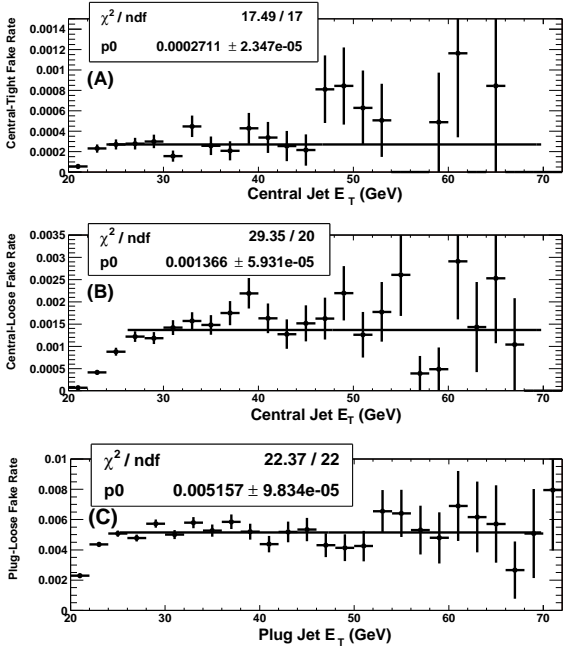}
\vspace{0in}
\caption{\emph{The rate of jets faking electrons for central-tight (A), central-loose (B), and
plug-loose (C) requirements in the 20 GeV dijet sample. The fake rate is 
measured for jets with $E_T^{jet}>24$~GeV for central-tight and plug-loose electrons, 
and $E_T^{jet}>26$~GeV for central-loose
electrons to take into account relative energy differences between jet clustering and
electron clustering. p0 is the value returned by the fit for a zeroth order
polynomial.}}
\label{f_fakerate}
\end{center}
\end{figure}

The dijet sample used to calculate the fraction of jets faking electrons 
(the single-jet fake rate) must pass the 20~GeV single-jet triggers, 
have two jets with $E_T > 20$~GeV, have $\met<10$~GeV, and have no more than one loose electron in the event.
Jets are clustered using a cone size of $\Delta R = 0.4$.
These requirements ensure that the electroweak 
contamination from $W$ and $Z$ decays to electrons is negligible. The fake rate is
defined as the fraction of jets in the sample which pass the electron-selection criteria. 
The measured rate is plotted as a function of jet $E_T$ in Fig.~\ref{f_fakerate}. 
Due to the bigger cluster size of jets compared to electrons, jet energies are larger than 
the corresponding fake-electron energies.
For example, when a jet of $E_T=25$~GeV fakes an electron, the
electron candidate is typically reconstructed with an $E_T$ of 20 GeV.
The single-jet fake rate with $E_T^{jet} > 25$~GeV (or $E_T^e > 20$~GeV) is 
measured to be $(2.7 \pm 0.2) \times 10^{-4}$ for central-tight electrons.
Note that jets used in this measurement are a mixture of both triggered and non-triggered jets.

The single-jet fake rate for non-triggered jets can be measured using jets with $E_T$ below 
trigger thresholds.   We use jets with $E_T < 45$~GeV in the jet 50~GeV data sample 
(50~GeV single jet triggers)
and $E_T < 95$~GeV in the jet 100~GeV data sample (100~GeV single jet triggers).  
In addition, Monte Carlo dijet events without trigger requirements are studied for comparison.
Table~\ref{t_fakerate} summarizes the rates in the 
four different samples. The fake rate for jets without trigger biases 
are roughly one half of the fake rate measured in the 20~GeV dijet sample.
The difference in rates is expected since the trigger requirements 
select jets that deposit a significant fraction of their energy into one tower and
hence look more ``electron-like.''

The number of dijet background events in the $\zgee$ sample is estimated 
by applying the average 
single-jet fake rate for a tight electron and a loose electron 
to every pair of jets in the 20~GeV dijet sample that pass the kinematic cuts.  
The systematic uncertainty on this estimation 
is determined by taking the simplified case where each event has two and only two jets, 
one triggered and one non-triggered. Let $f$ denote the fake rate without trigger biases and
$2f$ the average fake rate in the 20~GeV dijet sample (as described in the previous paragraph).
Then, the fake rate for the triggered jets is $3f$.  The event fake rate is estimated to be $4f^2$ if 
the average fake rate is applied to both jets (used for the estimation of the dijet background), 
and it is $3f^2$ if the rates for the non-triggered and triggered
jets are applied separately.  The difference between these two estimates, which is 25\%, 
is taken as a systematic uncertainty on the single-jet fake rate due to the jet trigger bias. 

\begin{figure}
\begin{center}
\includegraphics[width=8cm]{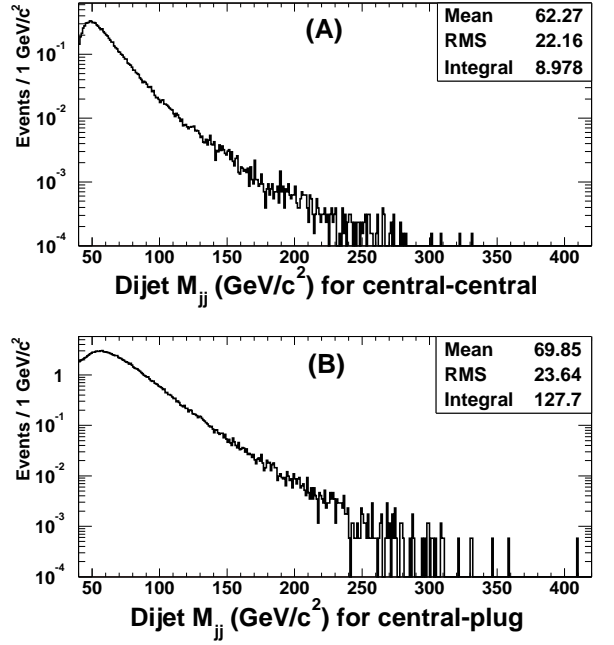}
\caption{\emph{The invariant-mass distribution measured in the 20 GeV dijet sample. 
The single-jet fake rate is applied to each jet in the dijet pair.
The jet energy has been corrected
to represent electron energies (see \rm{Sec.~\ref{s_dijet_fake}}).}}
\label{f_mass_qcd}
\end{center}
\end{figure}

The invariant-mass distribution of the dijet background is determined from 
the 20~GeV dijet sample where at least one of the two jets is found in the central region.
Jet energies are corrected to represent electron energies where the correction factors 
are determined from Gaussian fits to the $\frac{E_T^e}{E_T^{jet}} $
distributions for each of the three categories of electrons. 
(The Gaussian fits to $\frac{E_T^e}{E_T^{jet}} $ have means ranging 
from 0.88-0.92, and have widths ranging from 0.05-0.09 
depending on the type of electron.) 
Figure~\ref{f_mass_qcd} shows the
invariant-mass distribution of the central-central and central-plug dijet events, 
weighted by the probability of faking a dielectron 
candidate event.  Table~\ref{t_qcdafb} shows the number of dijet events 
expected in each invariant-mass bin.  The total number of dijet 
background candidates is estimated to be $9.0 \pm 3.2$ for the central-central topology, 
and $128 \pm 45$ for the central-plug topology. The invariant-mass distribution 
of the dijet background along with the Monte Carlo \zgee~prediction is
shown in Fig.~\ref{f_mass_tot}.

%%%%%%%%%%% QCD Control Sample TABLE %%%%%%%%%%%%%%%%%%%%%%%%%%%%%%%%%%%
\begin{table}
\begin{center}
\begin{tabular}{|c|c|c|c|c|}\hline
& Central-Tight & Central-Loose & Plug-Loose \\
&  Rate $(\times10^{-4})$ & Rate $(\times10^{-4})$ & Rate $(\times10^{-4})$\\ \hline \hline
20 GeV dijet   & $2.7\pm0.2 $ & $13.7\pm0.6 $ & $51.6\pm1.0$ \\ \hline
50 GeV dijet   & $1.3\pm0.2 $ & $6.5\pm0.4 $ & $30.2\pm1.2$ \\ \hline
100 GeV dijet  & $1.7\pm0.3 $ & $5.5\pm0.5 $ & $27.6\pm1.8$ \\ \hline
Monte Carlo    & $1.3\pm0.3 $ & $7.6\pm0.9 $ & $24.8\pm 3.1$ \\ \hline
\end{tabular}
\caption{\emph{\label{t_fakerate} The rate at which a jet fakes an electron. 
The jets in the 20~GeV dijet sample are a mixture of triggered and non-triggered jets.
The rate in the 50~GeV and 100~GeV dijet samples is measured
only for jets with $E_T$ below the trigger threshold, $E_T < $ 45 (95)~GeV 
for the 50 (100)~GeV dijet sample. The rate in Monte Carlo dijet events
is measured for jets with $E_T>$20 GeV.}}
\end{center}
\end{table}

%*********************************************************************************** 
\subsubsection{Check of dijet background using the charge of the dielectrons}
\label{s_dijet_check}

\begin{figure}
\begin{center}
\includegraphics[width=8cm]{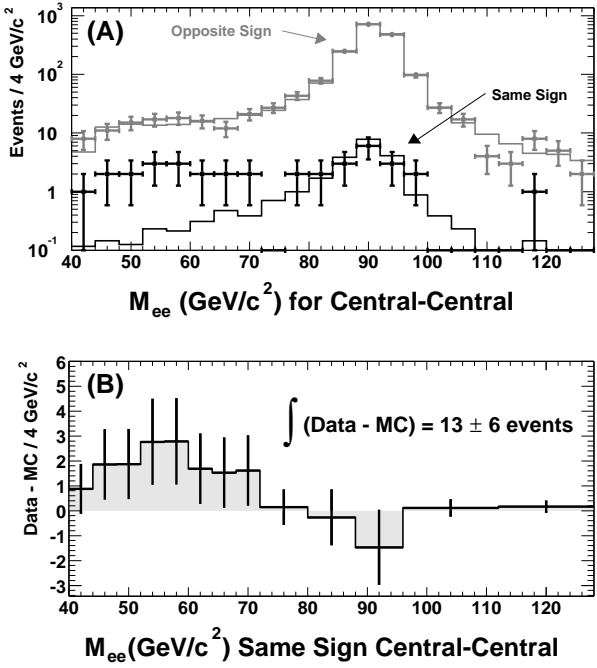}
\vspace{0in}
\caption{\emph{(A) invariant-mass distributions of data 
	 (points) and Monte Carlo events (histogram) in the central-central topology
	 where the Monte Carlo sample 
	 is normalized to the data using opposite-sign events in the $Z$ peak.
	 (B) invariant-mass distribution of same-sign data in the central-central topology 
	 after subtracting the same-sign MC distribution. The Monte Carlo subtracted same-sign 
	 distribution in (B) is a measurement of the dijet background.}}
\label{f_InvMass_SS_MC}
\end{center}
\end{figure}

We cross-check the estimation of the dijet background for the central-central topology 
using same-sign events in the data sample after correcting for the contribution from 
charge misidentification.  The dominant contribution to the 
charge misidentification of electrons comes from ``trident'' electrons except 
at very high energies where the curvature resolution is the dominant source 
(see Sec.~\ref{s_echarge}). The number of same-sign $\zgee$ events 
due to ``trident'' electrons is estimated from the Monte Carlo simulation. The Monte Carlo
samples are normalized such that the number of opposite-sign events in the $Z$ peak is
the same between the data and the Monte Carlo simulation. 
Figure~\ref{f_InvMass_SS_MC}A shows the resulting invariant-mass distributions in data 
and  Monte Carlo simulation for opposite-sign events and same-sign 
events. The number of same-sign events is found to be 36 in the data and
23 in the simulation, resulting in the estimation of 13 dijet background events for the central-central
topology. Figure~\ref{f_InvMass_SS_MC}B shows the invariant-mass distribution for the same-sign 
data events after the Monte Carlo same-sign events are subtracted. 
A systematic uncertainty of 4 events is estimated by repeating this procedure with 
the Monte Carlo samples generated with various amounts of material.
Hence the dijet background in the central-central topology is estimated to be 
$13 \pm 6{\rm (stat.)} \pm 4 {\rm (syst.)}$ events. This is in good 
agreement with the estimate obtained from the fake-rate method of $9.0 \pm 3.2$ events.

%*********************************************************************************** 
\subsection{Electroweak and top backgrounds}

\begin{figure}
\begin{center}
\includegraphics[width=8cm]{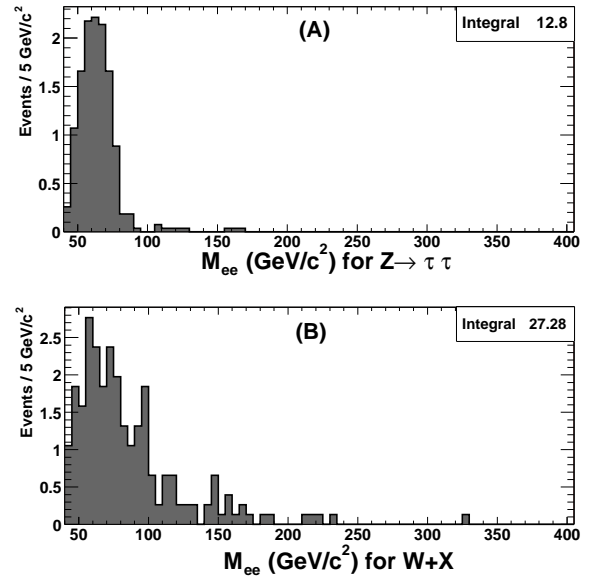}
\caption{\emph{Invariant-mass distributions for (A) $\zgtautau $ and 
(B) $W + X \rightarrow e\nu + X$ ($X = q, g$ or $\gamma$) backgrounds.}}
\label{f_ewk_bkgd}
\end{center}
\end{figure}

The electroweak and top background events are estimated using the Monte Carlo
simulation.  Table~\ref{t_ewbackground} shows the theoretical cross section 
and the number of events expected in the sample of 72 pb$^{-1}$ for each process. 
The systematic uncertainties on these background 
estimates reflect the 6\% uncertainty on the integrated luminosity, the 5\% 
uncertainty on the acceptances, and the uncertainties on the theoretical
production cross sections. The simulated events for each process which pass 
the selection requirements are used to
determine both the invariant-mass distributions and the expected forward-backward 
asymmetries. The invariant-mass distributions for $Z \rightarrow \tau^+\tau^-$ 
and $W+X\rightarrow e\nu+X$ are shown in Fig.~\ref{f_ewk_bkgd}.

The dominant electroweak background comes from the $W + X \rightarrow e\nu + X$ 
process where $X$ is a photon or hadronic jet and the $\zgtautau$ process where 
$\tau \rightarrow e\nu_{\tau}\nu_{e}$.  While the $\zgtautau\rightarrow\elel\nu_{\tau}\nu_{e}
\bar\nu_{\tau}\bar\nu_{e}$ background is reliably simulated by LO-based generators such 
as PYTHIA, the $W + X \rightarrow e\nu + X$ background estimated from PYTHIA requires 
cross-checks with the next-to-leading order (NLO) calculations.
PYTHIA,  using the parton-shower algorithms to generate photons and jets, is expected to 
give lower cross sections than the Standard Model prediction for high-$\pt$ photons and jets.
We cross-check the PYTHIA calculation with two matrix-element calculations, 
ALPGEN for $W$ + 1~parton production and WGAMMA for $W$ + 1~photon production. 
The number of $W+X\rightarrow e\nu+X$ background events estimated from PYTHIA is $27\pm5$.  
The combination of ALPGEN and WGAMMA estimates this number to be $24\pm3$, in good 
agreement with the PYTHIA expectation within the statistical uncertainties.  The 
difference between the PYTHIA and the ALPGEN/WGAMMA 
estimates is used as a systematic uncertainty on $\afb$.  Table~\ref{t_fakeparton} 
shows the sources of the electron candidates in the PYTHIA $W + X$ sample, indicating that 
the dominant source is from photons faking an electron (mostly in the plug region).  

\begin{table}
\begin{center}
\begin{tabular}{|c|c|c|c|}\hline
    Process & $\sigma \cdot BR$ (pb) & \multicolumn{2}{|c|}{Events Expected} \\ \cline{3-4}
                    &                                           & C-C & C-P \\ \hline
Dijet
	& N/A           &  $9.0 \pm 3.2$   & $128 \pm 45$ \\
$W+ jet/\gamma \rightarrow e\nu+jet/\gamma$
	& $ 2,690 \pm 100$ &$1.8 \pm 0.2$&$25.4 \pm 2.4$\\
$Z/\gamma^* \rightarrow \tau^+\tau^- $
	& $252 \pm 9$    & $5.6 \pm 0.5$ &$7.2 \pm 0.7$\\
$W^+W^- \rightarrow e^+e^-\nu_e \bar \nu_e$
	& $0.15 \pm 0.01$ & $1.5 \pm 0.1$ &$1.8 \pm 0.2$\\
$W^\pm Z$ where $\zee$
	& $0.15 \pm 0.01$ & $1.4 \pm 0.1$ &$1.7 \pm 0.2$\\
$\ttb \rightarrow \elel\nu\bar\nu + \bbb$  
	& $0.08 \pm 0.01$ & $1.1 \pm 0.1$ &$0.7 \pm 0.1$\\ \hline
\end{tabular}
\caption{\emph{\label{t_ewbackground} Summary of expected backgrounds. 
Cross sections for the electroweak and top processes are taken from the following 
references: \cite{Hamberg:1990np,Harlander:2002wh} for $W$ and $Z$, 
\cite{Campbell:1999ah} for $W^+W^-$ and $WZ$, and 
\cite{Cacciari:2003fi,Bonciani:1998vc,Catani:1996dj} for $t\bar t$.
Monte Carlo estimates are normalized to $72$~pb$^{-1}$.}}
\end{center}
\end{table}

%\clearpage
%*********************************************************************************** 
%*********************************************************************************** 
\section{Acceptance and corrections}
\label{s_acceptance}
	In order to correct the raw forward-backward asymmetry measured in $Z/\gamma$
decays and obtain \afbp, this analysis must account for any
acceptances and detector effects that can change the number of forward and backward 
events found in each \mee~bin.
Although there is very little in the detector and analysis that treats forward and
backward events differently, the angular distribution of the events, radiation, 
and detector resolution can each effectively change event acceptances 
and the reconstructed invariant mass so that net differences 
arise for forward and backward events.
The dominant contributions to the detector 
acceptances are kinematic acceptance ($a_{kin}^{\pm}$), geometric acceptance ($a_{geom}^{\pm}$), 
and electron identification efficiency ($\epsilon_{ID}^{\pm}$). The major effects
that contribute to event migration between $\mee$ bins are the energy 
resolution of the detector ($a_{res}^{\pm}$), and internal and external 
bremsstrahlung ($a_{rad}^{\pm}$).
Although the simulation is ultimately used to determine the corresponding corrections,
generator-level distributions are used to separate and understand 
the various effects (Secs.~\ref{s_fidkin},~\ref{s_eres},~and~\ref{s_rad}). 
These studies assume the Standard Model $\afb$ distribution
from PYTHIA, and are included to demonstrate how the acceptance and detector
effects independently change $\afbr$.

We present two unfolding methods to reconstruct $\afbp$. In Sec.~\ref{s_smear}, 
the simulation is used to find a parameterization to transform
$\afbp$ into $\afbr$. This parameterization is used to determine the
$\afbp$ that corresponds to our measured $\afbr$.
In Sec.~\ref{s_corrfac}, 
correction factors are calculated which transform the number of forward and
backward events in the candidate sample to the number of forward
and backward events that existed prior to being degraded by  detector acceptances and 
$\mee$ bin migration. The bias from the input $\afbp$ on the correction
factor calculation is also discussed. Both unfolding
analyses assume the Standard Model $\frac{d\sigma}{dM_{ee}}$ to be able to calculate 
the event migration effects.

%*********************************************************************************** 
\subsection{Fiducial and kinematic acceptance: $a^{\pm}_{geom}$ and $a^{\pm}_{kin}$}
\label{s_fidkin}
\begin{figure}
\begin{center}
\includegraphics[width=8cm]{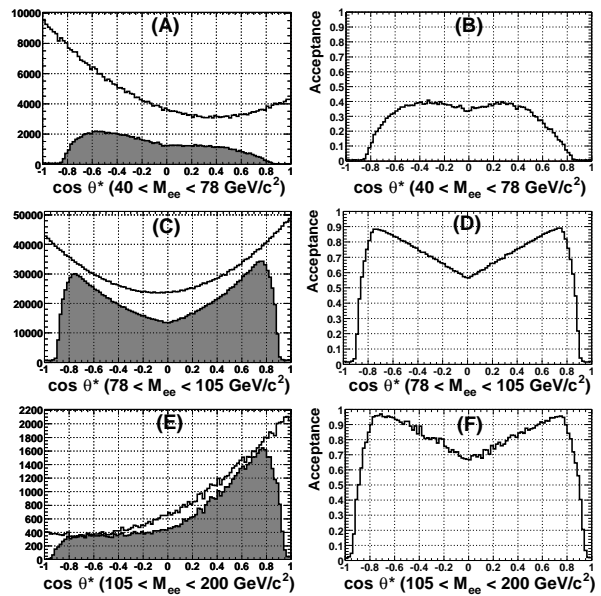}
\caption{\emph{For three mass ranges, the $\cost$ distributions 
(A, C, E) before (open histograms) and after 
(filled histograms) the geometric and kinematic cuts;
and the acceptance, $a^{\pm}_{geom} \cdot a^{\pm}_{kin}$, as a function of
$\cost$ (B, D, F).}}
\label{f_gen_cos_cuts}
\end{center}
\end{figure}

\begin{figure}
\begin{center}
\includegraphics[width=8cm]{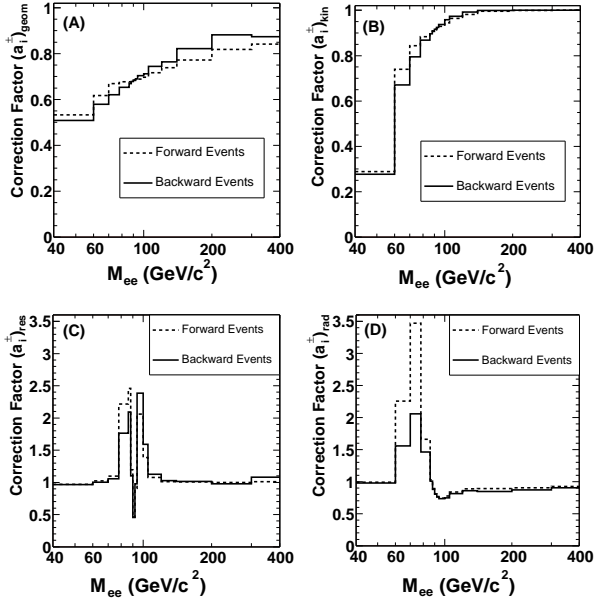}
\vspace {0in}
\caption{\emph{The $\mee$ dependence of acceptances; $a^{\pm}_{geom}$ (A), 
$a^{\pm}_{kin}$ (B), $a^{\pm}_{res}$ (C), and $a^{\pm}_{rad}$ (D). 
The kinematic acceptance, $a^{\pm}_{kin}$, is calculated after geometric cuts are applied, 
the acceptance due to resolution, $a^{\pm}_{res}$, is calculated after geometric and 
kinematic cuts are applied, and the acceptance due to radiation, $a^{\pm}_{rad}$,
is calculated after geometric and kinematic cuts and resolution smearing are applied.}}
\label{f_gen_accept_sum}
\end{center}
\end{figure}

	The kinematic and fiducial event requirements sculpt the polar angular 
distribution of the outgoing electrons and positrons, especially at high
$|\cost|$. The distributions in Fig.~\ref{f_gen_cos_cuts} show
this effect for Monte Carlo events prior to the simulation of detector and
radiation effects. The invariant-mass 
range is split into three different regions, the low-mass region ($40 < \mee < 
78$~GeV/$c^2$), the $Z$ pole region ($78 < \mee < 105$~GeV/$c^2$), and 
the high-mass region ($M_{ee} > 105$~GeV/$c^2$). 
These are the regions where $\afb$ is roughly at the low extremum, middle 
point, and high extremum, respectively. Although the acceptance for
these requirements is nearly symmetric for positive and negative $\cost$, the 
initial $\cost$ distributions are asymmetric, due to the forward-backward 
charge asymmetry.   When integrating the positive
and negative portions of $\cost$, the convolution of the Drell-Yan 
spectrum with the acceptance gives a different total acceptance
for forward and backward events.
For example, 
the detector has a low acceptance for events with very high $|\cost|$ because of 
the polar coverage, and in the case of high-mass events, this removes a greater percentage of 
forward events than backward events (see Fig.~\ref{f_gen_cos_cuts}E). 
The $M_{ee}$ dependence of the fiducial 
and kinematic acceptance is shown in Figs.~\ref{f_gen_accept_sum}A and \ref{f_gen_accept_sum}B, 
where the forward events have a higher acceptance than the 
backward events below the $Z$ pole, and vice versa above the $Z$ pole.

%*********************************************************************************** 
\subsection{Corrections for energy resolution: $a^{\pm}_{res}$}
\label{s_eres}
	The energy resolution of the calorimeter causes mismeasurements
of the invariant mass which can place events in the wrong invariant-mass bin.
If the asymmetry is changing with invariant mass, events placed in the 
wrong invariant-mass bin alter the measured asymmetry in that bin. 
The effect of the energy resolution on the measured asymmetry is therefore largest near 
the $Z$ pole where $\frac{d\afb}{d\mee}$ is largest and the bin sizes are smallest.  
As events from the $Z$ pole are smeared out, some events move
to higher masses where \afb~is more forward, and some events move to lower masses
where the \afb~is more backward. This event migration increases the correction
factor for backward events above the $Z$ pole and increases the correction factor
for forward events below the $Z$ pole.
The effect of the energy resolution on the correction factor is shown 
Fig.~\ref{f_gen_accept_sum}C. This plot is made by smearing the energy of 
the generator-level electron according to the measured detector response. 

%*********************************************************************************** 
\subsection{Corrections for bremsstrahlung: $a^{\pm}_{rad}$}
\label{s_rad}	
	The invariant mass can also be mis-measured due to the effects of
final state bremsstrahlung.  Most of these photons are  
emitted co-linearly with the electrons and deposit their energy into the
same calorimeter towers as the electron. In these cases, the energy of
the electron is recombined with its radiation products. 
Photons which are not recombined with electrons 
in the energy measurement lower the measured invariant mass and can cause a
candidate event to land in the 
wrong invariant-mass bin.  The effect is expected to be the most significant just below 
the $Z$ pole.  Figure~\ref{f_gen_accept_sum}D demonstrates the 
effect of internal bremsstrahlung on the acceptance at the generator level (no
external bremsstrahlung is included in this plot), when photons are not recombined 
with electrons. The correction factor below the $Z$ pole is large because of
events from the $Z$ pole that are being mis-measured due to radiation. The \afb~
at the $Z$ pole is larger, and boosts the correction factor for forward events.

\subsection{Parameterized acceptance and smearing}
\label{s_smear}
	In order to transform $\afbp$ into $\afbr$, a parameterized
function is needed to take into account the event loss and 
migration of events from one invariant-mass bin to another. From the Monte 
Carlo simulation, we compute the efficiency $\epsilon^{FF}_{ij}$ for an event that 
is forward and in \mee~bin $i$
to be found in the detector as forward in bin $j$. 
Likewise, $15\times15$ efficiency matrices are calculated for backward events
being found as backward ($\epsilon^{BB}_{ij}$), forward as backward ($\epsilon^{FB}_{ij}$), 
and backward as forward ($\epsilon^{BF}_{ij}$).
So the diagonal elements of $\epsilon^{FF}$ and $\epsilon^{BB}$ approximately represent the acceptance
for events in those bins, and the off-diagonal elements approximately represent the event bin migration.
The elements of $\epsilon^{BF}$ and $\epsilon^{FB}$ are much smaller and represent
those events that are reconstructed with a $\cost$ of the wrong sign. These efficiencies
have a small dependence on the underlying $\cost$ distribution, and so there is some residual
dependence on the input $\afbp$ used to calculate them. This might have a small bias on the result
if the $\cost$ distribution in the data is very different from the Standard Model. In Sec.~\ref{s_rawAfb}, the
$\cost$ distribution shows very good agreement between data and the Standard Model prediction.
The four matrices are defined as
\begin{eqnarray}
\epsilon^{FF}_{ij} &=& {\frac{N^F_j(sel)}{N^F_i(gen)}} ~;~
\epsilon^{BB}_{ij} = {\frac{N^B_j(sel)}{N^B_i(gen)}}  \\
\epsilon^{FB}_{ij} &=& {\frac{N^B_j(sel)}{N^F_i(gen)}} ~;~
\epsilon^{BF}_{ij} = {\frac{N^F_j(sel)}{N^B_i(gen)}}
\end{eqnarray}
where $N^{F}_{i}(gen) ~[N^{B}_{i}(gen)]$ is the number of forward [backward] events 
generated in the {\it i}-th $\mee$ bin, and $(N^{F}_{i}(sel) ~[N^{B}_{i}(sel)]$ is the number of 
forward [backward] events selected in the {\it i}-th $\mee$ bin.
By solving for $N^F$ or $N^B$ in Eq.~(\ref{e_afbdef}), we find the expected
number of forward and backward events for a given $\afb$,
\begin{eqnarray}
N^{F/B}_{i}(phys) = {N^{tot}_{i}(phys)\cdot(0.5 \pm 0.5\cdot(\afbp)_i)}.
\label{e_Ncalc}
\end{eqnarray}
$N^{tot}_{i}(phys)$ is the total number of events in bin $i$, and is computed from
the Standard Model $\frac{d\sigma}{dM_{ee}}$. 
Then by combining the expected number of events and their corresponding efficiencies, the
expected number of events found in the detector is written as
\begin{eqnarray}
N^{F}_{j}(raw) = \sum_i^{15}( \epsilon^{FF}_{ij}\cdot N^{F}_{i}(phys)+\epsilon^{BF}_{ij}\cdot N^{B}_{i}(phys)) \\
N^{B}_{j}(raw) = \sum_i^{15}( \epsilon^{BB}_{ij}\cdot N^{B}_{i}(phys)+\epsilon^{FB}_{ij}\cdot N^{F}_{i}(phys))
\end{eqnarray}
For a given  $\textbf{A}$ or 15 $\afbp$ values, $N^{F}_{j}(raw)$ and $N^{B}_{j}(raw)$ can be used together with 
Eq.~(\ref{e_afbdef}) to find
$(\afbr)_j$, which can be directly compared to the \afbr~measured in the detector. 
\begin{eqnarray}
(\afbr)_j = g(\textbf{A},j) = \frac{N^{F}_{j}(raw)-N^{B}_{j}(raw)} {N^{F}_{j}(raw)+N^{B}_{j}(raw)}
\end{eqnarray}

\subsection{Calculation of the correction factor $a^{\pm}_{cor}$}
\label{s_corrfac}

	The multiplicative correction factor $(a^{+}_{cor})_i$ 
[$(a^{-}_{cor})_i$], used to correct $\afb$ in Eq.~(\ref{e_afbcalc}),
is designed to be multiplied by the number of forward [backward] events
in the $i$-th $\mee$ bin. It corrects for event losses and for events 
that migrated into bin~$i$ from another bin with a different $\afb$. 
The full simulation is used to calculate the $a^{\pm}_{cor}$ 
which include the effects of fiducial and kinematic acceptance, energy resolution, 
bremsstrahlung, and electron selection efficiency. The combination 
of all these effects is defined as $(a^{\pm}_{cor})_{i}$ in Eq.~(\ref{e_accept}). 

\begin{eqnarray}
(a^{\pm}_{cor})_i & = & {\frac{N^{F/B}_{i}(sel)} {N^{F/B}_{i}(gen)}} 	\\
	&=& (a^{\pm}_{geom})_i \cdot (a^{\pm}_{kin})_i \cdot (a^{\pm}_{res})_i \cdot 
	(a^{\pm}_{rad})_i \cdot (\epsilon^{\pm}_{ID})_i, \nonumber
\label{e_accept}
\end{eqnarray}

\begin{figure}
\begin{center}
\includegraphics[width=8cm]{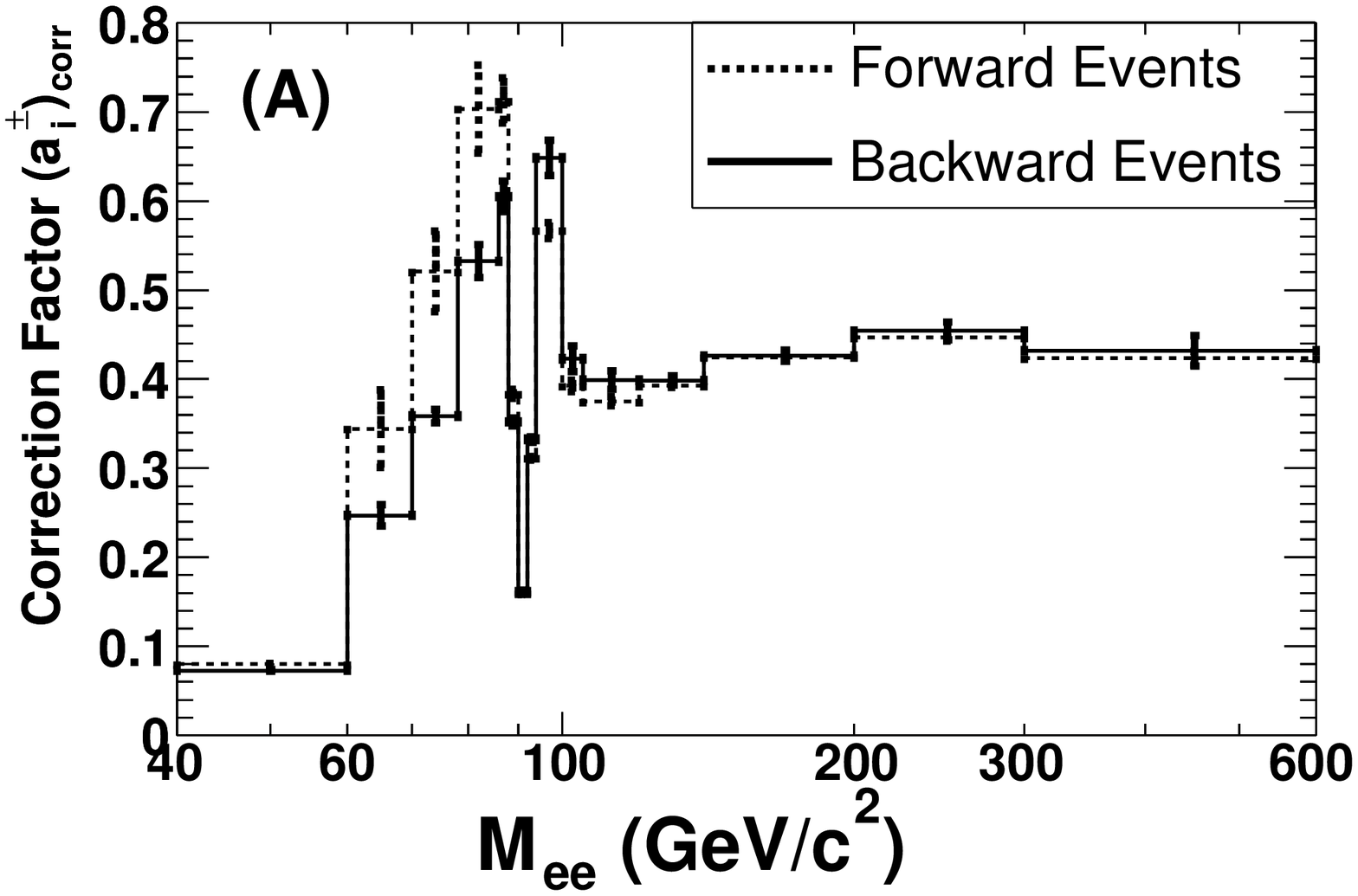}
\includegraphics[width=8cm]{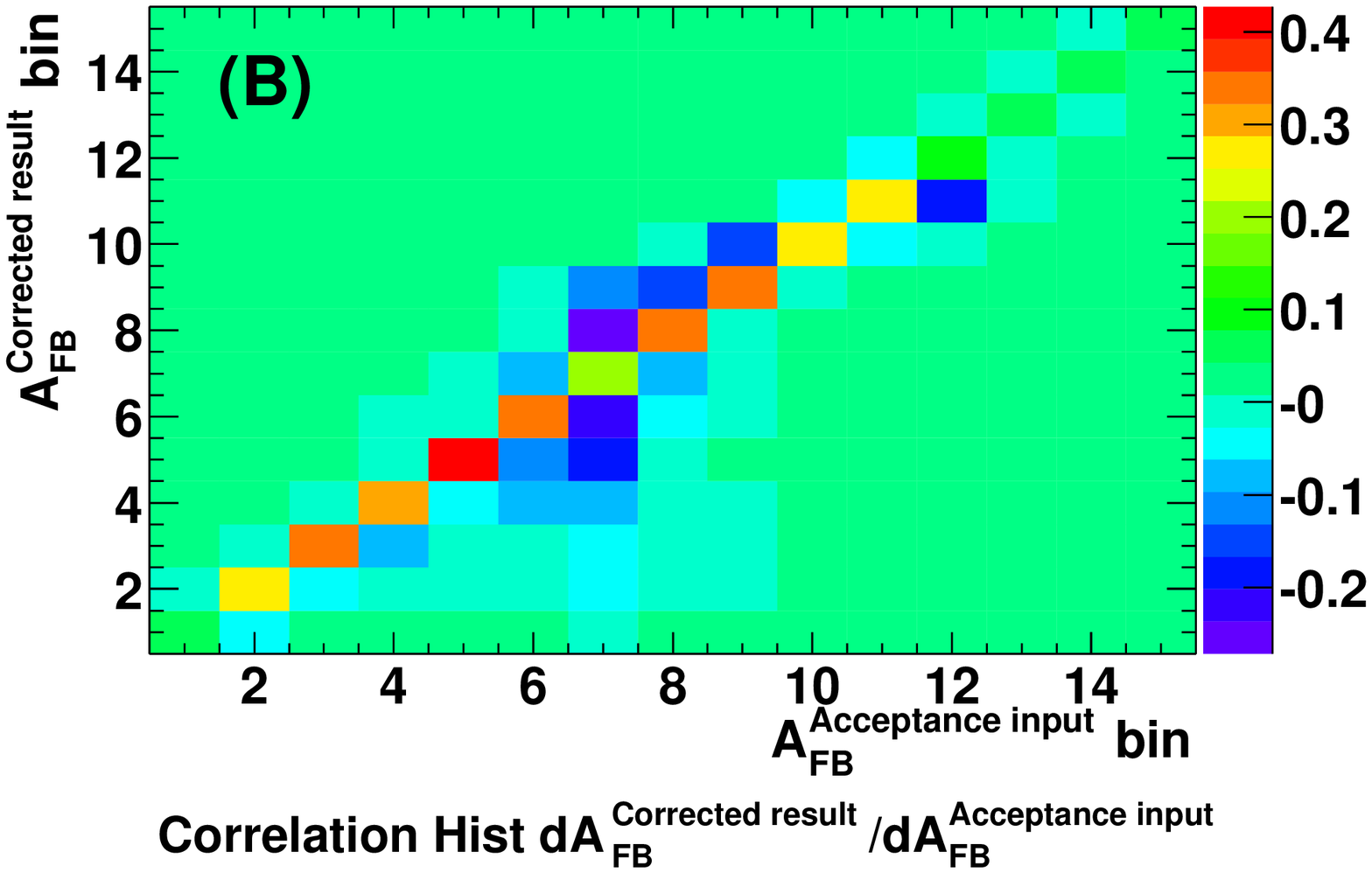}
\caption{\emph{
The correction factors, $a^{\pm}_{cor}$ (A).
The input $\afbp$ is taken from the $\afb$ couplings fit ({\rm Sec.~\ref{s_afb_couplings}}).
The correlation matrix between the input $\afbp$ to the correction factor calculation 
and the corresponding change
in the result due to Eq.~(\ref{e_afbcalc}) (B). The axes are labeled by the 
\mee~bin number where bin 1 is the lowest mass bin (40$<\mee<$60~GeV/$c^{2}$), bin 15 
is the highest mass bin (300$<\mee<$600~GeV/$c^{2}$), and bin 7 is the 
$Z$ pole (90$<\mee<$92~GeV/$c^{2}$).}}
\label{f_sim_tot_acc}
\end{center}
\end{figure}

In Fig.~\ref{f_sim_tot_acc}A and Table~\ref{t_sim_tot_acc}, the 
resulting correction factor is shown as a function of the dielectron invariant mass.  
This result can be compared to the studies done at the generator level for each individual
effect (see Fig.~\ref{f_gen_accept_sum}).  The overall acceptance is lower due to
the more detailed fiducial cuts and electron identification efficiencies in the 
full simulation. It is important to note that the correction for event migration
depends on the input $\afbp$ assumption used to calculate $(a^{\pm}_{cor})_i$.
The correction can only be calculated by assuming the difference in $\afbp$ between
the bin being corrected and the bins from which the events migrated. If the
$\afbp$ is significantly different in the two bins and there is a large amount of event 
migration, the result of the correction can be biased by this assumption.

	To measure the bias of the input $\afb$ on the final result, the
correlation matrix ($\frac{dA_{FB}^{\rm{Corrected~result}}}{dA_{FB}^{\rm{Acceptance~Input}}}$)
is measured by separately reweighting the
input $\afb$ in each bin in the Monte Carlo sample
used to calculate the correction factor. The modified correction
factors are then used to correct the $\afb$ from a high-statistics
pseudo-experiment using Eq.~(\ref{e_afbcalc}), and the change 
in the $\afb$ result in each $\mee$ bin is measured relative to 
the change in the input $\afb$ (Fig.~\ref{f_sim_tot_acc}B). 
The lowest mass bin and four highest mass bins are fairly independent 
of the $\afbp$ used to calculate the acceptance, whereas the bins near 
the $Z$ pole have strong correlations.

	To minimize the bias, one of the results uses the $\afbp$ 
from Sec.~\ref{s_afb_couplings}, calculated using Standard Model constraints, 
as the input to the $(a^{\pm}_{cor})_i$ calculation. 
In this approach, the uncertainties on the input $\afbp$ from 
the coupling fits are taken as systematic uncertainties on 
the acceptance measurement. 

\begin{table}
\begin{center}
\begin{tabular}{|c||c|c|}\hline
Mass Range
        & Forward & Backward \\ \hline\hline
$ 40 < M_{ee} < 60$ GeV/c$^2$ & 0.08 $\pm$ 0.00  &  0.07 $\pm$ 0.00   \\ \hline
$ 60 < M_{ee} < 70$ GeV/c$^2$ & 0.34 $\pm$ 0.05  &  0.25 $\pm$ 0.01   \\ \hline
$ 70 < M_{ee} < 78$ GeV/c$^2$ & 0.53 $\pm$ 0.05  &  0.36 $\pm$ 0.01   \\ \hline
$ 78 < M_{ee} < 86$ GeV/c$^2$ & 0.72 $\pm$ 0.06  &  0.53 $\pm$ 0.02   \\ \hline
$ 86 < M_{ee} < 88$ GeV/c$^2$ & 0.72 $\pm$ 0.03  &  0.60 $\pm$ 0.02   \\ \hline
$ 88 < M_{ee} < 90$ GeV/c$^2$ & 0.39 $\pm$ 0.01  &  0.35 $\pm$ 0.00   \\ \hline
$ 90 < M_{ee} < 92$ GeV/c$^2$ & 0.16 $\pm$ 0.00  &  0.16 $\pm$ 0.00   \\ \hline
$ 92 < M_{ee} < 94$ GeV/c$^2$ & 0.31 $\pm$ 0.00  &  0.33 $\pm$ 0.00   \\ \hline
$ 94 < M_{ee} < 100$ GeV/c$^2$ & 0.56 $\pm$ 0.01  &  0.65 $\pm$ 0.02   \\ \hline
$ 100 < M_{ee} < 105$ GeV/c$^2$ & 0.39 $\pm$ 0.00  &  0.43 $\pm$ 0.01   \\ \hline
$ 105 < M_{ee} < 120$ GeV/c$^2$ & 0.37 $\pm$ 0.00  &  0.40 $\pm$ 0.00   \\ \hline
$ 120 < M_{ee} < 140$ GeV/c$^2$ & 0.39 $\pm$ 0.00  &  0.40 $\pm$ 0.01   \\ \hline
$ 140 < M_{ee} < 200$ GeV/c$^2$ & 0.42 $\pm$ 0.00  &  0.43 $\pm$ 0.01   \\ \hline
$ 200 < M_{ee} < 300$ GeV/c$^2$ & 0.45 $\pm$ 0.00  &  0.45 $\pm$ 0.01   \\ \hline
$ 300 < M_{ee} < 600$ GeV/c$^2$ & 0.42 $\pm$ 0.00  &  0.43 $\pm$ 0.01   \\ \hline
\hline
\end{tabular}
\caption{\emph{\label{t_sim_tot_acc}
The correction factors, $a^{\pm}_{cor}$, for the
different mass bins using $\afb$ from the Standard Model coupling
fits of Sec.~\ref{s_afb_couplings}.
}}
\end{center}
\end{table}

%*********************************************************************************** 
%*********************************************************************************** 
\section{Systematic uncertainties in $A_{FB}$}
\label{s_sys}
	Systematic uncertainties due to energy scale, 
energy resolution, the amount of passive material in the detector, and
background estimation are considered. 
For a given source of uncertainty, a change is made to the corresponding parameter in
the simulation, and the impact on the asymmetry is evaluated after that change. 
The difference between the asymmetry with the changed parameter and the nominal one is taken
as the uncertainty from that source. The change in the input parameter is either
a one standard deviation (1$\sigma$) uncertainty on the variable in question
or a change in an assumption on that input. The systematic uncertainties depend 
on the method that is used to extract 
$\afbp$. There are separate tables for the systematic uncertainty
on $\afbp$ for the two $\afb$ measurements 
(Tables~\ref{t_sys_error_nonSM}~and~\ref{t_sys_error_SM}). 
The $\afb$ measurement based on the $a^{\pm}_{cor}$ correction
factors has an additional uncertainty due to the Standard Model
assumptions used the calculate these factors (see Table~\ref{t_sys_error_SM}).
The systematic uncertainties are quoted with a sign that represents 
the sign of the change in $\afb$ due to the $+1\sigma$ variation 
and is not used in the measurement of $\afb$. A positive variation
represents an increase in the energy scale and resolution, an increase
in the amount of material, and an increase in the amount of background.

The following systematic uncertainties have been investigated and have been 
found to have negligible effects on the measurement of $\afb$: fiducial 
acceptance, charge misidentification, $\afb$(dijet), and trigger efficiency. 
These effects are not included in the total systematic uncertainty.

\subsection{Systematic uncertainty from energy scale}
\begin{figure}
\begin{center}
\includegraphics[width=8cm]{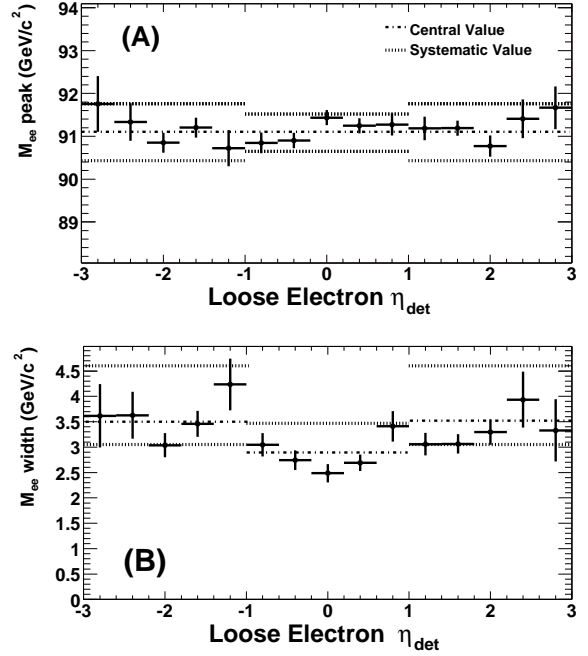}
\vspace{0in}
\caption{\emph{The peak (A) and width (B) of a Gaussian fit 
to the $\mee$ peak between 86 and 98 GeV/$c^2$ as a function of $\eta_{det}$ of loose
electrons.  The points are the data, and the lines are the 
chosen variations in the energy scale and resolution used 
to estimate the systematic uncertainties.}}
\label{f_eemass_eta_Esc_sys}
\end{center}
\end{figure}

	Variations in the calorimeter energy scale can lead to events being
placed in the wrong invariant-mass bins. For example, near the $Z$ pole where
the asymmetry is increasing monotonically with respect to the invariant mass, 
a positive variation in the energy scale will cause a systematic
decrease in the asymmetry (or a systematic increase for a negative variation in the 
energy scale). In general, a variation in the energy scale will have an 
effect only in the region where the bins are of the same order as the size of the 
variations (or smaller) and where the asymmetry is changing. In this analysis, 
uncertainties due to the energy scale are expected only in the region near the $Z$ pole.  
Figure~\ref{f_eemass_eta_Esc_sys} shows the Gaussian peak of the 
invariant mass as a function of the $\eta_{det}$ of the electron.
Based on the distribution of calculated masses, the central calorimeter scale is 
varied by 0.5\% and the plug calorimeter scale is varied by 
1\% to estimate the systematic uncertainties. 
The chosen energy scale 
variations are shown as lines in Fig.~\ref{f_eemass_eta_Esc_sys}. 
The corresponding shifts in $A_{FB}$ are shown in 
Tables~\ref{t_sys_error_nonSM}~and~\ref{t_sys_error_SM}.

%*********************************************************************************** 
\subsection{Systematic uncertainty from energy resolution}
	Variations in the energy resolution impact the forward-backward charge
asymmetry in much the same way as variations in the energy scale. Instead of
systematically shifting the events upwards or downwards, they tend to smear 
the forward-backward charge asymmetry to an average of
the bins around the bin in question. For example, a positive variation 
in the resolution near the $Z$ peak will cause a systematic decrease in the
measured asymmetry above the $Z$ peak and a systematic increase in the asymmetry below
the $Z$ peak. As in the case of the energy scale, only narrow bins in the region 
where the $\frac{d\afb}{d\mee}$ is large will be affected. 
	Figure~\ref{f_eemass_eta_Esc_sys} shows the Gaussian width of the 
invariant mass as a function of the $\eta_{det}$ of the electron.  
Based on the distribution of the widths, the central and plug 
calorimeter resolutions varied by 
0.5~GeV in the central calorimeter and 1.5~GeV in the plug calorimeter. In
the central calorimeter, only the change due to increasing the resolution
is used in calculating the systematic. The chosen
variations are shown as lines in Fig.~\ref{f_eemass_eta_Esc_sys}. 
The corresponding shifts $\afb$ are shown in 
Tables~\ref{t_sys_error_nonSM}~and~\ref{t_sys_error_SM}.

%*********************************************************************************** 
\subsection{Systematic uncertainty from amount of material}
The amount of material in front of the calorimeter affects the energy measurement of electrons.
The uncertainty in the amount of material 
is estimated to be less than 1.5\% of a radiation length ($X_0$) 
in the region between the interaction point and the tracking volume,
and less than ${\frac{1}{6}} X_0$ in the region between the collision point
and the plug calorimeter. 

	The systematic uncertainty on $\afb$ due to the material is estimated by changing
the amount of material in the simulation.  
The changes include adding or subtracting an extra $1.5\%X_0$ of copper 
in a cylinder at 34~cm (just before the COT) and ${\frac{1}{6}}X_0$ 
of steel on the face of the plug calorimeter. 
The corresponding shifts in $A_{FB}$ are shown in 
Tables~\ref{t_sys_error_nonSM}~and~\ref{t_sys_error_SM}.
The bins most sensitive to the amount of material are those just below the $Z$ pole;
this is more easily seen in the systematics on the fit result (Table~\ref{t_sys_error_nonSM}).

%*********************************************************************************** 
\subsection{Systematic uncertainty from background subtraction}
	The central values of $\afb$ are calculated after subtracting 
background events in the forward and backward regions separately. 
The number of background events in each $\mee$ bin are estimated in Sec.~\ref{s_backgrounds}. 
The systematic uncertainties due to the background estimates are taken as shifts in $A_{FB}$ 
when the estimated numbers of background events are varied by their uncertainties.  
The corresponding shifts $A_{FB}$ are shown in 
Tables~\ref{t_sys_error_nonSM}~and~\ref{t_sys_error_SM}.

%%%%%%%%%%% Summary of Systematic Uncertainties %%%%%%%%%%%%%%%%%%%%%%%%%%%%%%%%%%%
\begin{table}
\begin{center}
\begin{tabular}{|c||r|r|r|r||c|c|}\hline
Mass Range  & Energy & Energy & Material  &Bgrnd & Tot & Stat\\ 
(GeV/$c^2$) & \multicolumn{1}{|c|}{Scale} & \multicolumn{1}{|c|}{Resol.} & & & & \\ \hline\hline
$ 40 - 60$ & -0.003 &  0.010 & -0.035 & -0.034 &   0.050 &  0.127   \\ \hline
$ 60 - 70$ &  0.008 &  0.008 &  0.033 & -0.065 &   0.074 &  0.185   \\ \hline
$ 70 - 78$ & -0.018 &  0.019 &  0.040 & -0.017 &   0.051 &  0.186   \\ \hline
$ 78 - 86$ &  0.037 &  0.064 & -0.041 &  0.001 &   0.085 &  0.164   \\ \hline
$ 86 - 88$ & -0.080 &  0.112 &  0.135 & -0.001 &   0.193 &  0.211   \\ \hline
$ 88 - 90$ & -0.072 & -0.032 &  0.043 & -0.000 &   0.090 &  0.108   \\ \hline
$ 90 - 92$ &  0.058 &  0.042 & -0.024 &  0.000 &   0.076 &  0.064   \\ \hline
$ 92 - 94$ & -0.118 & -0.098 &  0.049 &  0.000 &   0.161 &  0.161   \\ \hline
$ 94 - 100$ &  0.031 & -0.077 &  0.013 &  0.000 &   0.084 &  0.168   \\ \hline
$ 100 - 105$ & -0.046 & -0.146 & -0.103 & -0.000 &  0.185 &  0.258   \\ \hline
$ 105 - 120$ & -0.001 &  0.014 &  0.045 &  0.014 &  0.049 &  0.138   \\ \hline
$ 120 - 140$ & -0.003 & -0.004 & -0.003 & -0.001 &  0.006 &  0.185   \\ \hline
$ 140 - 200$ &  0.002 &  0.003 &  0.012 &  0.004 &  0.013 &  0.165   \\ \hline
$ 200 - 300$ & -0.005 & -0.007 &  0.006 &  0.026 &  0.029 &  $^{+0.20}_{-0.27}$ \\ \hline
$ 300 - 600$ & -0.004 &  0.002 & -0.023 &  0.000 &  0.024 &  $^{+0.00}_{-0.51}$ \\ \hline
\hline
\end{tabular}

\caption{\emph{\label{t_sys_error_nonSM}
Sources of systematic uncertainty, the total systematic uncertainty (Tot), 
and the statistical uncertainty (Stat) on $\afbp$ when fitting using the parameterization
described in Sec.~\ref{s_smear} (see Sec.~\ref{s_fitAfb}). When a systematic shift is tested
in two directions ($+1\sigma$ and $-1\sigma$ of the variable in question), 
the larger shift is chosen.  The sign represents the sign of the change in $\afb$
due to the $+1\sigma$ variation and is not used in the measurement of $\afb$.}}
\end{center}
\end{table}

%%%%%%%%%%% Summary of Systematic Uncertainties %%%%%%%%%%%%%%%%%%%%%%%%%%%%%%%%%%%
\begin{table*}
\begin{center}
\begin{tabular}{|c||r|r|r|r|r||c|c|}\hline
Mass Range& Energy  & Energy & Material  &Bgrnd & Input & Tot & Stat\\ 
(GeV/$c^2$) & \multicolumn{1}{|c|}{Scale} &  \multicolumn{1}{|c|}{Resol.} & & & & & \\ \hline\hline
$ 40 - 60$ & -0.013 &  0.013 & -0.028 & -0.053 & -0.018 &  0.065 &  0.108   \\ \hline
$ 60 - 70$ &  0.012 &  0.012 &  0.027 & -0.038 &  0.074 &  0.089 &  0.095   \\ \hline
$ 70 - 78$ & -0.010 &  0.014 & -0.022 & -0.009 & -0.041 &  0.050 &  0.072   \\ \hline
$ 78 - 86$ & -0.015 &  0.033 &  0.012 & -0.001 & -0.053 &  0.066 &  0.043   \\ \hline
$ 86 - 88$ & -0.013 &  0.015 &  0.013 &  0.001 & -0.034 &  0.041 &  0.048   \\ \hline
$ 88 - 90$ & -0.011 &  0.005 &  0.011 & -0.000 & -0.014 &  0.021 &  0.035   \\ \hline
$ 90 - 92$ &  0.006 &  0.006 & -0.009 &  0.001 &  0.009 &  0.015 &  0.031   \\ \hline
$ 92 - 94$ & -0.009 & -0.005 & -0.010 &  0.000 & -0.004 &  0.014 &  0.033   \\ \hline
$ 94 - 100$ & -0.014 & -0.022 & -0.010 &  0.000 & -0.017 &  0.033 &  0.034   \\ \hline
$ 100 - 105$ & -0.017 & -0.080 & -0.045 &  0.002 & -0.015 &  0.095 &  0.099   \\ \hline
$ 105 - 120$ & -0.013 & -0.014 &  0.027 &  0.011 & -0.004 &  0.035 &  0.091   \\ \hline
$ 120 - 140$ & -0.004 & -0.004 &  0.008 &  0.023 & -0.008 &  0.026 &  $^{+0.14}_{-0.15}$  \\ \hline
$ 140 - 200$ &  0.004 &  0.004 &  0.011 &  0.006 & -0.007 &  0.016 &  $^{+0.14}_{-0.15}$  \\ \hline
$ 200 - 300$ & -0.008 & -0.008 &  0.017 &  0.033 & -0.005 &  0.039 &  $^{+0.18}_{-0.24}$  \\ \hline
$ 300 - 600$ & -0.017 &  0.017 & -0.032 &  0.000 &  0.000 &  0.040 &  $^{+0.00}_{-0.64}$   \\ \hline
\hline
\end{tabular}
\caption{\emph{\label{t_sys_error_SM}
Summary of uncertainties on $\afbp$ calculated using Eq.~(\ref{e_afbcalc}) and
the couplings fits for \afb~as input to the acceptance calculation. 
When a systematic shift is tested in two directions ($+1\sigma$ and $-1\sigma$ 
of the variable in question), the larger shift is chosen.  The sign represents 
the sign of the change in $\afb$ due to $+1\sigma$ variation and is not used in 
the measurement of $\afb$. The systematic uncertainty due to the Standard Model
assumptions used in the correction
factor calculation is labeled as ``Input.''}}
\end{center}
\end{table*}

%\clearpage
%*********************************************************************************** 
%*********************************************************************************** 
\section{Results}
\label{s_results}
%*********************************************************************************** 
\subsection{Comparisons of uncorrected distributions with the Standard Model}
\label{s_rawAfb}
\begin{figure}
\begin{center}
\includegraphics[width=8cm]{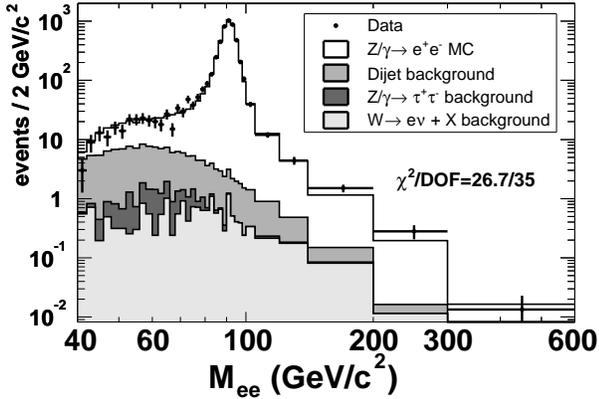}
\caption{\emph{Invariant-mass distribution of the data compared to the prediction 
for signal and background combined.
The points are the data, the histogram is the signal Monte Carlo sample, and the shaded 
histograms are the background predictions. The contributions are added or stacked.}}
\label{f_mass_tot}
\end{center}
\end{figure}

\begin{figure}
\begin{center}
\includegraphics[width=8cm]{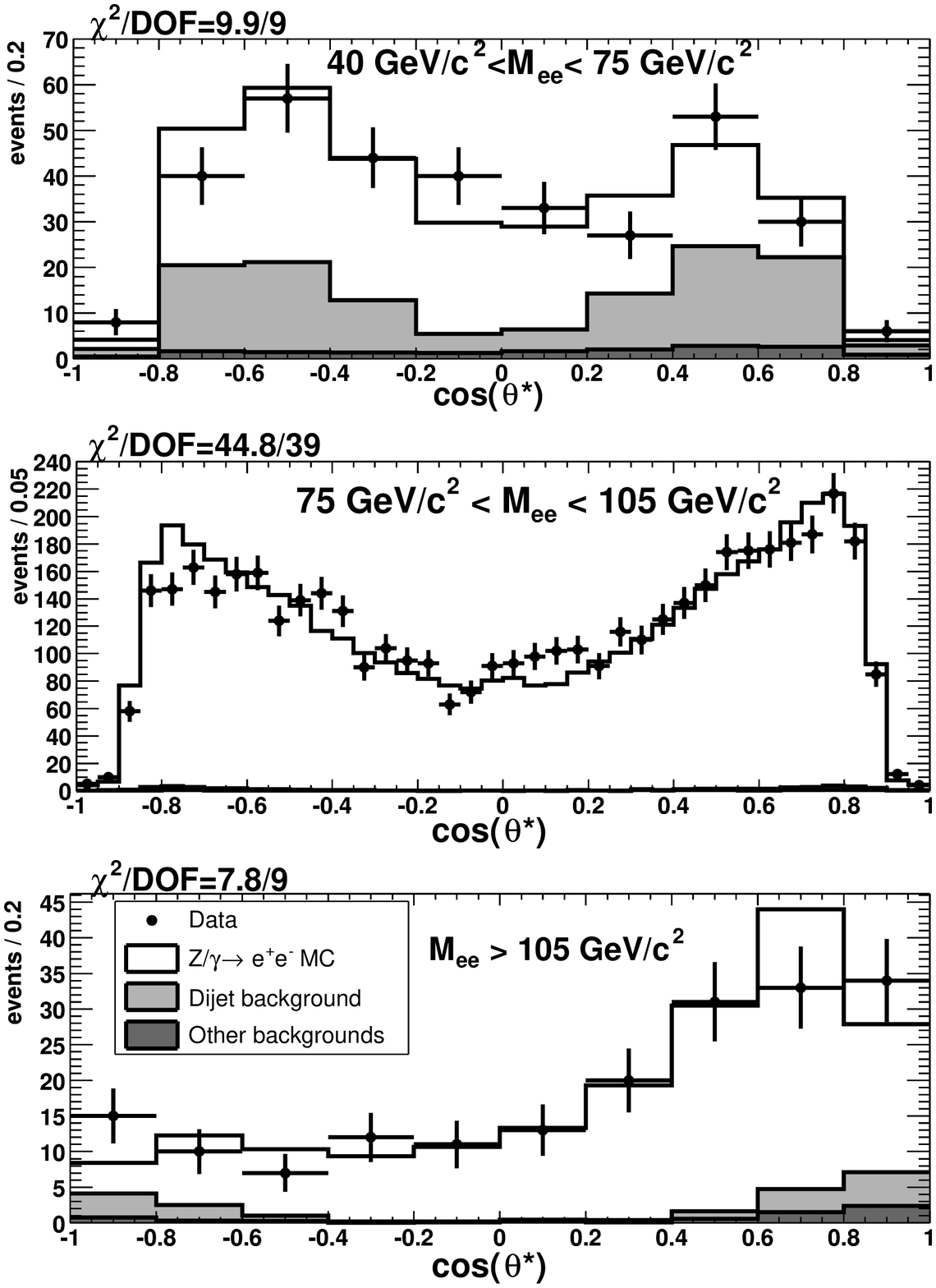}
\vspace{0in}
\caption{\emph{Distributions of $\cost$ for the three mass regions of the data 
compared to the predictions for the signal and background combined.
The points are the data, the open histograms are the predictions from 
signal Monte Carlo simulation, and the shaded 
histograms are the predictions from background. The contributions are added or stacked.}}
\label{f_costheta_tot}
\end{center}
\end{figure}

The best way to make a direct comparison of the data with the Standard Model is to 
use the simulation to compare the uncorrected data with simulated events. The background events are 
included using the predicted distributions as described in Sec.~\ref{s_backgrounds}.  
The distributions from the signal Monte Carlo simulation are normalized 
to the number of events in the data after subtracting the predicted background contribution.
Only statistical uncertainties have been included for the calculation of the $\chi^2$
comparisons. The uncorrected invariant-mass distribution from the data is compared to 
the signal and background predictions in Fig.~\ref{f_mass_tot}.
Since the energy scale and resolution in the simulation has been tuned to the data at the $Z$ peak 
(Sec.~\ref{s_ecorr}), the comparison of the $\mee$ lineshape gives a slightly better than 
expected $\chi^2$/DOF=26.7/35. The comparison is also made for the $\cost$ distribution (Fig.~\ref{f_costheta_tot}) 
in three mass regions where $\afb$ is at extremes; $40 < \mee < 75$~GeV/$c^2$ where $\afb$ is
large and negative (giving a $\chi^2$/DOF=9.9/9), $75 < \mee < 105$~GeV/$c^2$ where $\afb$ is small (giving
$\chi^2$/DOF=44.8/39), and $\mee > 105$~GeV/$c^2$ where $\afb$ is large and positive (giving $\chi^2$/DOF=7.8/9). 
Finally, the comparison of $\afbr$ in 15 $\mee$ bins (Table~\ref{t_events_sum}) with the Standard Model simulation gives
a $\chi^2$/DOF=15.7/15. The data shows excellent agreement with the Standard Model in
all of these distributions. The objective of the following sections will be to obtain 
the corrected $\afbp$ and $Z$ couplings that can be used without the CDF simulation.

%*********************************************************************************** 
\subsection{The Standard Model prediction}
\label{s_theory_band}
\begin{figure}
\begin{center}
\includegraphics[width=8cm]{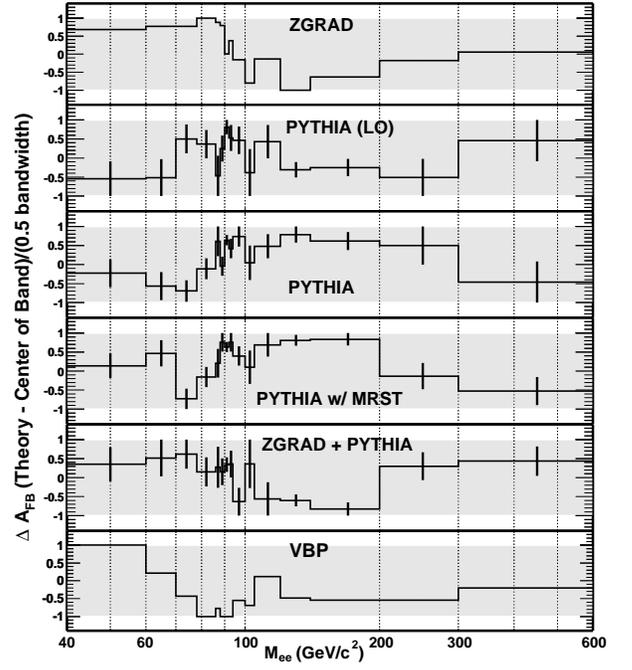}
\caption{\emph{Each of six theoretical calculations compared to the center and width of the 
theoretical band. The variation from different theoretical calculations is expressed as 
the width of the band which is determined by the highest and lowest values of $\afb$,
including the uncertainties, in each mass bin.
}}
\label{f_theoryband_relative}
\end{center}
\end{figure}

	Currently, there are a number of programs that generate Drell-Yan events 
produced in hadron collisions. PYTHIA generates events using leading-order (LO) cross sections 
and incorporates initial-state QCD radiation and initial- and final-state QED radiation 
via parton shower algorithms.
HERWIG uses LO cross sections with initial state QCD radiation via parton shower algorithms.  
ZGRAD~\cite{Baur:2001ze} includes 
full $\mathcal{O}(\alpha)$ electroweak corrections but no QCD corrections, resulting 
in $p_T^{Z/\gamma*} \simeq 0$.  The gluon resummation 
program VBP~\cite{Ellis:1997ii,Ellis:1997sc}, which does the gluon resummation in the $q_t$ space at low 
$p_T^{Z/\gamma*}$ and reduces to NLO
QCD at high $p_T^{Z/\gamma*}$, does not include any electroweak corrections. 
Unfortunately, there is no one program that includes both $\mathcal{O}(\alpha)$ 
electroweak and NLO QCD corrections. A calculation that includes $\mathcal{O}(\alpha)$
electroweak and some QCD corrections can be obtained by running ZGRAD with the parton showering
code in PYTHIA. 
Six Monte Carlo programs are used to constrain the possible values for the $\afb$ measurement. 
They are
PYTHIA, 
VBP,
ZGRAD,
ZGRAD + PYTHIA, and
PYTHIA with no QCD corrections with CTEQ5L parton distribution functions,
and PYTHIA with MRST2001 parton distribution functions.
In each mass bin, a band is constructed to
extend from the lowest to the highest values of the six $\afb$
calculations. The uncertainty on the theoretical prediction
due to different event generators is taken to be the
width of the band in any mass bin. In Fig.~\ref{f_theoryband_relative}, each calculation
is compared to the center and width of the band in each $\mee$ bin. 
A comparison of these Monte Carlo programs also demonstrates the extent
to which the Collin-Soper frame (Sec.~\ref{s_strategy}) minimizes the impact of the transverse 
momentum of the incoming quarks. The PYTHIA (LO) and ZGRAD programs, which have
no initial-state QCD radiation, can be compared to the VBP, ZGRAD+PYTHIA, and
PYTHIA programs which include initial-state QCD radiation. The difference
in $\afb$ between having and not having initial-state QCD radiation is negligible
compared to the $\afb$ measurement uncertainties (see Fig.~\ref{f_Afb_Results_fit}).

%*********************************************************************************** 
\subsection{$A_{FB}$ measurement without Standard Model constraints}
\label{s_fitAfb}

The forward-backward charge asymmetry is measured using
the maximum likelihood method and comparing $\afbr$ returned by the 
parameterized acceptance and smearing from Sec.~\ref{s_smear} to the
data at detector level. This method is unbiased since it does not make any
prior assumption about the values of \afb. The number of events in various 
invariant-mass bins between 40~GeV/$c^2$
and 600~GeV/$c^2$ and $\afbr$ are summarized in Table~\ref{t_events_sum}. 
 
	The probability for finding a forward or backward event for a given $\afb$ is
\begin{eqnarray}
P(^F_B) & = & {\frac{1}{2}(1\pm \afb)},
\end{eqnarray}
and the probability to find $N^F$ forward and $N^B$ backward events among
$N$ events is given by the binomial distribution:
\begin{eqnarray}
P(\afb,N^F,N^B) = && {(^N_{N^F})\cdot (1+\afb)^{N^F}} \nonumber\\
&&{\cdot (1-\afb)^{N^B}\cdot(\frac{1}{2})^{N^F+N^B}}.
\end{eqnarray}
Defining $\alpha$ as the negative logarithm of the likelihood function, we obtain:
\begin{eqnarray}
\alpha = \sum_i^{15}&&-N_i^F\cdot log(1+g(\textbf{A},i)) \nonumber \\
&&-N_i^B\cdot log(1-g(\textbf{A},i))+C,
\end{eqnarray}
\begin{eqnarray}
\mathrm{where}~{\textbf A}=\{(\afbp)_1,(\afbp)_2,\ldots,(\afbp)_{15}\}.
\end{eqnarray}

\begin{figure}
\begin{center}
\includegraphics[width=8cm]{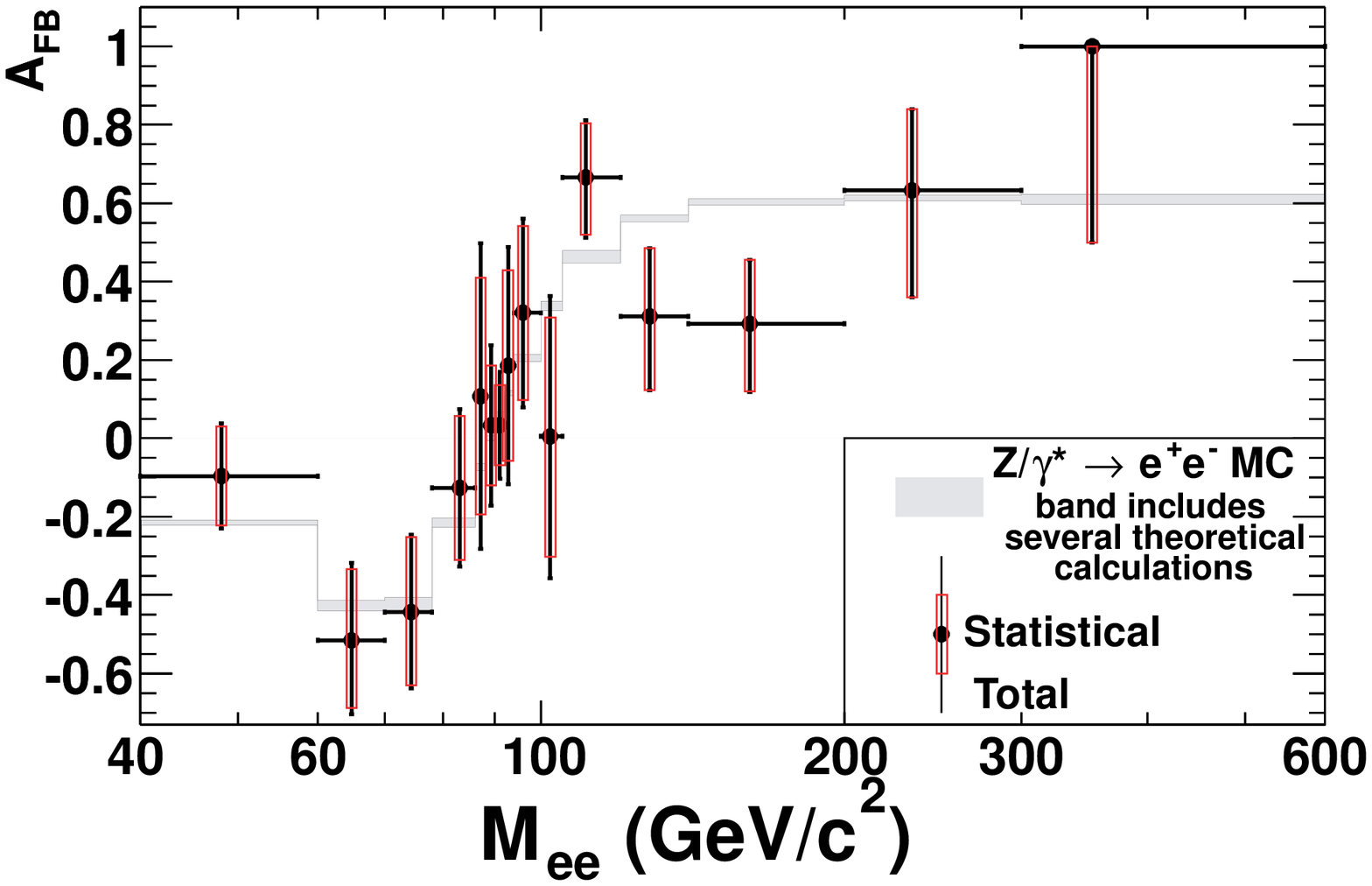}
\includegraphics[width=8cm]{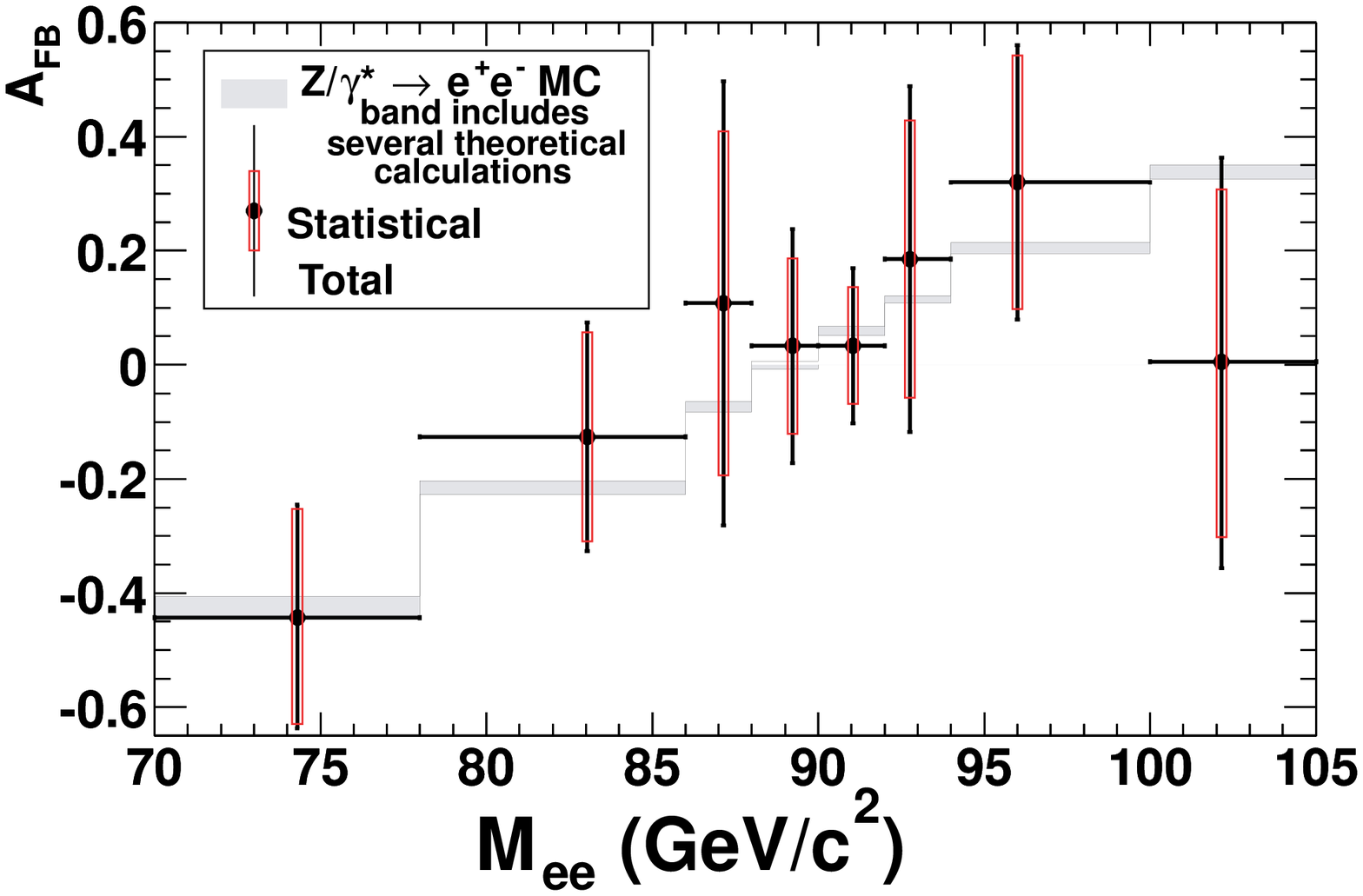}
\vspace{0in}
\caption{\emph{ Experimental results for $A_{FB}$ with statistical and systematic 
uncertainties (crosses), and theoretical predictions based on six independent
calculations as described in Sec.~\ref{s_theory_band} (bands). 
The experimental results for $\afb$ are measured by fitting to $\afbr$
with Tikhonov regularization (Sec.~\ref{s_fitAfb}). The bottom
figure shows the $Z$ pole region in more detail.}}
\label{f_Afb_Results_fit}
\end{center}
\end{figure}

The large correlations between invariant-mass bins near
the $Z$ pole (see Fig.~\ref{f_sim_tot_acc}B) lead to large uncertainties on $\afb$.
In this case, it is 
customary to add a regularization function to the likelihood~\cite{Cowan:2002in}
to reduce the variances. We choose
the Tikhonov regularization, which adds a function proportional to the 
second derivative of the true distribution.
\begin{eqnarray}
S[\textbf{A}_i] = {\int_{bin~i} \left(\frac{d^2 f_{true}(M_{ee},\textbf{A}_i)}{dM^2_{ee}}\right)^2 dM_{ee}}, \\
\mathrm{where}~{\textbf A}_i = \{(\afbp)_{i-1},(\afbp)_i,(\afbp)_{i+1}\},
\end{eqnarray}
and where the integral is performed over the entire range of bin $i$ in $\mee$.
Since the binning is not uniform, a parabolic form is assumed for $f_{true}(M_{ee},\textbf{A}_i)$,
\begin{eqnarray}
f_{true}(M_{ee},\textbf{A}_i) = a + b\cdot \mee + c \cdot M^2_{ee}.
\label{s_defF_true}
\end{eqnarray}
The parameters $a$, $b$, and $c$ are solved for by integrating $f$ over bins $i-1$, $i$, and
$i+1$, and setting the integrals equal to $\textbf{A}_i$. $S[\textbf{A}_i]$ can then be simplified
to
\begin{eqnarray}
S[\textbf{A}_i] = 4\cdot (c(\textbf{A}_i))^2 \cdot \Delta M^i_{ee}.
\end{eqnarray}
This function is only applied to bins 2 through 9, since those are the bins with large migration. 
Adding this to the likelihood, a new equation is made
\begin{widetext}
\begin{eqnarray}
\alpha' = {\sum_{i=1}^{15}\{-N_i^F\cdot log(1+g(\textbf{A},i))-N_i^B\cdot log(1-g(\textbf{A},i))\}
	 +\lambda\cdot\sum_{i=2}^{9} S[{\textbf A}_i]+C.}
\end{eqnarray}
\end{widetext}
The parameter $\lambda$ is called the regularization parameter, and is 
arbitrary. A very small $\lambda$ will have no effect on the fit, and a very large
$\lambda$ will dominate the likelihood. In this analysis $\lambda$ is chosen such that 
$\alpha'=\alpha+\frac{1}{2}$. The result is not sensitive to changes as large as 
a factor of 2 or $\frac{1}{2}$ in $\lambda$. The fitting package Minuit~\cite{James:1975dr} 
is used to minimize $\alpha'$, as a function of $\textbf{A}$. $\afbp$ obtained after Tikhonov 
regularization is shown in Fig.~\ref{f_Afb_Results_fit}
and in Table~\ref{t_afb_sum_fit} for different invariant-mass bins.
The statistical and systematic uncertainties of the measurement and the prediction
from PYTHIA are also given in  Table~\ref{t_afb_sum_fit}.

%%%%%%%%%%%% Summary of events and backgrounds %%%%%%%%%%%%%%%%%%%%%%%%%%%%%%%%%%%
\begin{table*}
\begin{center}
\begin{tabular}{|c||c|c|c|c|c|}\hline
Mass Range &\multicolumn{2}{|c|}{Observed Events}
&\multicolumn{2}{|c|}{Background}&   $\afbr$ \\ \cline{2-5}
(GeV/$c^2$) & $\cost>0$ & $\cost<0$ & $\cost>0$ & $\cost<0$ & $\pm$stat $\pm$ sys.  \\ \hline\hline
$ 40 < M_{ee} < 60$ & 76  &  78  &  37.8 $\pm$ 9.2  &  32.4 $\pm$ 9.2  &  -0.09 $\pm$ 0.11 $\pm$ 0.05   \\ \hline
$ 60 < M_{ee} < 70$ & 46  &  68  &  19.0 $\pm$ 4.6  &  17.4 $\pm$ 4.6  &  -0.31 $\pm$ 0.11 $\pm$ 0.04   \\ \hline
$ 70 < M_{ee} < 78$ & 69  &  98  &  12.4 $\pm$ 2.8  &  9.9 $\pm$ 2.7  &  -0.22 $\pm$ 0.08 $\pm$ 0.01   \\ \hline
$ 78 < M_{ee} < 86$ & 267  &  266  &  8.2 $\pm$ 2.0  &  7.4 $\pm$ 2.0  &   0.00 $\pm$ 0.04 $\pm$ 0.00   \\ \hline
$ 86 < M_{ee} < 88$ & 246  &  204  &  1.4 $\pm$ 0.4  &  1.6 $\pm$ 0.4  &   0.09 $\pm$ 0.05 $\pm$ 0.00   \\ \hline
$ 88 < M_{ee} < 90$ & 420  &  393  &  1.5 $\pm$ 0.4  &  1.3 $\pm$ 0.3  &   0.03 $\pm$ 0.04 $\pm$ 0.00   \\ \hline
$ 90 < M_{ee} < 92$ & 550  &  476  &  1.8 $\pm$ 0.4  &  1.4 $\pm$ 0.4  &   0.07 $\pm$ 0.03 $\pm$ 0.00   \\ \hline
$ 92 < M_{ee} < 94$ & 481  &  392  &  1.4 $\pm$ 0.3  &  1.3 $\pm$ 0.3  &   0.10 $\pm$ 0.03 $\pm$ 0.00   \\ \hline
$ 94 < M_{ee} < 100$ & 463  &  325  &  4.1 $\pm$ 0.9  &  3.1 $\pm$ 0.8  &   0.18 $\pm$ 0.04 $\pm$ 0.00   \\ \hline
$ 100 < M_{ee} < 105$ & 59  &  39  &  2.0 $\pm$ 0.5  &  1.6 $\pm$ 0.5  &   0.21 $\pm$ 0.10 $\pm$ 0.00   \\ \hline
$ 105 < M_{ee} < 120$ & 67  &  23  &  4.1 $\pm$ 0.9  &  3.2 $\pm$ 0.9  &   0.52 $\pm$ 0.09 $\pm$ 0.01   \\ \hline
$ 120 < M_{ee} < 140$ & 29  &  15  &  2.2 $\pm$ 0.6  &  2.2 $\pm$ 0.6  &   0.35 $\pm$ 0.15 $\pm$ 0.02   \\ \hline
$ 140 < M_{ee} < 200$ & 29  &  16  &  3.2 $\pm$ 0.6  &  1.9 $\pm$ 0.4  &   0.29 $\pm$ 0.15 $\pm$ 0.01   \\ \hline
$ 200 < M_{ee} < 300$ & 11  &  3  &  0.6 $\pm$ 0.2  &  0.5 $\pm$ 0.2  &   0.61 $\pm$ 0.22 $\pm$ 0.03   \\ \hline
$ 300 < M_{ee} < 600$ & 2  &  0  &  0.2 $\pm$ 0.1  &  0.0 $\pm$ 0.0  &   1.00$^{+0}_{-0.632}$ $\pm$ 0.00   \\ \hline
\hline
\end{tabular}
\caption{\emph{\label{t_events_sum}
Summary of observed events, estimated backgrounds, and $\afbr$ in the
dielectron sample from 72~pb$^{-1}$ of Run~II data. The systematic uncertainty
on $\afbr$ includes only the background subtraction.}}
\end{center}
\end{table*}

%%%%%%%%%%% Summary of Final Results %%%%%%%%%%%%%%%%%%%%%%%%%%%%%%%%%%%
\begin{table*}
\begin{center}
\begin{tabular}{|c||c||c|c|}\hline
Mass Range   & $<M_{ee}>$ & Measured $A_{FB}$ &PYTHIA  $A_{FB}$ \\
(GeV/$c^2$)  & (GeV/$c^2$) & & \\ \hline\hline
$ 40 \leq \mee < 60$ & 48.2 & -0.11 $\pm$ 0.13 $\pm$ 0.05  &  -0.214 $\pm$ 0.003   \\ \hline
$ 60 \leq \mee < 70$ & 64.9 & -0.51$^{+0.18}_{-0.17}$ $\pm$ 0.07  &  -0.420 $\pm$ 0.005   \\ \hline
$ 70 \leq \mee < 78$ & 74.3 & -0.45 $\pm$ 0.19 $\pm$ 0.05  &  -0.410 $\pm$ 0.005   \\ \hline
$ 78 \leq \mee < 86$ & 83.0 & -0.11 $\pm$ 0.17 $\pm$ 0.09  &  -0.214 $\pm$ 0.003   \\ \hline
$ 86 \leq \mee < 88$ & 87.1 & 0.07 $\pm$ 0.23 $\pm$ 0.19  &  -0.079 $\pm$ 0.004   \\ \hline
$ 88 \leq \mee < 90$ & 89.2 & 0.03 $\pm$ 0.13 $\pm$ 0.09  &  -0.001 $\pm$ 0.002   \\ \hline
$ 90 \leq \mee < 92$ & 91.0 & 0.047 $\pm$ 0.077 $\pm$ 0.076  &  0.054 $\pm$ 0.001   \\ \hline
$ 92 \leq \mee < 94$ & 92.8 & 0.15 $\pm$ 0.19 $\pm$ 0.16  &  0.112 $\pm$ 0.002   \\ \hline
$ 94 \leq \mee < 100$ & 96.0 & 0.35 $\pm$ 0.20 $\pm$ 0.08  &  0.198 $\pm$ 0.003   \\ \hline
$ 100 \leq \mee < 105$ & 102.2 & -0.02 $\pm$ 0.30 $\pm$ 0.19  &  0.338 $\pm$ 0.006   \\ \hline
$ 105 \leq \mee < 120$ & 110.7 & 0.67 $\pm$ 0.15 $\pm$ 0.05  &  0.454 $\pm$ 0.006   \\ \hline
$ 120 \leq \mee < 140$ & 128.2 & 0.32$^{+0.17}_{-0.19}$ $\pm$ 0.01  &  0.554 $\pm$ 0.002   \\ \hline
$ 140 \leq \mee < 200$ & 161.2 & 0.29$^{+0.16}_{-0.17}$ $\pm$ 0.01  &  0.598 $\pm$ 0.002   \\ \hline
$ 200 \leq \mee < 300$ & 233.6 & 0.65$^{+0.20}_{-0.27}$ $\pm$ 0.03  &  0.609 $\pm$ 0.004   \\ \hline
$ 300 \leq \mee < 600$ & 352.4 & 1.00$^{+0.00}_{-0.51}$$^{+0}_{-0.02}$  &  0.616 $\pm$ 0.007   \\ \hline
\hline
\end{tabular}
\caption{\emph{\label{t_afb_sum_fit}
Experimental results for $A_{FB}$ measured by fitting to $\afbr$ with 
a Tikhonov regularization function.
Statistical and systematic uncertainties are included along with predictions from PYTHIA 
using CTEQ5L. The uncertainties in the PYTHIA predictions are MC statistical errors.
$< M_{ee} >$ is the cross section weighted average of the invariant mass in each bin.}}
\end{center}
\end{table*}

%****************************************************************
\subsection{Study of the $Z$-quark couplings using the $A_{FB}$ measurement}
\label{s_afb_couplings}

As seen in Sec.~\ref{s_Intro}, the vector and axial-vector nature of the 
interaction
$\pbp \rightarrow Z/\gamma^* \rightarrow e^{\rm +}e^{\rm -}$ 
renders the $A_{FB}$ measurement a direct probe of the relative
strengths of the vector and axial-vector couplings between the quarks and the 
$Z$ boson. 
In this section, the effect of varying 
the $Z$-quark couplings on the forward-backward charge asymmetry of 
electron-positron pairs is investigated. 
\par 
The $\afbr$ measurement at detector level is compared to the following
theoretical parameterization:
 \begin{eqnarray}
\afbr(u_V,u_A,d_V,d_A) = &&\afbr_{LO}(u_V,u_A,d_V,d_A) \nonumber \\
&&- \afbr_{LO}(SM) \nonumber \\
&&+ \afbr_{\cal O(\alpha)}(SM), 
\label{eq:newpara}
\end{eqnarray} 
where \afbr$_{LO}$ is the predicted $A_{FB}$ at the leading order and
\afbr$_{\cal O(\alpha)}$ is the $A_{FB}$ calculated at the  
$\cal O(\alpha)$ electroweak corrections level with 
the ZGRAD generator described in Sec.~\ref{s_theory_band}. 
The quark couplings are changed at tree level assuming Standard Model 
couplings for the leptons.
Both \afbr$_{LO}$ and \afbr$_{\cal O(\alpha)}$ are obtained after
smearing the corresponding \afbp with 
the parameterization presented in Sec.~\ref{s_smear}. 
The parameterized $\afbr(u_V,u_A,d_V,d_A)$, function of the
$Z$-quark couplings, is compared to the measured $\afbr$, 
and the best match is used to extract the coupling constants. 
Table~\ref{tab:tabzgrad_sum} shows the 
coupling constants obtained after $\chi^2$ minimization, along with
the corresponding statistical and systematic uncertainties. 
The $\chi^2$ divided by the number of degrees of freedom ($\chi^2$/DOF) equals 
10.40/11.
\begin{table}[htb]
\begin{center}
\begin{tabular}{|c|c|c|c|} \hline
      & Coupling  & \multicolumn{2}{|c|}{Uncertainty} \\ \cline{3-4}
	&	& Statistical & Systematic \\ \hline\hline
u$_V$     &  0.399 & $^{+ 0.152}_{-0.188}$  &  0.066 \\\hline
u$_A$     &  0.441 & $^{+ 0.207}_{-0.173}$  &  0.067 \\\hline
d$_V$     & -0.226 & $^{+ 0.635}_{-0.290}$  &  0.090 \\\hline
d$_A$     & -0.016 & $^{+ 0.346}_{-0.536}$  &  0.091 \\\hline
\end{tabular}
\caption{\emph{Vector and axial-vector quark couplings with 
statistical and systematic errors obtained from the fit to the $A_{FB}$ 
measurement at detector level.}}
\label{tab:tabzgrad_sum}
\end{center}
\end{table}

\begin{figure}
 \begin{center}
    \includegraphics[width=8cm,clip=]{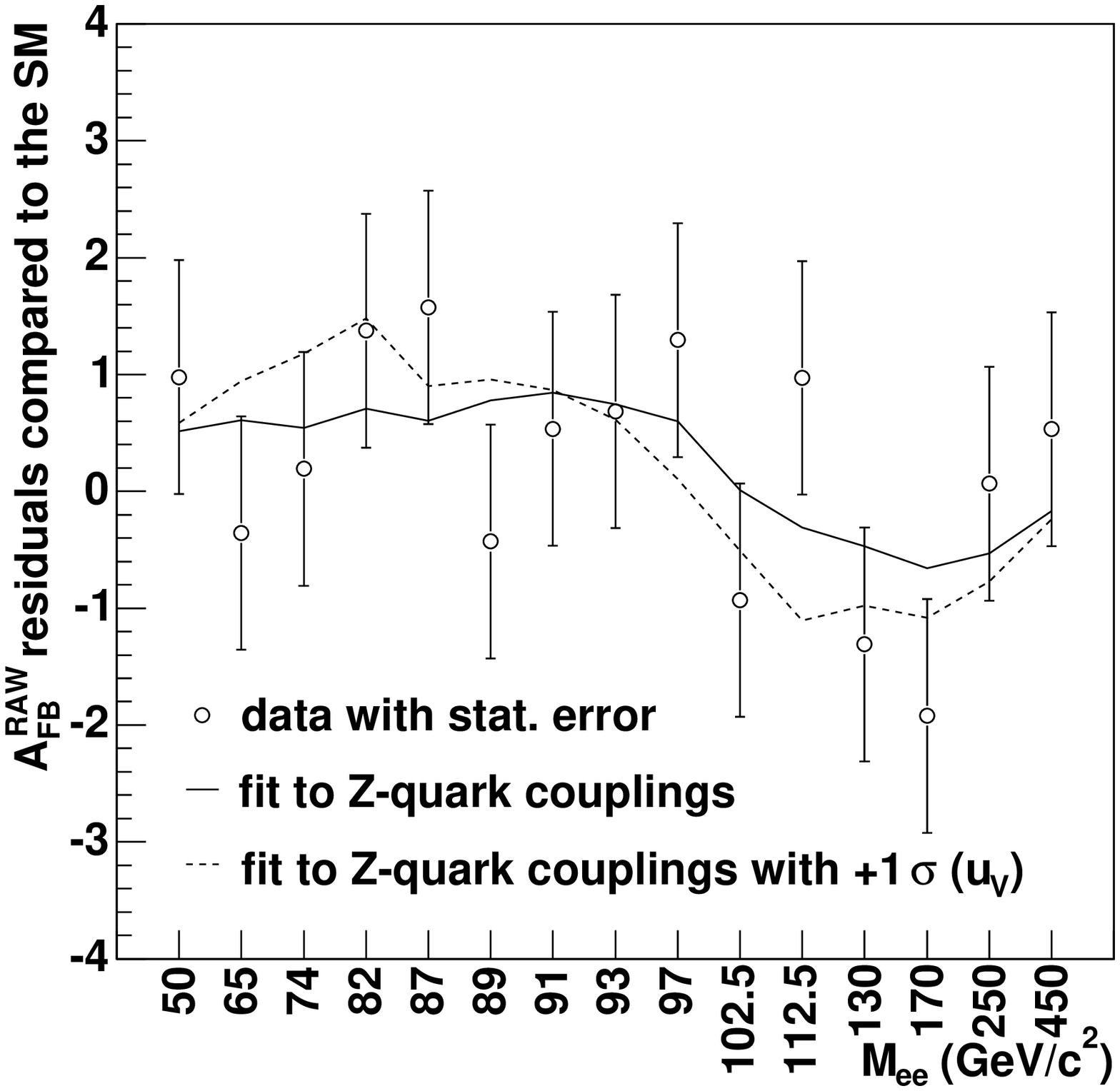}
    \caption{\emph{Residuals compared to the SM \afb~as a function of the 
invariant mass from the data (open marker) and from the fit (solid line). 
The dashed line represents the residuals with the $u_V$ coupling from the fit 
shifted by 1$\sigma$.}}
    \label{fig:residuals}
  \end{center}
\end{figure}

The contribution from the different sources of systematic uncertainties on the
quark coupling constants is given in Table~\ref{tab:sys_sum}, and the correlation matrix
is shown in Table~\ref{tab:correlmatrix}. 
For a given source of uncertainty,
the systematic error on the coupling is calculated by repeating the fit using the 
measured $\afbr$ shifted by its systematic uncertainty. The difference between the 
fitted coupling value obtained using the shifted $\afbr$ and the nominal one
is taken as the systematic uncertainty from that source. For the PDF uncertainty, we 
measured the couplings with four different parton distribution functions: MRST99, 
MRST2001, CTEQ6L and CTEQ5L (default). The uncertainty due to the PDF models is
determined by the largest difference in the fitted coupling values with respect to the 
couplings obtained with CTEQ5L.
The sign of the systematic uncertainty is defined in the same manner as in Table~\ref{t_sys_error_SM}. 
Figure~\ref{fig:residuals} displays
the residuals compared to the SM \afb~from the data (open markers) and 
from the fit (solid line) as a function of 
the invariant mass. The dashed
curve corresponds to the residuals with the $u_V$ quark coupling shifted
by 1$\sigma$.
\begin{table}[htb]
\begin{center}
\begin{tabular}{|c|c|c|c|c|c|} \hline
       & En. scale & Resol. & material & bckgrd. & PDF \\ \hline\hline
u$_V$   &  -0.056 & -0.023 &  -0.025 &   0.001 & -0.001 \\\hline
u$_A$   &  -0.013 & -0.048 &  -0.009 &  -0.044 &  0.003 \\\hline
d$_V$   &  -0.013 &  0.038 &  -0.076 &  -0.021 &  0.017 \\\hline
d$_A$   &  -0.059 &  0.062 &   0.006 &  -0.025 & -0.018 \\\hline
\end{tabular}
\caption{\emph{Contribution of the different sources of systematic uncertainty on
the quark coupling constants.}}
\label{tab:sys_sum}
\end{center}
\end{table}
\begin{table}[htb]
\begin{center}
\begin{tabular}{|c|r|r|r|r|} \hline
      & u$_V$ & u$_A$ & d$_V$ & d$_A$ \\ \hline\hline
u$_V$ &  1.000 & 0.454 & 0.303 & -0.037 \\ \hline
u$_A$ &  0.454 & 1.000 & 0.214 & 0.428  \\ \hline
d$_V$ &  0.303 & 0.214 & 1.000 & 0.548  \\ \hline
d$_A$ & -0.037 & 0.428 & 0.548 & 1.000  \\ \hline
\end{tabular}
\caption{\emph{The correlation matrix of the statistical errors in 
the measurement of the up and down quark couplings.}}
\label{tab:correlmatrix}
\end{center}
\end{table}

\begin{figure}
\begin{center}
    \includegraphics[width=8cm,clip=]{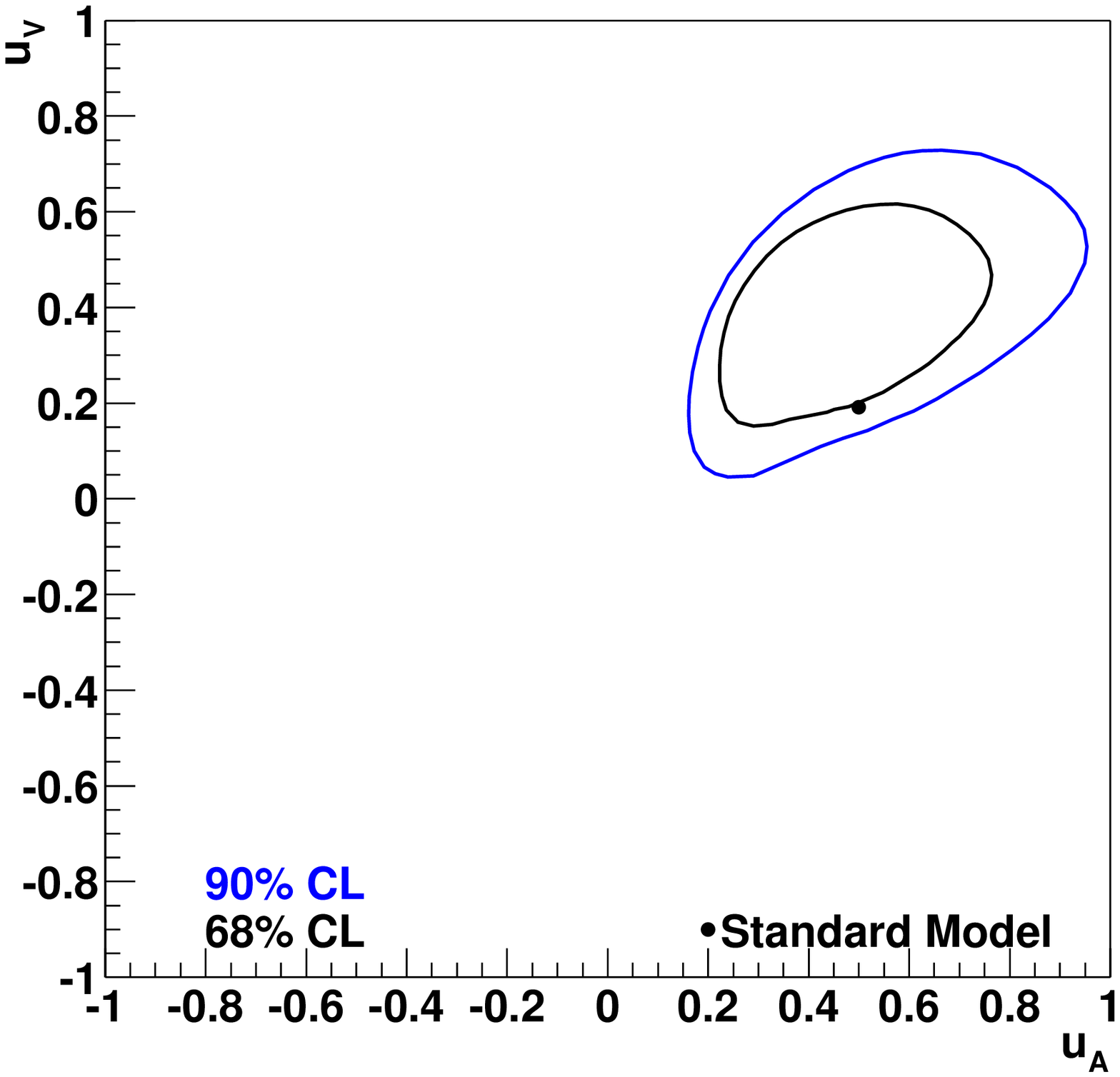}
    \includegraphics[width=8cm,clip=]{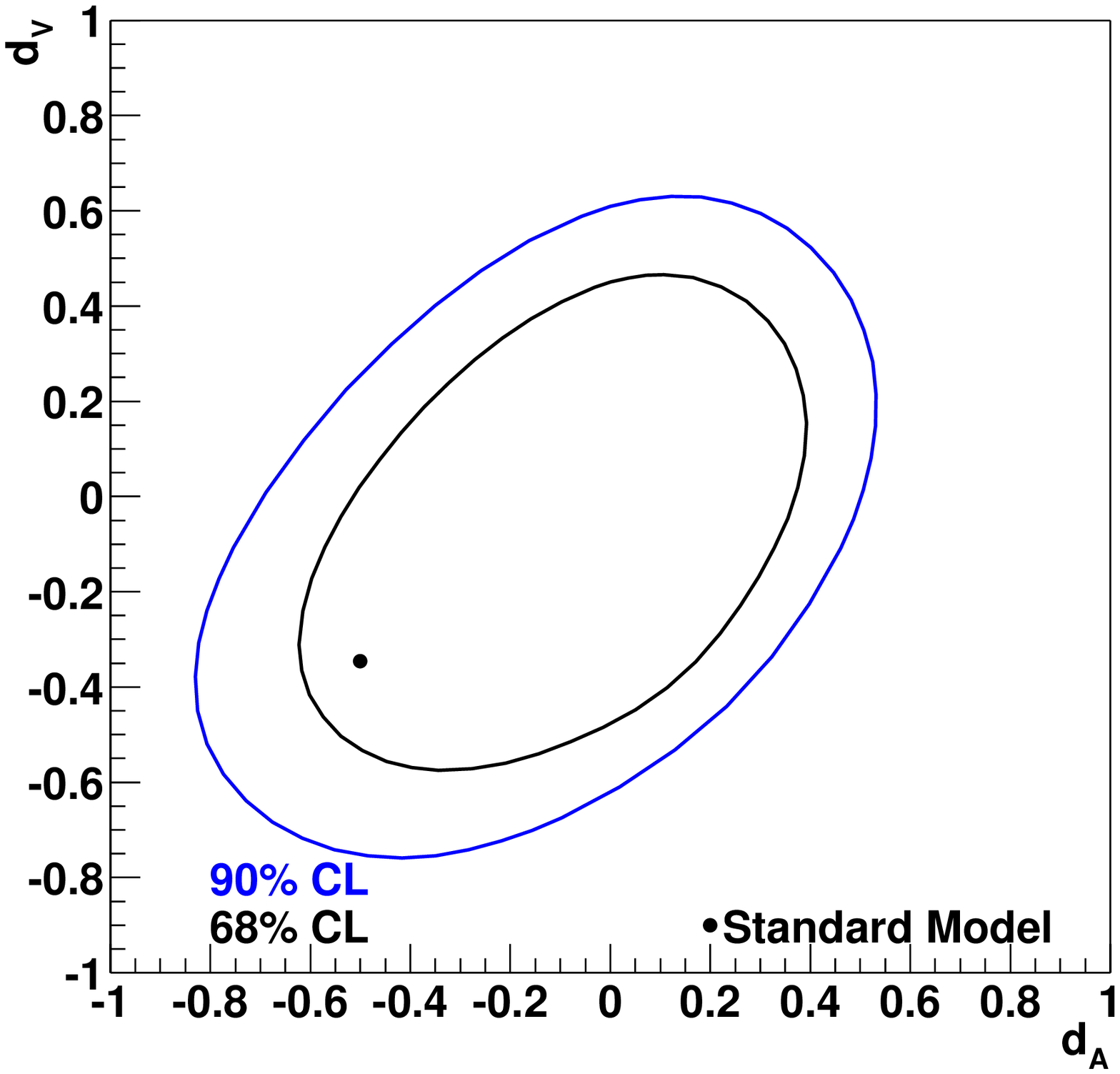}
%\begin{picture}(11.8,11.8)                 
%\put(-2.25,0.0){\epsfig{file=figures/PRD_u_2s_contour.eps,%
%            height=8.0cm,width=8.0cm,angle=0}}
%\put(6.0,0.0){\epsfig{file=figures/PRD_d_2s_contour.eps,%
%            height=8.0cm,width=8.0cm,angle=0}}
%\end{picture}
\end{center}
\caption{\emph{68\% and 90\% confidence level contours for the $u$ (top) and $d$ (bottom) 
quark couplings to the $Z$ boson. The Standard Model predictions for these couplings are 
indicated by closed markers.}}
    \label{fig:contour_plot}
  
\end{figure}

The CDF sensitivity, with $\sim$72~$pb^{-1}$ of analyzed data, is 
limited, but the values of the couplings are consistent with the Standard 
Model. Figure~\ref{fig:contour_plot} shows the contours at 68\% and 90\% confidence level
for the $u$ (left) and $d$ (right) quark couplings to the $Z$ boson in the vector-, axial-vector
basis. The closed markers correspond to the Standard Model predictions.

\begin{table}[htb]
\begin{center}
\begin{tabular}{|c|r|c|c|c|} \hline
      & This study    & Exp. values~\cite{Hagiwara:2002fs} & SM prediction~\cite{Hagiwara:2002fs} \\ \hline
u$_L$ & 0.419 $^{+0.131}_{-0.167}$  & 0.330 $\pm$~0.016 & 0.3459 $\pm$~0.0002  \\ \hline
d$_L$ &-0.116 $^{+0.418}_{-0.352}$  &-0.439 $\pm$~0.011 &-0.4291 $\pm$~0.0002 \\ \hline
u$_R$ & 0.020 $^{+0.145}_{-0.150}$  &-0.176 $^{+0.011}_{-0.006}$ &-0.1550 $\pm$~0.0001  \\ \hline
d$_R$ & 0.105 $^{+0.128}_{-0.315}$ &-0.023 $^{+0.070}_{-0.047}$ &0.0776 \\ \hline
\end{tabular}
\caption{\emph{Effective left- and right-handed coupling constants obtained from this study 
and compared to current experimental values and the Standard Model prediction 
from~\cite{Hagiwara:2002fs}.}}
\label{tab:sum}
\end{center}
\end{table} 

\par Table~\ref{tab:sum} summarizes the effective left- and right-handed
coupling constants obtained from this study and compares them to the
Standard Model prediction as well as the current experimental values of the effective couplings 
determined from `model independent' fits to neutral-current data~\cite{Hagiwara:2002fs}.
As the present study is dominated by the statistical uncertainties, the 
sensitivity of these measurements will improve with higher integrated luminosities.

The $Z$-quark couplings returned by the fit are subsequently used to compute the $\afbp$ which 
is shown for different invariant-mass bins in Table~\ref{tab:residuals}. The uncertainties 
in the resulting $\afbp$ are the total (statistical and systematic)
uncertainties from the fit of the $Z$ couplings.
 $\afbp$ is then
used to calculate the acceptance as discussed in Sec.~\ref{s_corrfac}.
\begin{table}[htb]
\begin{center}
\begin{tabular}{|c|c|c|} \hline
Bin  &Mass Range & Fitted \afbp    \\ \hline
0 & $ 40 < M_{ee} < 60$ GeV/c$^2$ & -0.170 $\pm$ 0.074   \\ \hline
1 & $ 60 < M_{ee} < 70$ GeV/c$^2$ & -0.355 $\pm$ 0.125   \\ \hline
2 & $ 70 < M_{ee} < 78$ GeV/c$^2$ & -0.373 $\pm$ 0.109   \\ \hline
3 & $ 78 < M_{ee} < 86$ GeV/c$^2$ & -0.183 $\pm$ 0.124   \\ \hline
4 & $ 86 < M_{ee} < 88$ GeV/c$^2$ & -0.044 $\pm$ 0.083   \\ \hline
5 & $ 88 < M_{ee} < 90$ GeV/c$^2$ & 0.028 $\pm$ 0.053   \\ \hline
6 & $ 90 < M_{ee} < 92$ GeV/c$^2$ & 0.088 $\pm$ 0.027   \\ \hline
7 & $ 92 < M_{ee} < 94$ GeV/c$^2$ & 0.140 $\pm$ 0.039   \\ \hline
8 & $ 94 < M_{ee} < 100$ GeV/c$^2$ & 0.223 $\pm$ 0.060   \\ \hline
9 & $ 100 < M_{ee} < 105$ GeV/c$^2$ & 0.342 $\pm$ 0.087   \\ \hline
10 & $ 105 < M_{ee} < 120$ GeV/c$^2$ & 0.429 $\pm$ 0.101   \\ \hline
11 & $ 120 < M_{ee} < 140$ GeV/c$^2$ & 0.493 $\pm$ 0.135   \\ \hline
12 & $ 140 < M_{ee} < 200$ GeV/c$^2$ & 0.506 $\pm$ 0.159   \\ \hline
13 & $ 200 < M_{ee} < 300$ GeV/c$^2$ & 0.501 $\pm$ 0.171   \\ \hline
14 & $ 300 < M_{ee} < 600$ GeV/c$^2$ & 0.498 $\pm$ 0.175   \\ \hline
\end{tabular}
\caption{\emph{Forward-Backward asymmetry calculated with the $Z$-quark coupling values 
returned by the fit.
The uncertainties in $A_{FB}$ are based on the total (statistical and systematic)
uncertainties on the $Z$ coupling values returned from the fit.}}
\label{tab:residuals}
\end{center}
\end{table} 

Assuming Standard Model $Z$-quark couplings, we can also determine the vector
and axial-vector couplings between the electron and the $Z$-boson, using the
same method. Table~\ref{tab:tabzgrad_sum2}
shows the fitted electron coupling values along with statistical 
and systematic uncertainties. Combined LEP and SLD data~\cite{LEP:2003ih} as 
well as the 
Standard Model prediction~\cite{Hagiwara:2002fs} are shown for comparison. 
The $\chi^2$/DOF of the fit equals to 13.14/13.
For the present measurement, the uncertainties are dominated by the 
statistical uncertainties.
The contributions of the different
systematic uncertainties are given in Table~\ref{tab:sys_sum2}. Note that the
correlation coefficient between the vector and axial-vector couplings is
0.78.
\begin{table*}[htb]
\begin{center}
\begin{tabular}{|c|c|c|c|c|c|c|} \hline
      & coupling  & Stat. & Syst. & Total & SLD+LEP meas. & SM prediction\\
      &           & err. & err.  & err.  & &  \\ \hline 
e$_V$ & -0.058 & 0.016 & 0.007 & 0.017 & -0.03816 $\pm$0.00047& -0.0397 $\pm$0.0003\\ \hline
e$_A$ & -0.528 & 0.123 & 0.059 & 0.136 & -0.50111 $\pm$0.00035& -0.5064 $\pm$0.0001 \\ \hline
\end{tabular}

\caption{\emph{Vector and axial-vector electron couplings with 
statistical and systematics uncertainties obtained from a fit to the $A_{FB}$ 
measurement. Combined LEP and SLD data~\cite{LEP:2003ih} as well as the standard 
model prediction~\cite{Hagiwara:2002fs} are also given.}}
\label{tab:tabzgrad_sum2}
\end{center}
\end{table*}
\begin{table}
\begin{center}
\begin{tabular}{|c|c|c|c|c|c|} \hline
       & En. scale & Resol. & material & bckgrd. & PDF \\ \hline
e$_V$   &  -0.005 & 0.005 & 0.001 &  -0.001 &  0.001 \\\hline
e$_A$   &  -0.002 & 0.055 & 0.006 &  -0.005 &  0.019 \\\hline
\end{tabular}
\caption{\emph{Contribution of the different sources of systematic uncertainty on
the electron coupling constants.}}
\label{tab:sys_sum2}
\end{center}
\end{table}
\par 
Finally a fit where  the quark and electron couplings to 
the $Z$ boson are expressed as a function of $\sin^2\theta_W^{eff}$ gives:
\begin{eqnarray}
\sin^2\theta_W^{eff} = 0.2238 \pm 0.0040(\rm stat) \pm 0.0030(\rm syst), 
\end{eqnarray}
with a $\chi^2$/DOF equals to 12.50/14.
The present CDF sensitivity on $\sin^2\theta_W^{eff}$ is provided by the 
$Z$-electron couplings and is expected to improve with higher statistics.

%*********************************************************************************** 
\subsection{$A_{FB}$ measurement assuming the fitted Standard Model couplings in the acceptance calculation}
\label{s_afb_measurement}
\begin{figure}
\begin{center}
\includegraphics[width=8cm]{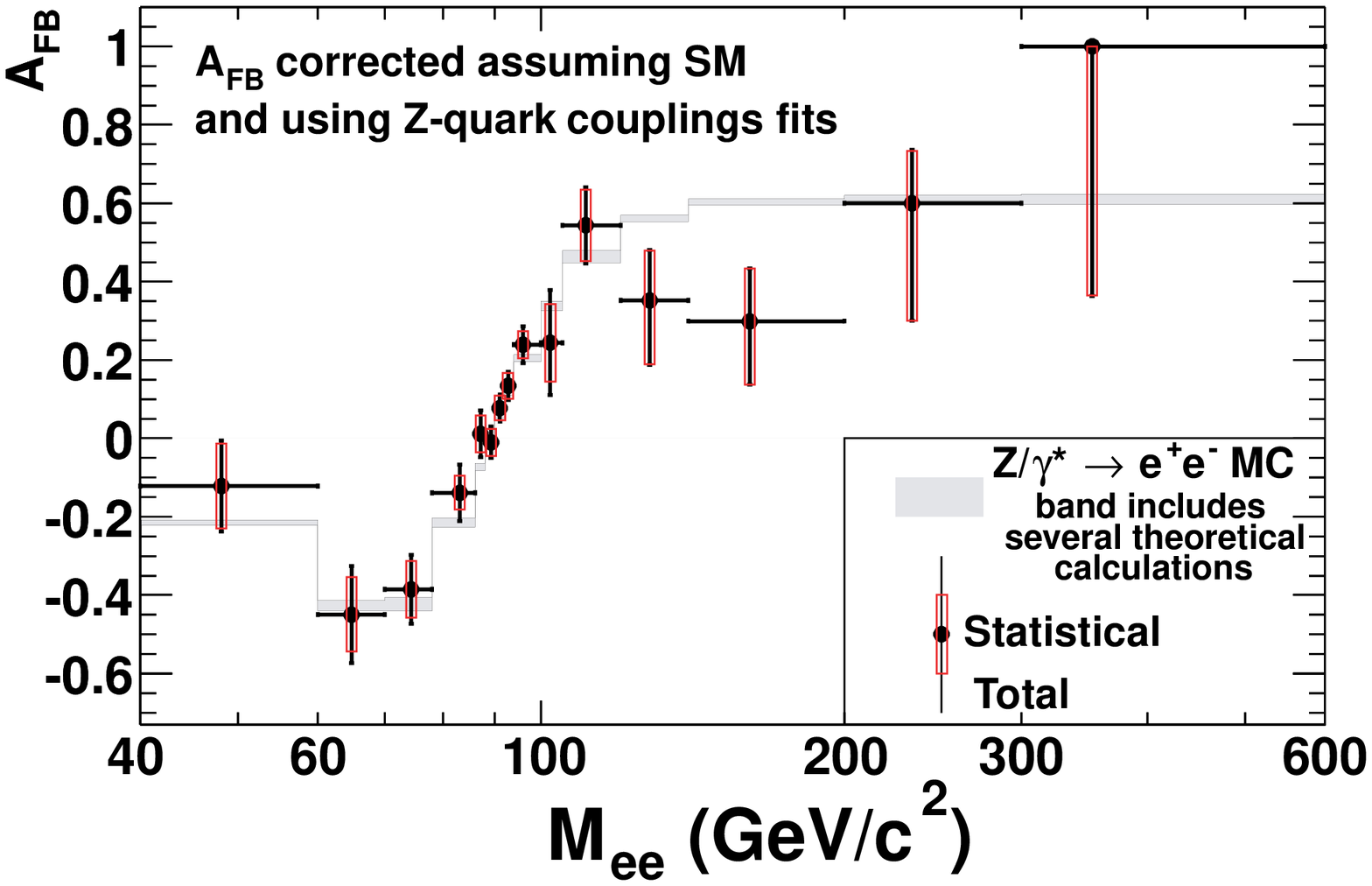}
\includegraphics[width=8cm]{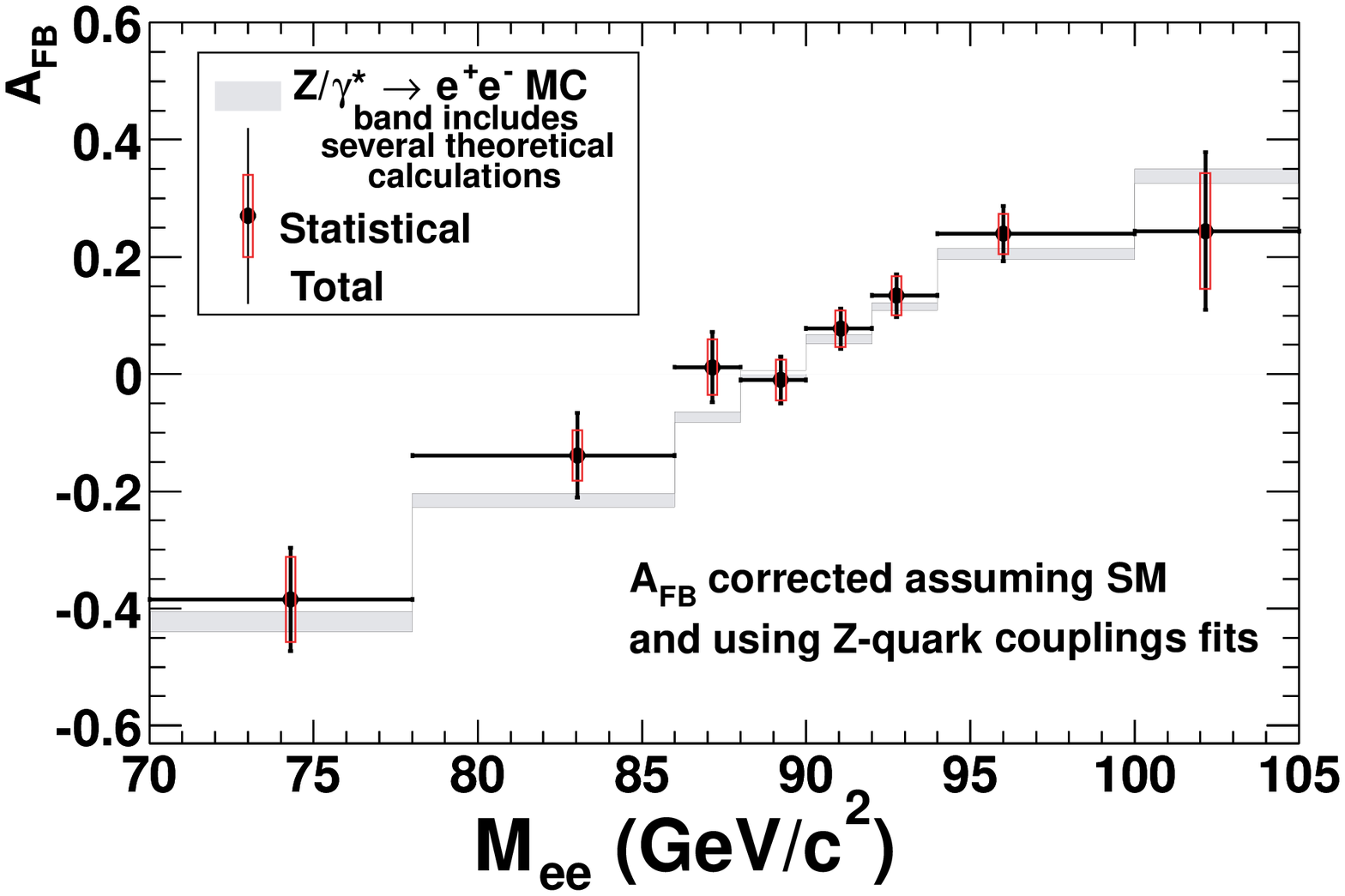}
\vspace{0in}
\caption{\emph{Experimental results for $A_{FB}$ along with statistical and systematic 
uncertainties~(crosses), and theoretical predictions based on the six
calculations as described in {\rm Sec.~\ref{s_theory_band}} (bands). 
The measured $A_{FB}$ values are corrected for acceptance, efficiency,
resolution, and bremsstrahlung. The correction assumes the Standard Model
and uses the derived $\afbp$ from the $Z$-quark coupling values. The bottom
figure shows the $Z$ pole region in more detail.}}
\label{f_Afb_Results}
\end{center}
\end{figure}

The total acceptance, $a^{\pm}_{cor}$, is calculated for each bin using the 
$\afbp$ obtained from the $Z$ coupling fits (Table~\ref{tab:residuals}). 
The $A_{FB}$ measurements are corrected
for acceptance, efficiency, resolution and bremsstrahlung using 
Eq.~(\ref{e_afbcalc}). The measured $\afb$ and 
the prediction from PYTHIA using CTEQ5L parton distribution functions are listed in 
Table~\ref{t_afb_sum}, and the measurements are compared with the
Standard Model theoretical calculations 
(see Sec.~\ref{s_theory_band}) in Fig.~\ref{f_Afb_Results}. This technique,
which is biased by the Standard Model input, is 
similar to the Run I analysis~\cite{Affolder:2001ha}.
	
%%%%%%%%%%% Summary of Final Results %%%%%%%%%%%%%%%%%%%%%%%%%%%%%%%%%%%
\begin{table*}
\begin{center}
\begin{tabular}{|c||c||c|c|}\hline
Mass Range   & $<M_{ee}>$ & Measured $A_{FB}$ &PYTHIA  $A_{FB}$ \\
(GeV/$c^2$)  & (GeV/$c^2$) & & \\ \hline\hline
$ 40 \leq \mee < 60$ & 48.2 & -0.131 $\pm$ 0.108 $\pm$ 0.065  &  -0.214 $\pm$ 0.003   \\ \hline
$ 60 \leq \mee < 70$ & 64.9 & -0.447 $\pm$ 0.095 $\pm$ 0.089  &  -0.420 $\pm$ 0.005   \\ \hline
$ 70 \leq \mee < 78$ & 74.3 & -0.400 $\pm$ 0.072 $\pm$ 0.050  &  -0.410 $\pm$ 0.005   \\ \hline
$ 78 \leq \mee < 86$ & 83.0 & -0.154 $\pm$ 0.043 $\pm$ 0.066  &  -0.214 $\pm$ 0.003   \\ \hline
$ 86 \leq \mee < 88$ & 87.1 & 0.002 $\pm$ 0.048 $\pm$ 0.041  &  -0.079 $\pm$ 0.004   \\ \hline
$ 88 \leq \mee < 90$ & 89.2 & -0.015 $\pm$ 0.035 $\pm$ 0.021  &  -0.001 $\pm$ 0.002   \\ \hline
$ 90 \leq \mee < 92$ & 91.0 & 0.078 $\pm$ 0.031 $\pm$ 0.015  &  0.054 $\pm$ 0.001   \\ \hline
$ 92 \leq \mee < 94$ & 92.8 & 0.138 $\pm$ 0.033 $\pm$ 0.014  &  0.112 $\pm$ 0.002   \\ \hline
$ 94 \leq \mee < 100$ & 96.0 & 0.247 $\pm$ 0.034 $\pm$ 0.033  &  0.198 $\pm$ 0.003   \\ \hline
$ 100 \leq \mee < 105$ & 102.2 & 0.248 $\pm$ 0.099 $\pm$ 0.095  &  0.338 $\pm$ 0.006   \\ \hline
$ 105 \leq \mee < 120$ & 110.7 & 0.545 $\pm$ 0.091 $\pm$ 0.035  &  0.454 $\pm$ 0.006   \\ \hline
$ 120 \leq \mee < 140$ & 128.2 & 0.36$^{+0.14}_{-0.15}$ $\pm$ 0.03  &  0.554 $\pm$ 0.002   \\ \hline
$ 140 \leq \mee < 200$ & 161.2 & 0.30$^{+0.14}_{-0.15}$ $\pm$ 0.02  &  0.598 $\pm$ 0.002   \\ \hline
$ 200 \leq \mee < 300$ & 233.6 & 0.62$^{+0.18}_{-0.24}$ $\pm$ 0.04  &  0.609 $\pm$ 0.004   \\ \hline
$ 300 \leq \mee < 600$ & 352.4 & 1.000$^{+0.000}_{-0.64}$$^{+0}_{-0.04}$  &  0.616 $\pm$ 0.007   \\ \hline
\hline
\end{tabular}

\caption{\emph{\label{t_afb_sum}Experimental results for $A_{FB}$ along with statistical and systematic 
uncertainties and predictions from PYTHIA with CTEQ5L. The uncertainties
on the PYTHIA predictions are MC statistical errors.
The measured $A_{FB}$ values are corrected for acceptance, efficiency,
resolution, and bremsstrahlung. The correction assumes the Standard Model
and uses the $\afbp$ derived from the fitted $Z$-quark coupling values. 
$< \mee >$ is the cross section weighted average of the invariant mass in each bin.}}
\end{center}
\end{table*}

%***********************************************************
\section{Conclusions}
We report a measurement of the forward-backward charge asymmetry
($\afb$) of electron pairs resulting from the process
$\pbp \rightarrow Z/\gamma^* + X$ where $Z/\gamma^* \rightarrow
e^+e^-$.  The data are collected with the CDF II detector between March 2002 and
January 2003, corresponding to about 72~pb$^{-1}$. 

	Comparisons have been made between the data and a simulated Standard Model
prediction of the uncorrected $\mee$ line shape, $\cost$ distributions in three $\mee$ 
regions, and $\afb$ distribution. All comparisons give excellent agreement,
with the uncorrected $\afb$ distribution giving a $\chi^2$/DOF=15.7/15.
The first principal result is a measurement of the corrected $\afb$ 
in 15 $\mee$ bins using an unfolding analysis that doesn't 
assume a prior Standard Model $\afb$ distribution. It has large
uncertainties near the $Z$ pole because the $\afb$ in those bins have large
correlations. In the current dataset, there is no evidence of deviations 
from the Standard Model in the high $\mee$ bins that might 
indicate non--Standard Model physics.
The second principal result is a measurement of three 
sets of parameters:  the $Z$-quark couplings, 
the $Z$-electron couplings, and $\sin^2\theta_W$. All three
couplings measurements yield results consistent with the Standard Model.
It may be possible to improve our understanding of the $Z$-quark couplings
with a much larger dataset.

%%%%%%%%%%%%%%%%%%%%%%%%%%%%%%%%%%%%%%%%%%%%%%%%%%%%%%%%%%%%%%%%%%%%%%%%%%%%%%%%
% Specify following sections are appendices. Use \appendix* if there
% only one appendix.
%\appendix
%\section{}
%%%%%%%%%%%%%%%%%%%%%%%%%%%%%%%%%%%%%%%%%%%%%%%%%%%%%%%%%%%%%%%%%%%%%%%%%%%%%%%%

% If you have acknowledgments, this puts in the proper section head.
\begin{acknowledgments}

We thank the Fermilab staff and the technical staffs of the participating institutions for 
their vital contributions. This work was supported by the U.S. Department of Energy and National 
Science Foundation; the Italian Istituto Nazionale di Fisica Nucleare; the Ministry of Education, 
Culture, Sports, Science and Technology of Japan; the Natural Sciences and Engineering Research 
Council of Canada; the National Science Council of the Republic of China; the Swiss National 
Science Foundation; the A.P. Sloan Foundation; the Research Corporation; the Bundesministerium 
fuer Bildung und Forschung, Germany; the Korean Science and Engineering Foundation and the 
Korean Research Foundation; the Particle Physics and Astronomy Research Council and the Royal 
Society, UK; the Russian Foundation for Basic Research; the Comision Interministerial de 
Ciencia y Tecnologia, Spain; and in part by the European Community's Human Potential Programme 
under contract HPRN-CT-2002-00292, Probe for New Physics.
\end{acknowledgments}

% Create the reference section using BibTeX:
\bibliography{AfbPRD_revtex_final}

\begin{thebibliography}{43}
\expandafter\ifx\csname natexlab\endcsname\relax\def\natexlab#1{#1}\fi
\expandafter\ifx\csname bibnamefont\endcsname\relax
  \def\bibnamefont#1{#1}\fi
\expandafter\ifx\csname bibfnamefont\endcsname\relax
  \def\bibfnamefont#1{#1}\fi
\expandafter\ifx\csname citenamefont\endcsname\relax
  \def\citenamefont#1{#1}\fi
\expandafter\ifx\csname url\endcsname\relax
  \def\url#1{\texttt{#1}}\fi
\expandafter\ifx\csname urlprefix\endcsname\relax\def\urlprefix{URL }\fi
\providecommand{\bibinfo}[2]{#2}
\providecommand{\eprint}[2][]{\url{#2}}

\bibitem[{\citenamefont{Drell and Yan}(1970)}]{Drell:1970wh}
\bibinfo{author}{\bibfnamefont{S.~D.} \bibnamefont{Drell}} \bibnamefont{and}
  \bibinfo{author}{\bibfnamefont{T.-M.} \bibnamefont{Yan}},
  \bibinfo{journal}{Phys. Rev. Lett.} \textbf{\bibinfo{volume}{25}},
  \bibinfo{pages}{316} (\bibinfo{year}{1970}).

\bibitem[{\citenamefont{Drell and Yan}(1971)}]{Drell:1970yt}
\bibinfo{author}{\bibfnamefont{S.~D.} \bibnamefont{Drell}} \bibnamefont{and}
  \bibinfo{author}{\bibfnamefont{T.-M.} \bibnamefont{Yan}},
  \bibinfo{journal}{Ann. Phys.} \textbf{\bibinfo{volume}{66}},
  \bibinfo{pages}{578} (\bibinfo{year}{1971}).

\bibitem[{\citenamefont{Rosner}(1996)}]{Rosner:1995ft}
\bibinfo{author}{\bibfnamefont{J.~L.} \bibnamefont{Rosner}},
  \bibinfo{journal}{Phys. Rev.} \textbf{\bibinfo{volume}{D54}},
  \bibinfo{pages}{1078} (\bibinfo{year}{1996}).

\bibitem[{\citenamefont{Zeller et~al.}(2002)}]{Zeller:2001hh}
\bibinfo{author}{\bibfnamefont{G.~P.} \bibnamefont{Zeller}}
  \bibnamefont{et~al.} (\bibinfo{collaboration}{NuTeV}),
  \bibinfo{journal}{Phys. Rev. Lett.} \textbf{\bibinfo{volume}{88}},
  \bibinfo{pages}{091802} (\bibinfo{year}{2002}).

\bibitem[{\citenamefont{Bennett and Wieman}(1999)}]{Bennett:1999pd}
\bibinfo{author}{\bibfnamefont{S.~C.} \bibnamefont{Bennett}} \bibnamefont{and}
  \bibinfo{author}{\bibfnamefont{C.~E.} \bibnamefont{Wieman}},
  \bibinfo{journal}{Phys. Rev. Lett.} \textbf{\bibinfo{volume}{82}},
  \bibinfo{pages}{2484} (\bibinfo{year}{1999}).

\bibitem[{\citenamefont{Affolder et~al.}(2001)}]{Affolder:2001ha}
\bibinfo{author}{\bibfnamefont{T.}~\bibnamefont{Affolder}} \bibnamefont{et~al.}
  (\bibinfo{collaboration}{CDF}), \bibinfo{journal}{Phys. Rev. Lett.}
  \textbf{\bibinfo{volume}{87}}, \bibinfo{pages}{131802}
  (\bibinfo{year}{2001}).

\bibitem[{\citenamefont{Abe et~al.}(1988)}]{Abe:1988me}
\bibinfo{author}{\bibfnamefont{F.}~\bibnamefont{Abe}} \bibnamefont{et~al.}
  (\bibinfo{collaboration}{CDF}), \bibinfo{journal}{Nucl. Instr. Meth.}
  \textbf{\bibinfo{volume}{A271}}, \bibinfo{pages}{387} (\bibinfo{year}{1988}).

\bibitem[{\citenamefont{Abe et~al.}(1994)}]{Abe:1994st}
\bibinfo{author}{\bibfnamefont{F.}~\bibnamefont{Abe}} \bibnamefont{et~al.}
  (\bibinfo{collaboration}{CDF}), \bibinfo{journal}{Phys. Rev.}
  \textbf{\bibinfo{volume}{D50}}, \bibinfo{pages}{2966} (\bibinfo{year}{1994}).

\bibitem[{\citenamefont{Blair et~al.}(1996)}]{Blair:1996kx}
\bibinfo{author}{\bibfnamefont{R.}~\bibnamefont{Blair}} \bibnamefont{et~al.}
  (\bibinfo{collaboration}{CDF-II}) (\bibinfo{year}{1996}),
  \bibinfo{note}{\uppercase{f}ERMILAB-PUB-96-390-E}.

\bibitem[{\citenamefont{Affolder et~al.}(2004)}]{Affolder:2003ep}
\bibinfo{author}{\bibfnamefont{T.}~\bibnamefont{Affolder}} \bibnamefont{et~al.}
  (\bibinfo{collaboration}{CDF}), \bibinfo{journal}{Nucl. Instrum. Meth.}
  \textbf{\bibinfo{volume}{A526}}, \bibinfo{pages}{249} (\bibinfo{year}{2004}).

\bibitem[{\citenamefont{Balka et~al.}(1988)}]{Balka:1987ty}
\bibinfo{author}{\bibfnamefont{L.}~\bibnamefont{Balka}} \bibnamefont{et~al.}
  (\bibinfo{collaboration}{CDF}), \bibinfo{journal}{Nucl. Instrum. Meth.}
  \textbf{\bibinfo{volume}{A267}}, \bibinfo{pages}{272} (\bibinfo{year}{1988}).

\bibitem[{\citenamefont{Hahn et~al.}(1988)}]{Hahn:1987tx}
\bibinfo{author}{\bibfnamefont{S.~R.} \bibnamefont{Hahn}} \bibnamefont{et~al.}
  (\bibinfo{collaboration}{CDF}), \bibinfo{journal}{Nucl. Instr. Meth.}
  \textbf{\bibinfo{volume}{A267}}, \bibinfo{pages}{351} (\bibinfo{year}{1988}).

\bibitem[{\citenamefont{Yasuoka et~al.}(1988)\citenamefont{Yasuoka, Mikamo,
  Kamon, and Yamashita}}]{Yasuoka:1987ar}
\bibinfo{author}{\bibfnamefont{K.}~\bibnamefont{Yasuoka}},
  \bibinfo{author}{\bibfnamefont{S.}~\bibnamefont{Mikamo}},
  \bibinfo{author}{\bibfnamefont{T.}~\bibnamefont{Kamon}}, \bibnamefont{and}
  \bibinfo{author}{\bibfnamefont{A.}~\bibnamefont{Yamashita}}
  (\bibinfo{collaboration}{CDF NW Wedge Group}), \bibinfo{journal}{Nucl. Instr.
  Meth.} \textbf{\bibinfo{volume}{A267}}, \bibinfo{pages}{315}
  (\bibinfo{year}{1988}).

\bibitem[{\citenamefont{Wagner et~al.}(1988)}]{Wagner:1987mf}
\bibinfo{author}{\bibfnamefont{R.~G.} \bibnamefont{Wagner}}
  \bibnamefont{et~al.} (\bibinfo{collaboration}{CDF}), \bibinfo{journal}{Nucl.
  Instrum. Meth.} \textbf{\bibinfo{volume}{A267}}, \bibinfo{pages}{330}
  (\bibinfo{year}{1988}).

\bibitem[{\citenamefont{Devlin et~al.}(1988)}]{Devlin:1987mf}
\bibinfo{author}{\bibfnamefont{T.}~\bibnamefont{Devlin}} \bibnamefont{et~al.}
  (\bibinfo{collaboration}{CDF}), \bibinfo{journal}{Nucl. Instrum. Meth.}
  \textbf{\bibinfo{volume}{A267}}, \bibinfo{pages}{24} (\bibinfo{year}{1988}).

\bibitem[{\citenamefont{Bertolucci et~al.}(1988)}]{Bertolucci:1987zn}
\bibinfo{author}{\bibfnamefont{S.}~\bibnamefont{Bertolucci}}
  \bibnamefont{et~al.} (\bibinfo{collaboration}{CDF}), \bibinfo{journal}{Nucl.
  Instrum. Meth.} \textbf{\bibinfo{volume}{A267}}, \bibinfo{pages}{301}
  (\bibinfo{year}{1988}).

\bibitem[{\citenamefont{de~Barbaro et~al.}(1995)}]{deBarbaro:1995ci}
\bibinfo{author}{\bibfnamefont{P.}~\bibnamefont{de~Barbaro}}
  \bibnamefont{et~al.}, \bibinfo{journal}{IEEE Trans. Nucl. Sci.}
  \textbf{\bibinfo{volume}{42}}, \bibinfo{pages}{510} (\bibinfo{year}{1995}).

\bibitem[{\citenamefont{Winer}(2001)}]{Winer:2001gj}
\bibinfo{author}{\bibfnamefont{B.~L.} \bibnamefont{Winer}},
  \bibinfo{journal}{Int. J. Mod. Phys.} \textbf{\bibinfo{volume}{A16S1C}},
  \bibinfo{pages}{1169} (\bibinfo{year}{2001}).

\bibitem[{\citenamefont{Anikeev et~al.}(2001)}]{Anikeev:2001pc}
\bibinfo{author}{\bibfnamefont{K.}~\bibnamefont{Anikeev}} \bibnamefont{et~al.}
  (\bibinfo{collaboration}{On behalf of the CDF}), \bibinfo{journal}{Comput.
  Phys. Commun.} \textbf{\bibinfo{volume}{140}}, \bibinfo{pages}{110}
  (\bibinfo{year}{2001}).

\bibitem[{\citenamefont{Thomson et~al.}(2002)}]{Thomson:2002xp}
\bibinfo{author}{\bibfnamefont{E.~J.} \bibnamefont{Thomson}}
  \bibnamefont{et~al.}, \bibinfo{journal}{IEEE Trans. Nucl. Sci.}
  \textbf{\bibinfo{volume}{49}}, \bibinfo{pages}{1063} (\bibinfo{year}{2002}).

\bibitem[{\citenamefont{Sjostrand et~al.}(2001)}]{Sjostrand:2000wi}
\bibinfo{author}{\bibfnamefont{T.}~\bibnamefont{Sjostrand}}
  \bibnamefont{et~al.}, \bibinfo{journal}{Comput. Phys. Commun.}
  \textbf{\bibinfo{volume}{135}}, \bibinfo{pages}{238} (\bibinfo{year}{2001}).

\bibitem[{\citenamefont{Corcella et~al.}(2001)}]{Corcella:2000bw}
\bibinfo{author}{\bibfnamefont{G.}~\bibnamefont{Corcella}}
  \bibnamefont{et~al.}, \bibinfo{journal}{JHEP} \textbf{\bibinfo{volume}{01}},
  \bibinfo{pages}{010} (\bibinfo{year}{2001}).

\bibitem[{\citenamefont{Lai et~al.}(2000)}]{Lai:1999wy}
\bibinfo{author}{\bibfnamefont{H.~L.} \bibnamefont{Lai}} \bibnamefont{et~al.}
  (\bibinfo{collaboration}{CTEQ}), \bibinfo{journal}{Eur. Phys. J.}
  \textbf{\bibinfo{volume}{C12}}, \bibinfo{pages}{375} (\bibinfo{year}{2000}).

\bibitem[{\citenamefont{Affolder et~al.}(2000)}]{Affolder:1999jh}
\bibinfo{author}{\bibfnamefont{T.}~\bibnamefont{Affolder}} \bibnamefont{et~al.}
  (\bibinfo{collaboration}{CDF}), \bibinfo{journal}{Phys. Rev. Lett.}
  \textbf{\bibinfo{volume}{84}}, \bibinfo{pages}{845} (\bibinfo{year}{2000}).

\bibitem[{\citenamefont{Baur and Berger}(1990)}]{Baur:1989gk}
\bibinfo{author}{\bibfnamefont{U.}~\bibnamefont{Baur}} \bibnamefont{and}
  \bibinfo{author}{\bibfnamefont{E.~L.} \bibnamefont{Berger}},
  \bibinfo{journal}{Phys. Rev.} \textbf{\bibinfo{volume}{D41}},
  \bibinfo{pages}{1476} (\bibinfo{year}{1990}).

\bibitem[{\citenamefont{Mangano et~al.}(2003)\citenamefont{Mangano, Moretti,
  Piccinini, Pittau, and Polosa}}]{Mangano:2002ea}
\bibinfo{author}{\bibfnamefont{M.~L.} \bibnamefont{Mangano}},
  \bibinfo{author}{\bibfnamefont{M.}~\bibnamefont{Moretti}},
  \bibinfo{author}{\bibfnamefont{F.}~\bibnamefont{Piccinini}},
  \bibinfo{author}{\bibfnamefont{R.}~\bibnamefont{Pittau}}, \bibnamefont{and}
  \bibinfo{author}{\bibfnamefont{A.~D.} \bibnamefont{Polosa}},
  \bibinfo{journal}{JHEP} \textbf{\bibinfo{volume}{07}}, \bibinfo{pages}{001}
  (\bibinfo{year}{2003}).

\bibitem[{\citenamefont{Agostinelli et~al.}(2003)}]{Agostinelli:2002hh}
\bibinfo{author}{\bibfnamefont{S.}~\bibnamefont{Agostinelli}}
  \bibnamefont{et~al.} (\bibinfo{collaboration}{GEANT4}),
  \bibinfo{journal}{Nucl. Instrum. Meth.} \textbf{\bibinfo{volume}{A506}},
  \bibinfo{pages}{250} (\bibinfo{year}{2003}).

\bibitem[{\citenamefont{Was}(2001)}]{Was:2000st}
\bibinfo{author}{\bibfnamefont{Z.}~\bibnamefont{Was}}, \bibinfo{journal}{Nucl.
  Phys. Proc. Suppl.} \textbf{\bibinfo{volume}{98}}, \bibinfo{pages}{96}
  (\bibinfo{year}{2001}).

\bibitem[{\citenamefont{Collins and Soper}(1977)}]{Collins:1977iv}
\bibinfo{author}{\bibfnamefont{J.~C.} \bibnamefont{Collins}} \bibnamefont{and}
  \bibinfo{author}{\bibfnamefont{D.~E.} \bibnamefont{Soper}},
  \bibinfo{journal}{Phys. Rev.} \textbf{\bibinfo{volume}{D16}},
  \bibinfo{pages}{2219} (\bibinfo{year}{1977}).

\bibitem[{\citenamefont{Albrow et~al.}(2002)}]{Albrow:2001jw}
\bibinfo{author}{\bibfnamefont{M.~G.} \bibnamefont{Albrow}}
  \bibnamefont{et~al.} (\bibinfo{collaboration}{CDF}), \bibinfo{journal}{Nucl.
  Instrum. Meth.} \textbf{\bibinfo{volume}{A480}}, \bibinfo{pages}{524}
  (\bibinfo{year}{2002}).

\bibitem[{\citenamefont{Hamberg et~al.}(1991)\citenamefont{Hamberg, van
  Neerven, and Matsuura}}]{Hamberg:1990np}
\bibinfo{author}{\bibfnamefont{R.}~\bibnamefont{Hamberg}},
  \bibinfo{author}{\bibfnamefont{W.~L.} \bibnamefont{van Neerven}},
  \bibnamefont{and} \bibinfo{author}{\bibfnamefont{T.}~\bibnamefont{Matsuura}},
  \bibinfo{journal}{Nucl. Phys.} \textbf{\bibinfo{volume}{B359}},
  \bibinfo{pages}{343} (\bibinfo{year}{1991}).

\bibitem[{\citenamefont{Harlander and Kilgore}(2002)}]{Harlander:2002wh}
\bibinfo{author}{\bibfnamefont{R.~V.} \bibnamefont{Harlander}}
  \bibnamefont{and} \bibinfo{author}{\bibfnamefont{W.~B.}
  \bibnamefont{Kilgore}}, \bibinfo{journal}{Phys. Rev. Lett.}
  \textbf{\bibinfo{volume}{88}}, \bibinfo{pages}{201801}
  (\bibinfo{year}{2002}).

\bibitem[{\citenamefont{Campbell and Ellis}(1999)}]{Campbell:1999ah}
\bibinfo{author}{\bibfnamefont{J.~M.} \bibnamefont{Campbell}} \bibnamefont{and}
  \bibinfo{author}{\bibfnamefont{R.~K.} \bibnamefont{Ellis}},
  \bibinfo{journal}{Phys. Rev.} \textbf{\bibinfo{volume}{D60}},
  \bibinfo{pages}{113006} (\bibinfo{year}{1999}).

\bibitem[{\citenamefont{Cacciari et~al.}(2004)\citenamefont{Cacciari, Frixione,
  Mangano, Nason, and Ridolfi}}]{Cacciari:2003fi}
\bibinfo{author}{\bibfnamefont{M.}~\bibnamefont{Cacciari}},
  \bibinfo{author}{\bibfnamefont{S.}~\bibnamefont{Frixione}},
  \bibinfo{author}{\bibfnamefont{M.~L.} \bibnamefont{Mangano}},
  \bibinfo{author}{\bibfnamefont{P.}~\bibnamefont{Nason}}, \bibnamefont{and}
  \bibinfo{author}{\bibfnamefont{G.}~\bibnamefont{Ridolfi}},
  \bibinfo{journal}{JHEP} \textbf{\bibinfo{volume}{04}}, \bibinfo{pages}{068}
  (\bibinfo{year}{2004}).

\bibitem[{\citenamefont{Bonciani et~al.}(1998)\citenamefont{Bonciani, Catani,
  Mangano, and Nason}}]{Bonciani:1998vc}
\bibinfo{author}{\bibfnamefont{R.}~\bibnamefont{Bonciani}},
  \bibinfo{author}{\bibfnamefont{S.}~\bibnamefont{Catani}},
  \bibinfo{author}{\bibfnamefont{M.~L.} \bibnamefont{Mangano}},
  \bibnamefont{and} \bibinfo{author}{\bibfnamefont{P.}~\bibnamefont{Nason}},
  \bibinfo{journal}{Nucl. Phys.} \textbf{\bibinfo{volume}{B529}},
  \bibinfo{pages}{424} (\bibinfo{year}{1998}).

\bibitem[{\citenamefont{Catani et~al.}(1996)\citenamefont{Catani, Mangano,
  Nason, and Trentadue}}]{Catani:1996dj}
\bibinfo{author}{\bibfnamefont{S.}~\bibnamefont{Catani}},
  \bibinfo{author}{\bibfnamefont{M.~L.} \bibnamefont{Mangano}},
  \bibinfo{author}{\bibfnamefont{P.}~\bibnamefont{Nason}}, \bibnamefont{and}
  \bibinfo{author}{\bibfnamefont{L.}~\bibnamefont{Trentadue}},
  \bibinfo{journal}{Phys. Lett.} \textbf{\bibinfo{volume}{B378}},
  \bibinfo{pages}{329} (\bibinfo{year}{1996}).

\bibitem[{\citenamefont{Baur et~al.}(2002)\citenamefont{Baur, Brein, Hollik,
  Schappacher, and Wackeroth}}]{Baur:2001ze}
\bibinfo{author}{\bibfnamefont{U.}~\bibnamefont{Baur}},
  \bibinfo{author}{\bibfnamefont{O.}~\bibnamefont{Brein}},
  \bibinfo{author}{\bibfnamefont{W.}~\bibnamefont{Hollik}},
  \bibinfo{author}{\bibfnamefont{C.}~\bibnamefont{Schappacher}},
  \bibnamefont{and}
  \bibinfo{author}{\bibfnamefont{D.}~\bibnamefont{Wackeroth}},
  \bibinfo{journal}{Phys. Rev.} \textbf{\bibinfo{volume}{D65}},
  \bibinfo{pages}{033007} (\bibinfo{year}{2002}).

\bibitem[{\citenamefont{Ellis and Veseli}(1998)}]{Ellis:1997ii}
\bibinfo{author}{\bibfnamefont{R.~K.} \bibnamefont{Ellis}} \bibnamefont{and}
  \bibinfo{author}{\bibfnamefont{S.}~\bibnamefont{Veseli}},
  \bibinfo{journal}{Nucl. Phys.} \textbf{\bibinfo{volume}{B511}},
  \bibinfo{pages}{649} (\bibinfo{year}{1998}).

\bibitem[{\citenamefont{Ellis et~al.}(1997)\citenamefont{Ellis, Ross, and
  Veseli}}]{Ellis:1997sc}
\bibinfo{author}{\bibfnamefont{R.~K.} \bibnamefont{Ellis}},
  \bibinfo{author}{\bibfnamefont{D.~A.} \bibnamefont{Ross}}, \bibnamefont{and}
  \bibinfo{author}{\bibfnamefont{S.}~\bibnamefont{Veseli}},
  \bibinfo{journal}{Nucl. Phys.} \textbf{\bibinfo{volume}{B503}},
  \bibinfo{pages}{309} (\bibinfo{year}{1997}).

\bibitem[{\citenamefont{Cowan}(2002)}]{Cowan:2002in}
\bibinfo{author}{\bibfnamefont{G.}~\bibnamefont{Cowan}} (\bibinfo{year}{2002}),
  \bibinfo{note}{prepared for Conference on Advanced Statistical Techniques in
  Particle Physics, Durham, England, 18-22 Mar 2002}.

\bibitem[{\citenamefont{James and Roos}(1975)}]{James:1975dr}
\bibinfo{author}{\bibfnamefont{F.}~\bibnamefont{James}} \bibnamefont{and}
  \bibinfo{author}{\bibfnamefont{M.}~\bibnamefont{Roos}},
  \bibinfo{journal}{Comput. Phys. Commun.} \textbf{\bibinfo{volume}{10}},
  \bibinfo{pages}{343} (\bibinfo{year}{1975}).

\bibitem[{\citenamefont{Hagiwara et~al.}(2002)}]{Hagiwara:2002fs}
\bibinfo{author}{\bibfnamefont{K.}~\bibnamefont{Hagiwara}} \bibnamefont{et~al.}
  (\bibinfo{collaboration}{Particle Data Group}), \bibinfo{journal}{Phys. Rev.}
  \textbf{\bibinfo{volume}{D66}}, \bibinfo{pages}{010001}
  (\bibinfo{year}{2002}).

\bibitem[{LEP(2003)}]{LEP:2003ih}
 (\bibinfo{year}{2003}), \eprint{hep-ex/0312023}.

\end{thebibliography}

\end{document}